\documentclass[ejsv2,noshowframe]{imsart}

\RequirePackage[authoryear]{natbib}
\RequirePackage[colorlinks,citecolor=blue,urlcolor=blue]{hyperref}
\RequirePackage{graphicx}

\usepackage{microtype}
\usepackage{mathtools}
\usepackage{amsfonts}
\usepackage{bbm}
\usepackage{tikz}
\usepackage{float}
\usepackage{booktabs}
\usepackage{threeparttable}
\usepackage{array}

\arxiv{2507.11833}

\startlocaldefs
\graphicspath{ {./images/} }

\theoremstyle{plain}
\newtheorem{theorem}{Theorem}
\newtheorem{corollary}{Corollary}
\newtheorem{proposition}{Proposition}

\newcommand{\cov}{\operatorname{Cov}}
\newcommand{\cor}{\operatorname{Cor}}
\newcommand{\diag}{\operatorname{diag}}
\newcommand{\var}{\operatorname{Var}}
\newcommand{\mbe}{\mathbb{E}}

\newcommand{\given}{\, | \,}

\newcommand{\normal}{\operatorname{Normal}}

\newcommand{\dirichlet}{\operatorname{Dirichlet}}

\newcommand{\betadist}{\operatorname{Beta}}
\newcommand{\betaprime}{\operatorname{BetaPrime}}
\newcommand{\gammadist}{\operatorname{Gamma}}

\newcommand{\invgammadist}{\operatorname{InvGamma}}

\DeclareMathOperator*{\argmin}{arg\,min}

\newcommand{\rmse}{\text{RMSE}}

\newcommand{\elpd}{\text{ELPD}}

\definecolor{Javiercolour}{HTML}{7B3B1D}
\definecolor{Davidcolour}{HTML}{CCBB44}

\definecolor{revcolour}{HTML}{000000}

\newcommand{\mygithub}{\url{https://github.com/jear2412/GroupR2priors}}

\usetikzlibrary{bayesnet}
\usetikzlibrary{shapes,decorations,arrows,calc,arrows.meta,fit,positioning}
\tikzset{
    -Latex,auto,node distance =1 cm and 1 cm,semithick,
    state/.style ={circle, draw, minimum width = 1.0 cm},
    point/.style = {circle, draw, inner sep=0.04cm,fill,node contents={}},
    vmissing/.style={
    draw=none,
    scale=1,
    text height=0.111cm,
    execute at begin node=\color{black}$\vdots$
    },
    hmissing/.style={
    draw=none,
    scale=1,
    text height=0.111cm,
    execute at begin node=\color{black}$...$
    },
    bidirected/.style={Latex-Latex,dashed},
    el/.style = {inner sep=2pt, align=left, sloped}
}
\endlocaldefs

\begin{document}
\begin{frontmatter}
\title{R2 priors for Grouped Variance Decomposition in High-dimensional Regression}
\runtitle{Group-R2 priors}

\begin{aug}
\author[A]{\fnms{Javier Enrique}~\snm{Aguilar}\ead[label=e1]{javier.aguilarr@icloud.com}},
\author[B]{\fnms{David}~\snm{Kohns}},
\author[B]{\fnms{Aki}~\snm{Vehtari}}
\and
\author[A]{\fnms{Paul-Christian}~\snm{B\"urkner}}

\address[A]{Department of Statistics,
TU Dortmund University\printead[presep={,\ }]{e1}}

\address[B]{Department of Computer Science,
Aalto University}

\runauthor{Aguilar et al.}
\end{aug}

\begin{abstract}

We introduce the Group-R2 decomposition prior, a hierarchical shrinkage prior that extends R2-based priors to structured regression settings with known groups of predictors. By decomposing the prior distribution of the coefficient of determination $R^2$ in two stages, first across groups, then within groups, the prior enables interpretable control over model complexity and sparsity. We derive theoretical properties of the prior, including marginal distributions of coefficients, tail behavior, and connections to effective model complexity. We also prove posterior consistency when the number and the size of the groups grow with the sample size, and posterior contraction of the fitted mean when there are more predictors than observations. Through simulation studies, we evaluate the conditions under which grouping improves predictive performance and parameter recovery compared to priors that do not account for groups, and we compare the Group-R2 prior with a wide range of recent grouped shrinkage priors. Our results provide practical guidance for prior specification and highlight both the strengths and limitations of incorporating grouping into R2-based shrinkage priors.

\end{abstract}



\end{frontmatter}

\section{Introduction}

In regression problems involving high-dimensional or structured predictors, covariates often have natural groupings that reflect scientific knowledge. Examples include genes in the same biological pathway \citep{li_network_2008, sun_network-regularized_2014}, indicators of socioeconomic time-series \citep{mogliani_bayesian_2021,kohns2025flexible}, or survey items that measure the same latent construct \citep{bentler_significance_1980}. In such cases, incorporating group structure into the prior distribution can improve estimation, interpretation, and predictive performance. However, many existing group-aware priors are motivated by marginal shrinkage properties and offer limited control over interpretable model summaries, such as the proportion of variance explained.

The coefficient of determination, $R^2$, is one such interpretable quantity which summarizes how well the model fits the data and is considered to be understood in many applied fields \citep{gelman_bayesian_2013}. Shrinkage priors built around a prior distribution on $R^2$, such as the R2D2 prior and its extensions, allow users to express beliefs about model complexity and sparsity \citep{zhang_bayesian_2020, aguilar_intuitive_2023, Yanchenko-R2D2-glm, kohns_arr2_2024, aguilar_generalized_2025}. As R2-type priors start from defining the prior distribution on $R^2$, they are predictively consistent priors in the sense of \citet{riha2026select}. This means that the prior predictive distribution does not substantially change when more model components are added or model complexity increased. However, R2-based priors typically assume exchangeability across coefficients and do not consider structural information such as grouping. This can be suboptimal for parameter inference as well as prediction in structured regression settings where variance is expected to concentrate in specific subsets of predictors. 

In this work, we introduce the Group-R2 prior, a class of continuous shrinkage priors that generalizes the R2-based framework to explicitly account for known grouping structures among predictors. Our approach builds on the idea of decomposing the total prior variance implied by an $R^2$ distribution, and extends it through a two-stage hierarchical allocation: first, the total variance is distributed across groups, and then within each group across individual coefficients. This leads to a prior that allows for targeted control over sparsity and signal dispersion at both the group and coefficient levels.

We show that the Group-R2 prior retains the desirable properties of global-local shrinkage priors \citep{polson_local_2012}, including heavy-tailed marginals, concentration at zero, and robustness to large signals. We further prove posterior consistency of the Group-R2 prior when the number and the size of the groups grow with the sample size, without assuming sparsity of the true coefficients. When there are more predictors than observations, we show that the fitted mean contracts around the true mean. Such guarantees exist for the GIGG prior \citep{boss_group_2023}, the spike-and-slab group lasso \citep{bai_group_lassos_2022} and the R2D2 prior without groups \citep{zhang_bayesian_2020}. To our knowledge, Section~\ref{subsec:posterior_consistency} gives the first such results for an R2-based prior with grouped predictors. We additionally show how the shrinkage behavior of the prior depends on a small number of interpretable hyperparameters, and we propose  strategies for setting them based on theoretical and shrinkage properties. These influence a measure of implicit  model sizes, which the user can use to tune the prior. 

Many previous approaches to setting priors on known groups of coefficients have been similarly motivated from the global-local class of priors for exchangeable coefficients, conditional on hyperparameters \citep{raman_bayesian_2009,casella_penalized_2010,babacan_bayesian_2014,xu_bayesian_2015,xu_bayesian_2016,greenlaw_bayesian_2017,boss_group_2023,graspgroupedregressionadaptive2025}, or place discrete mass at zero for whole groups or single coefficients \citep{stingo_incorporating_2011,zhang_bayesian_2014,xu_bayesian_2015,chen_bayesian_2016,zhao_structured_2019,agarwal_bayesian_2025}. Like the group horseshoe \citep{xu_bayesian_2016} and GRASP \citep{graspgroupedregressionadaptive2025}, the Group Inverse-Gamma Gamma (GIGG) prior of \citet{boss_group_2023} belongs to the global-local family of priors where the prior consists of three scales: a global, common to all coefficients, a group-wise scale shared by all members of a group, and one individual to each coefficient. As we will show, under certain hyperparameter coupling, the GIGG and our proposed group prior can yield the same prior for the group-wise scale. However, a defining feature of the Group-R2 is that it further decomposes the group scale in order to control the $R^2$. More generally, none of these priors is derived from a prior on $R^2$. Hence their prior induced on $R^2$ depends on their hyperparameters in a way that may not be straightforward and changes as predictors are added, making them not predictively consistent. Lastly, while we were conducting this research, \citet{yanchenko2025groupr2d2shrinkageprior} independently proposed a related group R2D2 prior that also employs a hierarchical decomposition of explained variance across groups and within groups. Compared to them, we further derive prior properties, posterior consistency as well as new tools to calibrate the prior, which make it accessible in practice.

We study the Group-R2 prior in simulations from two angles. First, to understand whether and when grouping information improves inference, we compare grouped and nongrouped versions of the R2 prior across scenarios that vary in signal strength, sparsity pattern and dimensionality. This isolates the effect of the additional decomposition and shows when grouping helps and when it does not. Second, we compare the Group-R2 prior with a wide range of recent grouped priors, both continuous and with point masses at zero. For the GIGG, we calibrate its hyperparameters so that it implies the same prior on $R^2$ as the Group-R2 prior, so that the two differ only in how the explained variance is decomposed. 
These comparisons show that the Group-R2 prior is competitive with existing grouped priors while giving users direct control over the prior on $R^2$, the quantity that most influences its performance, and that it shrinks real signals less than those priors that allow exact zeros. Its posterior is also sampled more reliably than those of all competing priors.



\section{Methods}

In this section, we develop the Group-R2 prior, a hierarchical shrinkage prior that decomposes explained variance across and within predictor groups. We begin by detailing the construction of the prior in Section~\ref{subsec:group_r2_intro}. We introduce the concept of group-wise explained variance and related quantities in Section \ref{subsec:group_wise_r2}. We then derive key theoretical properties, including marginal and joint distributions, behavior near the origin and in the tails, and implications for shrinkage at both the group and coefficient levels in Section~\ref{subsec:marginal_dist}. We analyze the shrinkage behavior induced by the prior through the study of shrinkage factors and the effective number of non-zero coefficients in Section~\ref{subsec:shrinkage_properties}, and provide practical recommendations for hyperparameter specification in Section~\ref{subsec:hyperparameter_specification}. 

\subsection{Group-R2 Decomposition priors }
\label{subsec:group_r2_intro}

Our main interest lies in performing inference for the linear regression model in a situation in which the user already has knowledge about relevant groups of predictors. We will use the available group information to construct a multivariate continuous joint shrinkage prior that accounts for the grouping structures through the variance. Importantly, we do not assume the user has information about the dependency structure within the groups, only about which predictors belong to each mutually exclusive group \citep{nonparametric_bda}. 

Let $y$ denote the vector of $n$ observations $(y_1, \dots, y_n)'$, and let $X$ be the design matrix of dimension $n \times p$. Suppose that we have prior knowledge of $G$ groups. For each group $g=1, \dots, G$, let $X_g$ be the matrix of predictors for that group, of dimension $n \times p_g$, where $p_g$ denotes the column dimension of the $g$th group, and let the corresponding regression coefficients be $b_g = (b_{g1}, \dots, b_{g p_g})'$. The overall design matrix can be written as $X = (X_1 \dots X_G)$, with associated coefficient vector $b = (b_1' \dots b_G')'$. The linear regression model under consideration is
\begin{equation}
    \label{eq::linreg}
    y = b_0 J  + \sum_{g = 1}^G X_g b_g + \varepsilon,
\end{equation}
where $J$ is a vector of ones, $b_0 \in \mathbb{R}$ is the intercept and $\varepsilon = (\varepsilon_1 \ldots \varepsilon_n)'$ is such that $\varepsilon_i \sim \mathcal{N}(0, \sigma^2)$.

This model can be equivalently expressed observation-wise as
\begin{equation}
    \label{eq::linreg_2}
    y_i = b_0 + \sum_{g = 1}^G  \sum_{j = 1}^p  x_{ij} b_j \mathbf{1}_{\{ j \in \textrm{ group } g \}}  + \varepsilon_i, \ \ i = 1,..., n,
\end{equation}
 where $\mathbf{1}$ is the indicator function. Priors that decompose the proportion of explained variance $R^2$ first define a global $R^2$ measure, which is then related to the variance of the linear predictor $x'b$ via a one-to-one transformation. Consequently, probabilistic statements about $R^2$ directly translate to statements about the total variance. This total variance is subsequently decomposed to determine the scales of the coefficients $b$ using an appropriate distribution, such as a normal or double exponential, to express the variability attributed to these coefficients.

We begin by defining the global $R^2$ measure as
\begin{equation}
\label{eq::r2_def}
    R^2 \coloneqq \cor^2(y, x'b) = \frac{\var(x'b)}{\var(x'b)+\sigma^2}.
\end{equation}
In the following assume that $\mathbb{E}(x)=0$ and $\var(x)= \Sigma_x$, where $\Sigma_x$ has a diagonal of ones. We further assume the prior for $b$ satisfies $\mbe(b)=0$ and $\var(b)= \sigma^2\Lambda$, with $\Lambda$ being a diagonal matrix whose entries are $\lambda_1^2, \dots, \lambda_p^2$. Calculating the variance of the linear predictor yields $\var(x'b) = \sigma^2 \sum_{i=1}^p \lambda_i^2$. The quantity $\tau^2 \coloneqq \sum_{i=1}^p \lambda_i^2$ is denoted as the total variance. This implies the following one-to-one relationship between $R^2$ and $\tau^2$
\begin{equation}
\label{eq::tau_def}
    R^2 = \frac{\tau^2}{\tau^2+1}.
\end{equation}
Therefore probabilistic statements about the proportion of explained variance $R^2$ can be translated to the total variance $\tau^2$ and vice versa. To incorporate group structures, we further decompose $\Lambda$ into a block-diagonal matrix with entries $\Lambda_1, \dots, \Lambda_G$, where $\Lambda_g = \diag(\lambda_{g1}^2, \dots, \lambda_{g p_g}^2)$. We can rewrite the total variance $\tau^2$ as
\begin{equation}
    \tau^2 = \sum_{g=1}^G \sum_{l=1}^{p_g} \lambda_{gl}^2 = \sum_{g=1}^G \tau^2_{g}, 
\end{equation}
where $\tau_g^2 = \sum_{l=1}^{p_g} \lambda_{gl}^2  $ represents the group-wise total variance attributable to the $g$th group. We construct the Group-R2 prior through the following two-stage decomposition:

\paragraph{Stage I: Group variance decomposition.}
Decompose the total variance $\tau^2$ across groups using groupwise proportions $\phi = (\phi_1, \dots, \phi_G)$ such that $\phi_g \geq 0$ and $\sum_{g=1}^G \phi_g = 1$. The variance contribution of each group is then $\tau_g^2 = \phi_g \tau^2$.

\paragraph{Stage II: Within group decomposition.}

For each group $g$, decompose its total variance $\tau_g^2$ into the variances of its coefficients. Denote the within-group proportions by $\varphi_{g}= (\varphi_{g1}, \dots, \varphi_{gp_g})$, with $\varphi_{gl} \geq 0$ and $\sum_{l=1}^{p_g} \varphi_{gl} = 1$. The variances of the individual coefficients $b_{gl}$ are then given by
\begin{equation}
\var(b_{gl} \mid \varphi_g, \phi, \tau^2, \sigma^2)= \lambda_{gl}^2 \sigma^2= \varphi_{gl} \tau_g^2 \sigma^2 = \varphi_{gl} \phi_g \tau^2 \sigma^2, \ \ l= 1,...,p_g, \ \ g = 1, \dots, G.    
\end{equation}
To complete the specification of the joint prior, we assume that the $R^2$ prior follows a beta distribution with prior mean $\mu_{R^2}$ and precision $\nu_{R^2}$. This facilitates an intuitive incorporation of domain knowledge into the R2 prior, as $\mu_{R^2}$ and $\nu_{R^2}$ can be directly communicated with users. From Equation~\eqref{eq::tau_def}, it follows that $\tau^2$ is distributed according to a Beta Prime distribution with shape parameters $a_1, a_2$ \citep{bai_beta_2021}, where $(a_1, a_2)$ denote the canonical parameters of the Beta distribution. Their relationship between the mean and precision is $\mu_{R^2} = a_1/(a_1 + a_2)$ and $\nu_{R^2} = a_1 + a_2$. We will make use of both parametrizations in the paper. 
\begin{figure}[t!]
    \centering
    \includegraphics[width=1\linewidth]{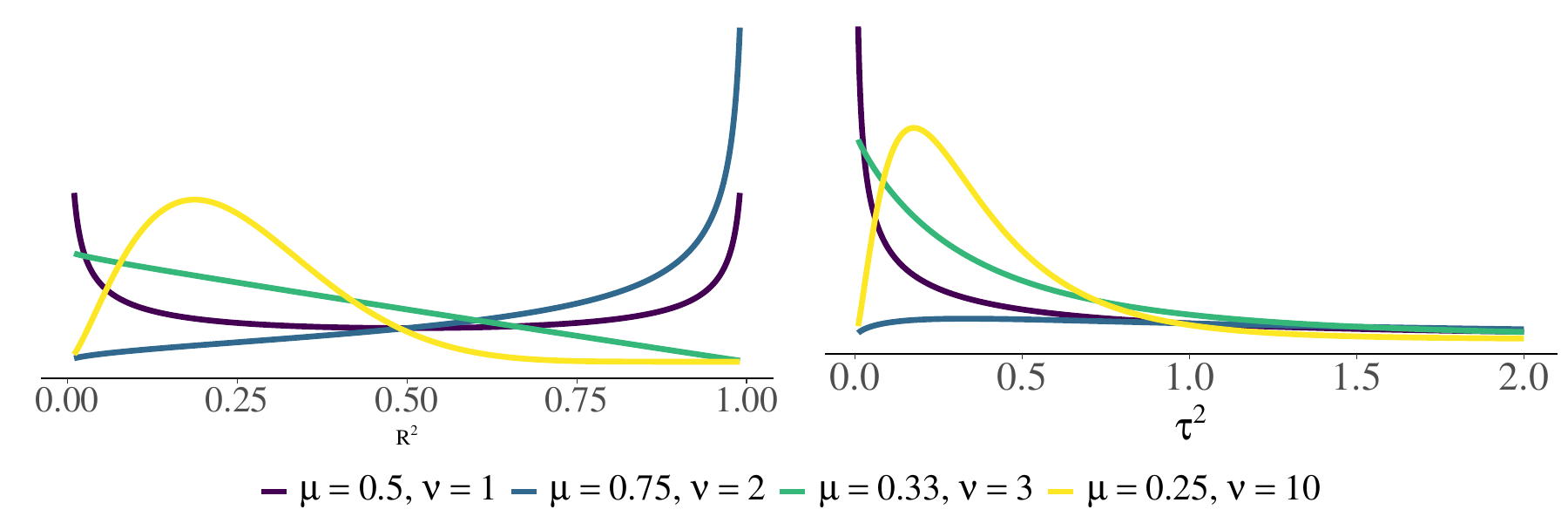}
    \vspace{-0.5cm}  
    \caption{Beta and Beta Prime densities for several values of the prior mean $\mu_{R^2}$ and precision $\nu_{R^2}$. }
    \label{fig:betapriors}
\end{figure}

Figure~\ref{fig:betapriors} illustrates the flexibility of Beta distributions in encoding prior beliefs about $R^2$. It also shows the implied Beta Prime prior on $\tau^2$. When the parameters are set to $(\mu_{R^2}, \nu_{R^2}) = (0.5, 1)$, the resulting prior on $R^2$ is bathtub-shaped, placing most of its mass near 0 and 1. This reflects a belief that the model is likely to explain either very little or nearly all of the variance—indicating an expectation of either predominantly noise or strong signal. 

The proportions of group variance $\phi$ and within-group variance $\varphi_g$ lie on the simplex. The canonical distribution for the simplex is the Dirichlet distribution, parameterized by the concentration vector, $\alpha$ \citep{bhattacharya_dirichletlaplace_2015, lin_dirichlet_2016}. This vector, $\alpha$, fully determines the mean and correlation structure of the Dirichlet distribution. A parsimonious choice within this family is the symmetric Dirichlet distribution, where $\alpha = (a_\pi, \ldots, a_\pi)$ with $a_\pi > 0$. We denote this distribution as $\phi \sim \dirichlet s(a_\pi)$, where  $s(\cdot)$ emphasizes the symmetry of the concentration parameters. This choice greatly simplifies the number of hyperparameters to specify and facilitates the development of theoretical properties of the prior.  \footnote{For an alternative prior distribution for the simplex, see \citet{aguilar_generalized_2025} who use the more flexible, logistic normal distribution. We leave further theoretical investigations with the logistic normal for future research.}


To fully specify the joint hierarchical prior, we require a kernel for the coefficients $b_{gl}$. Consistent with the approaches of \cite{aguilar_intuitive_2023, aguilar_generalized_2025} and \cite{kohns_arr2_2024}, we model $b_{gl}$ as normally distributed with conditional variance $\var(b_{gl} \mid \lambda_{gl}^2, \sigma^2) = \sigma^2 \lambda_{gl}^2$.\footnote{An alternative is presented in \citet{zhang_bayesian_2020}, who use a Laplace kernel to induce strong marginal concentration around zero. However, \cite{logscale_priors} show that the asymptotic properties of this choice are equivalent to those obtained using a normal kernel.} The overall procedure to construct the Group-R2 prior is illustrated in Figure~\ref{fig:group-r2-construction}.

We place a half Student-$t$ prior on $\sigma$ with $\eta_\sigma$ degrees of freedom and scale parameter $\nu_\sigma$, whose mean, when $\eta_\sigma > 1$ is proportional to $\nu_\sigma$, and whose variance is proportional to $\nu_\sigma^2$ when $\eta_\sigma>2$ \citep{gelman_prior_2006, goodrich_rstanarm_2020}. If a Gibbs sampler is preferred, an inverse gamma prior for $\sigma^2$ may be used as a computationally convenient alternative. The intercept $b_0$ is typically assigned either a normal prior or an improper uniform prior to ensure flexibility in modeling the overall mean.
    
The complete Group-R2 prior has the following form: 
\begin{equation}
\label{eq:group_r2_prior}
\begin{aligned}
    b_{gl} \mid \varphi_g, \phi, \tau^2, \sigma^2 &\sim \normal \left(0, \varphi_{gl} \phi_g \tau^2 \sigma^2 \right), \\
    \phi &\sim \dirichlet(\cdot), \qquad \varphi_g \sim \dirichlet(\cdot), \\
    R^2 &\sim \betadist(a_1, a_2), \qquad \tau^2 = \frac{R^2}{1-R^2}, \qquad \sigma \sim \pi(\sigma),
\end{aligned}
\end{equation}
where $a_1 = \mu_{R^2} \nu_{R^2}$ and $a_2 = (1-\mu_{R^2}) \nu_{R^2}$. Equivalently, $\tau^2 \sim \betaprime(a_1, a_2)$.

\begin{figure}[t!]%
    \centering
\begin{tikzpicture}[]
  \node[latent]                  (b)      {$b_{gl}$} ; %
  \node[latent, above=of b, xshift= -1cm]    (tau2) {$\tau^2$};
  \node[latent, above=of b, xshift= 1cm] (phi)  {$\phi$}; %
  \node[latent, right=of b, xshift= 1cm, yshift = 0cm] (varphig)  {$\varphi_g$}; 
  \node[latent, left=of b, xshift= -1cm, yshift = 0cm] (sigma)  {$\sigma$}; 
  
  \edge{varphig}  {b};
  \edge{tau2} {b};
  \edge {phi}  {b} ; 
  \edge {sigma}  {b} ; 

  
  \node[const, above=of tau2, xshift= 1cm] (a1tau2) {$a_1$};
  \node[const, above=of tau2, xshift= -1cm] (a2tau2) {$a_2$};

  \node[const, above=of phi] (nuphi) {$a_G$};

  \node[const, right=of varphig] (nuvarphig) {$c_g$};

  \node[const, left=of sigma] (nusigma) {$\nu_{\sigma}$};

  \edge {a1tau2}  {tau2} ; %
  \edge {a2tau2}  {tau2} ; %
  \edge {nuphi}  {phi} ; %
  \edge {nuvarphig}  {varphig} ; %
  \edge {nusigma}  {sigma} ; %
  
  \plate {plate1} { %
    (b) %
  } {$\forall l, 1 \leq l  \leq p_g$}; %
  
  \plate {} { %
    (plate1) %
    (nuvarphig)
    (varphig)
  } {$\forall g, 1 \leq g  \leq G$}; %
\end{tikzpicture}
  \caption{ Directed acyclic graph (DAG) representing the prior structure of the Group-R2 decomposition prior. A global scale parameter $\tau^2$, governed by hyperparameters $a_1$ and $a_2$, controls the total variance allocated across groups. The vector $\phi$ represents a group-level variance decomposition, drawn from a distribution over the simplex parameterized by $a_G$. For each group $g$, a within-group allocation vector $\varphi_g$ is drawn from another simplex distribution with parameter $c_g$. Each coefficient $b_{gl}$ inherits its prior variance through the product $\varphi_{gl} \phi_g  \tau^2 \sigma^2$. Plates indicate replication over groups ($g = 1,\dots,G$) and coefficients within groups ($l = 1,\dots,p_g$). This illustrates how variance is hierarchically allocated across and within groups.}
  \label{fig:group-r2-construction}
\end{figure}%


\subsection{Group-wise explained variances}
\label{subsec:group_wise_r2}

In the following we introduce and characterize group-wise explained variances $R^2_g$, which quantify the share of total explained variance attributable to each predictor group. This helps interpret how the Group-R2 prior distributes variance across groups, complementing the global $R^2$ perspective. We derive properties of $R^2_g$ and related quantities, such as the marginal distributions of group variances $\tau_g^2$ and coefficient specific variances $\lambda_{gl}^2$. These results will help us in our discussion on theory, shrinkage behavior, and hyperparameter specification shown thereafter.

Under the assumptions presented in Section \ref{subsec:group_r2_intro}, we define a group-specific explained variance metric $R^2_g$ as:
\begin{equation}
\label{eq:r2_pergroup_def}
R^2_g = \frac{\var(x_g'b_g) }{ \var(x'b) + \sigma^2 } = \frac{\tau_g^2 }{\tau^2+1}, \ \ g=1, \ldots, G.
\end{equation}
Equivalently,
\begin{equation}
R^2_g = \frac{\tau_g^2}{\tau^2 + 1} = \frac{\phi_g \tau^2}{\tau^2 + 1} = \phi_g R^2.
\end{equation}
The quantity $R^2_g$ can be interpreted as the proportion of the total variance explained by the predictors in the $g$th group. Proposition~\ref{prop:properties_r2g} characterizes the joint distribution of $R^2_g$ and shows that different group contributions $R^2_{g}, R^2_{h}, h \neq g$ will exhibit negative dependencies conditional on $R^2$. Even though we could expect this since $\sum_{g=1}^G R^2_g = R^2$, mixing over random totals can change signs of the covariance \citep{kruijer_adaptive_2010}. All proofs can be found in the Appendix. 

\begin{proposition}(Joint distribution of $R^2_g$)
\label{prop:properties_r2g}
If $\phi \sim \dirichlet(\alpha)$ is independent of $R^2$, with $G \geq 2$ then the following holds.
\begin{enumerate}
    \item The joint distribution of $(R^2_1/ R^2, \dots,  R^2_G/ R^2 ) = \phi \sim \dirichlet(\alpha)$.
    \item $R^2_g/ R^2 = \phi_g \sim \betadist \left(  \alpha_g, \sum_{j \neq g} \alpha_j\right)$.
    \item The conditional distribution $\left( R_1^2, \ldots, R_G^2 \right) \mid (R^2 = r^2) \sim r^2 \dirichlet(\alpha)$. Given $g \neq h$ then   
\begin{equation}
\cov\left( R_g^2, R_h^2  \mid R^2 = r^2 \right)  = (r^2)^2 \cov(\phi_g, \phi_h) = - (r^2)^2  \frac{\alpha_g \alpha_h}{\alpha_0^2 (\alpha_0 + 1)}, 
\end{equation}

    where $\alpha_0 = \sum_{i=1}^G \alpha_i$.
\end{enumerate}
\end{proposition}

When $\phi \sim \dirichlet(\alpha)$, the distribution of $R^2_g = \phi_g R^2$ corresponds to the product of two Beta-distributed random variables. A closed-form expression exists, however it is typically too cumbersome to be useful in practice, so we do not present it here. Its full derivation can be found in \cite{coelho_distribution_2021}. However, the $k$th moment of $R^2_g$ can be found under a closed form and the set of them will completely determine their distribution \citep{schmudgen_ten_2020}: 

\begin{equation}
\label{eq:moments_r2_g}
    \begin{aligned}
        \mathbb{E}[(R^2_g)^k] = \mathbb{E}[(\phi_g)^k] \mathbb{E}[(R^2)^k] = \left( \prod_{i=0}^{k-1} \frac{\alpha_g + i}{ \alpha_0 + i} \right) \left(\prod_{j=0}^{k-1} \frac{a_1 + j}{a_1 + a_2 + j} \right), \ \ k \in \mathbb{N}.
    \end{aligned}
\end{equation}

For instance if $\phi_g \sim \betadist(\alpha_g, 1)$ and independently $ R^2 \sim \betadist(\alpha_g + 1/2, 1)$, then $R_g^2$ has the distribution of $Z^2$ with $Z \sim \betadist(2\alpha_g, 2)$, as follows by comparing the moments above \citep{schmudgen_ten_2020}.

We characterize the marginal distributions of the total variance of the $g$th group $\tau^2_g$ and the variances $\lambda_{gl}^2$ in Proposition~\ref{prop:properties_lambda_gl}. These results will be used in Section \ref{subsec:marginal_dist} to establish properties of our prior. Additionally, it will be useful for hyperparameter specification in Section~\ref{subsec:hyperparameter_specification}. 
\begin{proposition}(Marginals of $\tau^2_g$ and $\lambda_{gl}^2$)
\label{prop:properties_lambda_gl}
Let $\tau^2 \sim \betaprime(G a_G, a_2)$, $\phi \sim \dirichlet s(a_G)$, and $\varphi_g \sim \dirichlet s(c_g)$ be independent, then:
\begin{enumerate}
    \item The group-wise total variance $\tau_g^2 = \phi_g \tau^2 \sim \betaprime(a_G, a_2)$ marginally. The $\tau_g^2$ are independent Gamma variables conditionally on a common rate $\xi \sim \gammadist(a_2, 1)$, and hence marginally dependent.
    \item If $p_g c_g = a_G$, then the coefficient-specific variances $\lambda_{gl}^2 = \varphi_{gl} \tau^2_g \sim \betaprime(c_g, a_2)$. 
\end{enumerate}
\end{proposition}
Proposition~\ref{prop:properties_lambda_gl} demonstrates that aligning the concentration parameters across the two decomposition levels yields a closed-form expression for the marginal distribution of the coefficient specific variances $\lambda_{gl}^2$. Under this formulation, $c_g \leq a_G$, indicating that the within-group shrinkage will be stronger than the group-level shrinkage \citep{aguilar_generalized_2025}. Note that specifying the second-level concentration parameter $c_g$ becomes immediate once the first-level parameter $a_G$ is chosen. In other words, if we have an idea of the desired group-level shrinkage, the corresponding coefficient-level shrinkage follows directly. Conversely, if we instead have information about the within-group concentration parameters $c_g$, then $a_G$ satisfies the restriction as long as all products $p_g c_g$ are equal. Alternatively, an asymmetric group Dirichlet with shapes $\alpha_g = p_g c_g$ and $a_1= \sum_g \alpha_g$ yields the analogous marginal identities.

\begin{proposition}[Flat Dirichlets]
\label{prop:flat_reduction}
If $\phi\sim\dirichlet(a_1^*,\ldots,a_G^*)$ and independently $\varphi_g\sim\dirichlet(c_{g1},\ldots,c_{gp_g})$, with $a_g^*=\sum_lc_{gl}$, then the concatenated weights $(\phi_g\varphi_{gl})_{g,l}$ have a single Dirichlet distribution with coordinate shapes $(c_{gl})_{g,l}$.
\end{proposition}

Therefore setting $a_G = p_g c_g$ produces a flat Dirichlet allocation vector. If the groups have the same size with a common $c_g$, the prior treats the partitions equally. The general construction can express additional group dependencies by choosing the two allocation concentrations separately, which we show in the following. 


\begin{figure}[t]
    \centering
    \includegraphics[width=\linewidth]{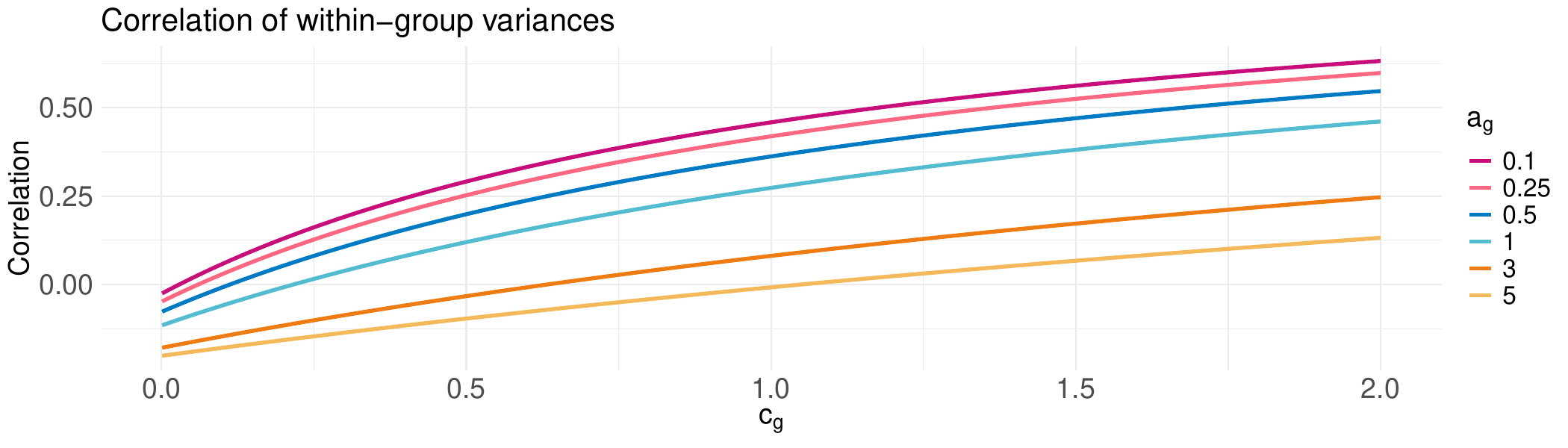}
   \caption[Correlation of within-group variance allocations]{Correlation between within-group variance allocations $\phi_g \varphi_{gj}$ and $\phi_g \varphi_{gk}$ $(j\neq k)$ as a function of the within-group concentration $c_g$, for different values of the group-level concentration $a_G$. Positive correlations correspond to more distributed signal within a group, whereas negative correlations correspond to more concentrated signal. Results are shown for $G=25$ groups with $p_g=5$ covariates per group.}
    \label{fig:log_cor_analytical}
\end{figure}

Proposition~\ref{prop:covariance_lambdas} characterizes the dependence structure induced by the Group-R2 prior on the marginal variance proportions $\phi_g \varphi_{gj}$. In particular, it quantifies both the covariance and correlation of their logarithms, $\log(\phi_g \varphi_{gj})$ and $\log(\phi_g \varphi_{gk})$, for $j \ne k$. These log-scale quantities have been highlighted by \cite{logscale_priors} and \cite{graspgroupedregressionadaptive2025} as useful to understand the concentration and tail behavior of shrinkage priors. We extend the results of \citet{graspgroupedregressionadaptive2025} to the case in which dependencies are considered for the within-group scales, allowing us to capture richer covariance structures than priors with independent scales. 
\begin{proposition}[Within-group variance dependencies]
\label{prop:covariance_lambdas}
Assume that $\phi$ and $\varphi_g$ are independent, $\phi$ is symmetric, and $p_g \geq 2$. Let $\phi_g$ denote the proportion of total variance allocated to group $g$, and let $\varphi_g = (\varphi_{g1}, \dots, \varphi_{gp_g})$ denote the within-group allocation proportions for the $p_g$ coefficients in group $g$. Then:
\begin{enumerate}
    \item The correlation between the log-scaled variances of two coefficients within the same group satisfies
\begin{equation}
    \cor \left( \log(\phi_g \varphi_{gj}),\ \log(\phi_g \varphi_{gk}) \right)
    = \frac{ \var(\log \phi_g) + \cov(\log \varphi_{gj}, \log \varphi_{gk}) }{ \var(\log \phi_g) + \var(\log \varphi_{g\cdot}) }.
\end{equation}
    \item Suppose $\tau^2 \sim \betaprime(G a_G, a_2)$, $\phi \sim \dirichlet s(a_G)$, and $\varphi_g \sim \dirichlet s(c_g)$ independently. Then:
    \begin{itemize}
        \item The correlation of log-variances is given by
        \begin{equation}
        \cor \left( \log(\phi_g \varphi_{gj}),\ \log(\phi_g \varphi_{gk}) \right) = \frac{ \psi_1(a_G) - \psi_1(G a_G) - \psi_1(p_g c_g) }{ \psi_1(a_G) - \psi_1(G a_G) + \psi_1(c_g) - \psi_1(p_g c_g) },
        \end{equation}
        where $\psi_1(x) = d^2 \log \Gamma(x) / dx^2$ is the trigamma function and $\Gamma(x)$ is the Gamma function \citep{olver_nist_2010}. 
        \item The covariance of the raw allocation proportions is
        \begin{equation}
        \cov \left( \phi_g \varphi_{gj},\ \phi_g \varphi_{gk} \right)
        = \frac{a_G(a_G + 1)}{G a_G (G a_G + 1)} \cdot \frac{c_g}{p_g(c_g p_g + 1)} - \frac{1}{G^2 p_g^2}.
        \end{equation}
       \end{itemize}
\end{enumerate}
\end{proposition}

Our framework introduces structured dependencies among the variances assigned to coefficients by using Dirichlet priors for both $\phi$ and $\varphi_g$. The second decomposition stage, parameterized by $c_g$, plays a key role in controlling these within-group correlations. As shown in Figure~\ref{fig:log_cor_analytical}, the correlation between $\phi_g \varphi_{gj}$ and $\phi_g \varphi_{gk}$ increases with $c_g$, with the speed of this increase controlled by the group-level parameter $a_G$. Notably, for small values of $c_g$ the correlation can be negative. This cannot happen under GIGG or GRASP \citep{graspgroupedregressionadaptive2025}, where the group scales and the local scales are conditionally independent given the global scale, so that the correlation of the log-variances within a group equals $\var(\log \gamma_g^2)/\{\var(\log \gamma_g^2) + \var(\log \lambda_{gl}^2)\} \geq 0$. In the Group-R2 prior the scales are exchangeable but not independent, since the within-group proportions are normalized.

\subsection{Marginal and joint distributions}
\label{subsec:marginal_dist}

In this section, we explore the marginal distribution of the coefficients $b_{gl}$ and the joint distribution of the coefficients within a group when $\tau^2_g$ is specified. We demonstrate that the marginal coefficients exhibit polynomial tails and high concentration near the origin. Furthermore, we show that the joint distribution of the coefficients within each group displays a pronounced spike at zero. We also show that our prior can represent the horseshoe prior as a special case \citep{carvalho_horseshoe_2010}, hence expecting it to inherit the desirable properties from the latter. 

For simplicity, we set $\sigma^2 = 1$ in the following derivations, although we note that all results are conditional on $\sigma^2$, consistent with common practice in the analysis of marginal distributions in shrinkage priors \citep{bhattacharya_dirichletlaplace_2015,van_der_pas_theoretical_2021,bai_beta_2021}. We denote the Beta and Gamma functions by $\text{B}(\cdot, \cdot)$ and $\Gamma(\cdot)$, respectively, and use $U(\eta, \nu, z)$ to represent the confluent hypergeometric function of the second kind \citep{Zwillinger, olver_nist_2010}. In the following we assume that $a_G > 0$, $R^2 \sim \betadist(G a_G, a_2)$, $\phi \sim \dirichlet s(a_G)$, and $\varphi_g \sim \dirichlet s(c_g)$ with $c_g = a_G / p_g$.


\begin{proposition}[Marginal distributions of $b_{gl}$]
\label{prop:marginalpriors}
The marginal prior density of $b_{gl}$ for any $g = 1, \dots, G$ and $l = 1, \dots, p_g$ is
\begin{equation}
    p(b_{gl}) = \frac{1}{\sqrt{2\pi} \, \operatorname{B}(c_g, a_2)} \Gamma(\eta) U(\eta, \nu_g, z_{gl}),
\end{equation}
where $\eta = a_2 + 1/2$, $\nu_g = 3/2 - c_g$, and $z_{gl} = |b_{gl}|^2/2$. 
\end{proposition}

\begin{proposition}[Marginal distributions of $b_{gl}$ near the origin]
\label{prop:originprior}
As $|b_{gl}| \to 0$, with $0 < c_g \leq 1/2$ and $a_2 > 0$ the marginal prior densities are unbounded and exhibit a singularity at zero. Specifically, they satisfy
\begin{equation}
    p(b_{gl}) \sim 
    \begin{cases}
        c_1 \, |b_{gl}|^{2c_g - 1} + \mathcal{O}\left( |b_{gl}|^{2 c_g + 1} \right), & c_g < \tfrac{1}{2}, \\
        -c_2 \, \ln (b_{gl}^2) + \mathcal{O}\left( b_{gl}^2 \ln (b_{gl}^2) \right), & c_g = \tfrac{1}{2},
    \end{cases}
\end{equation}
where $c_1, c_2 > 0$. Conversely, when $c_g > \tfrac{1}{2}$, the marginal prior densities are bounded and continuous at zero. Furthermore, they are differentiable at $b_{gl} = 0$ for all $c_g > 1$.
\end{proposition}

Proposition~\ref{prop:originprior} shows that the interplay between the two levels of decomposition directly governs the behavior of the marginal priors near the origin, which in turn determines the amount of shrinkage imposed on the coefficients. Specifically, since the coupling used in the proposition implies $c_g \leq a_G$, it suggests that the within-group shrinkage is stronger than the group-level shrinkage. This ensures sufficient prior mass near zero, a crucial property for shrinkage priors to effectively detect and suppress irrelevant signals \citep{pas_horseshoe_2014, pas_uncertainty_2017, van_der_pas_theoretical_2021}.

As $c_g$ decreases from $1/2$ to zero, the prior shifts an increasing amount of mass toward the origin, resulting in posterior distributions that provide sparse estimates of the coefficient vector $b$. Conversely, when $c_g$ exceeds $1/2$, the marginal priors become bounded at zero, favoring less sparse coefficient estimates. Thus, the parameter $c_g$ offers a flexible mechanism for translating prior beliefs about sparsity (or the lack thereof) into the prior model, controlling the degree of shrinkage applied to the coefficients.

The tails of the prior play a crucial role in determining whether large signals can be detected and how strongly these signals are pulled back toward zero \citep{armagan_posterior_2013,pas_conditions_2016, van_der_pas_theoretical_2021}. Ideally, priors with heavy tails are favored because they protect substantial signals from being overly shrunk \citep{carvalho_horseshoe_2010, piironen_sparsity_2017}. This results in priors of bounded influence, meaning that truly large coefficients remain relatively unshrunken by the prior, i.e., if $X = I$, then $\mbe[b_i \mid y_i] \approx  y_i $ when $y_i$ is greater than a threshold. This phenomenon, also known as tail robustness \citep{carvalho_horseshoe_2010, van_der_pas_theoretical_2021}, is essential in sparse settings to selectively shrink only the smallest coefficients while leaving larger, genuine signals largely untouched. We show in Proposition~\ref{prop:tailprior} that the Group-R2 prior attains heavier tails than the Cauchy distribution for $a_2 < 1/2$ and then establish in Proposition~\ref{prop:r2d2-horseshoe} that it can represent the horseshoe prior and is therefore of bounded influence. 

\begin{proposition}[Tails of the marginal prior]
\label{prop:tailprior}
As $|b_{gl}| \to \infty$, the marginal prior density of $b_{gl}$ satisfies
\begin{equation}
    p(b_{gl}) \sim \mathcal{O}\left( \frac{1}{|b_{gl}|^{2a_2 + 1}} \right).
\end{equation}
In particular, when $0 < a_2 < 1/2 $, the marginal priors have heavier tails than the Cauchy distribution and have the same polynomial order when $a_2 = 1/2$.
\end{proposition}

\begin{proposition}[Horseshoe type behavior]
\label{prop:r2d2-horseshoe}
If $c_g = a_2 = \tfrac{1}{2}$, then $\lambda_{gl} \sim \textrm{Cauchy}^+(0,1)$. The marginal distribution of $b_{gl}$ is the horseshoe marginal with global scale one. This implies that the Group-R2 prior has bounded influence.
\end{proposition}

The following proposition shows that we can still achieve poles at zero and Cauchy-order tails even without the coupling. However, this does not lead to exact horseshoe marginals in general.

\begin{proposition}[Marginal behavior without coupling]
\label{prop:uncoupled_marginals}
Assume $a_1=Ga_G$, $a_G,a_2,c_g>0$, and fix $\sigma>0$. For $p_g>1$, the density of $b_{gl}\mid\sigma$ is finite at zero if and only if both $a_G>1/2$ and $c_g>1/2$, otherwise it diverges as $|b_{gl}|\to0$. Moreover,
\begin{equation}
    p(b_{gl} )\sim \frac{2^{a_2}\Gamma(a_2+1/2) \mbe(\varphi_{gl}^{a_2})} {\sqrt\pi B(a_G,a_2)}|b_{gl}|^{-2a_2-1}.
\end{equation}
For a singleton group, $\varphi_{g1}=1$ and finiteness at the origin requires only $a_G>1/2$.
\end{proposition}

We can find the joint distribution of the coefficients within a group by integrating over the group variance parameter $\tau_g^2 = \phi_g \tau^2$. This is a similar approach to the idea presented by \cite{boss_group_2023}, where integrating out variance parameters leads to conditional densities for grouped coefficients.

\begin{proposition}[Joint distribution of group coefficients]
\label{prop:jointdist_group}
Assume that $\tau_g^2 \sim \betaprime(a_G,a_2)$ independently of $\varphi_g$. Conditional on $\tau_g^2, \varphi_g$, and $\sigma$, let 
\begin{equation}
b_g \mid \tau_g^2,\varphi_g  \sim \normal_{p_g} \left( 0, \tau_g^2 \operatorname{diag}(\varphi_{g1},\ldots,\varphi_{gp_g})\right).  
\end{equation}
Then the joint density of the coefficients in group $g$, conditional on $\varphi_g$, is
\begin{equation}
\label{eq:jointdist_group}
p(b_g\mid\varphi_g) =\frac{(2\pi)^{-p_g/2}}{B(a_G,a_2)}
\left(\prod_l\varphi_{gl}\right)^{-1/2}
\Gamma(\eta_g)U(\eta_g,\nu_g,z_g),
\end{equation}
where $\eta_g =a_2+p_g/2, \nu_g=1+p_g/2-a_G, z_g=1/2\sum_l b_{gl}^2/\varphi_{gl}$. Moreover, if $\varphi_g$ is in the interior of the simplex, the joint density diverges at the group origin whenever $ a_G < p_g/2$. When $a_G=p_g/2$, the divergence is logarithmic, and is finite when $a_G>p_g/2$.
\end{proposition}

The divergence of the joint density at zero for $c_g \leq 1/2$ indicates that the prior allocates substantial mass near zero, resulting in an infinite spike at the origin. This implies that we are able to shrink entire groups of coefficients if they are deemed as noise. Notably, when $p_g = 1$, the group consists of a single coefficient, and we recover the marginal prior distributions presented in Proposition~\ref{prop:marginalpriors}. Equation~\eqref{eq:jointdist_group} does not exhibit a traditional dependence structure, at least not one that is immediately apparent from its analytical form. To properly analyze the dependency structures implied by this joint prior, one should move beyond conventional multivariate dependency metrics and consider alternative representations, such as the implied copula.


\subsection{Shrinkage properties}
\label{subsec:shrinkage_properties}

 In this section, we examine how the prior influences shrinkage at both the coefficient and group levels. We first analyze the distribution of group shrinkage factors $\kappa_{g}$ and marginal shrinkage factors $\kappa_{gl}$, which determine how much each group is jointly regularized and how much each coefficient is pulled toward zero under the normal means model. We then study the effective number of nonzero coefficients, $m_{\text{eff}}$, as a global measure of sparsity. These results show how hyperparameter choices translate into different regularization patterns.

\subsubsection{Shrinkage Factors}
\label{subsubsec:shrinkage_factors}

As is common in the shrinkage literature, we can characterize the effect of the prior through the properties of the posterior distribution. In particular, analyzing the posterior mean and variance of each coefficient $b_{gl}$ provides insight into how the prior induces shrinkage. Assume the normal means setting, where $X = I$ and $p = n$. Then the posterior distribution of $b_{gl}$ given $y_{gl}$ and its prior variance $\lambda_{gl}^2 \sigma^2 = \varphi_{gl} \phi_g \tau^2 \sigma^2$ is:
\begin{equation}
\label{eq:posterior_normal_means}
b_{gl} \mid y_{gl}, \lambda_{gl}^2 \sim \normal \left( \frac{\lambda_{gl}^2}{1+\lambda_{gl}^2} y_{gl}, \frac{\lambda_{gl}^2}{1+\lambda_{gl}^2} \sigma^2 \right) .
\end{equation}
In this context, the maximum likelihood estimate (MLE) of $b_{gl}$ is simply $y_{gl}$ \citep{castillo_needles_2012}. The quantity $\kappa_{gl} = \tfrac{1}{1 + \lambda_{gl}^2} \in (0,1)$ measures the degree of shrinkage imposed on the MLE. When the variance $\lambda_{gl}^2$ is large, little shrinkage is applied and $\kappa_{gl} \to 0$, whereas small variances lead to stronger shrinkage with $\kappa_{gl} \to 1$. The posterior mean of $b_{gl}$ is directly related to $\kappa_{gl}$ via
\begin{equation}
\label{eq:posterior_normal_means_expectation}
\mbe \left(  b_{gl} \mid y_{gl} \right) = \left( 1- \mbe (\kappa_{gl} \mid y_{gl}) \right) y_{gl} .
\end{equation}
Hence, studying the distribution of $\kappa_{gl}$ is key to understanding when the posterior of $b_{gl}$ will be centered near zero or not. This connection makes $\kappa_{gl}$ a useful diagnostic for assessing effective shrinkage, especially in hierarchical or group-structured priors \citep{castillo_bayesian_2015,pas_horseshoe_2014, pas_conditions_2016, zhang_variable_2018, bai_large-scale_2019}.

Let $\kappa_g = (\kappa_{g1}, \dots, \kappa_{gp_g})'$ denote the vector of shrinkage factors for the $g$th group. Assume $\sigma^2 = 1$ and $R^2 \sim \betadist(G a_G, a_2)$  such that $\tau_g^2 = \phi_g \tau^2 \sim \betaprime(a_G, a_2)$. We use the parametrization $\lambda_{gl}^2 = \varphi_{gl} \tau_g^2$, with $\varphi_g = (\varphi_{g1}, \dots, \varphi_{gp_g})$ distributed as a symmetric Dirichlet with parameter $c_g$.

Note that the local scale of each coefficient is bounded above by the group scale, since $\varphi_{gl} \in (0,1)$ and $\sum_{l=1}^{p_g} \varphi_{gl} = 1$ implies $\lambda_{gl}^2 \leq \tau_g^2$. As a consequence, even when $\tau_g^2$ is moderately large, some $\lambda_{gl}^2$ may remain small due to the Dirichlet structure, leading to persistent shrinkage for certain coefficients. In particular, if $\tau_g^2$ is small, all $\lambda_{gl}^2$ will be small as well, and the group will experience strong shrinkage uniformly across its elements.

Conditional on $\tau_g^2 = t$, the vector $\kappa_g$ lies on
\begin{equation}
    \mathcal{K}(t) = \left\lbrace \kappa_g \in (0,1)^{p_g} \mid \sum_{l=1}^{p_g} \tfrac{1-\kappa_{gl}}{\kappa_{gl} } = t \right\rbrace .
\end{equation}
The shrinkage vector $\kappa_g$ is the pushforward distribution of Dirichlet vector $\varphi_g$ and has no density with respect to the Lebesgue measure on the cube $(0,1)^{p_g}$ conditionally on $\tau^2_g$. For a singleton group $\kappa_{gl} = (1+t)^{-1}$ deterministically.  For $p_g \geq 2$ the moments are given by 
\begin{equation}
\label{eq:moments_kappagl}
\mbe\left( \kappa_{gl}^m \mid t \right) = {}_2F_1\left( m, c_g, c_g p_g, -t  \right),  \ \ m\geq 0 ,
\end{equation}
where ${}_2F_1(\alpha,\beta, \gamma, z)$ denotes the Gauss hypergeometric function \citep{Zwillinger}.

Even though we do not have access to the joint density of $\kappa_g$, we can still study the dependence structure of its elements. The following proposition shows that conditional competition among the weights coexists with dependence induced by the common group scale $\tau_g^2$. Increasing the shrinkage of one component forces other components to be less shrunk.

\begin{proposition}[Dependence of shrinkage factors]
\label{prop:kappa_dependence}
For $p_g\ge2$, $c_g>0$ and $t>0$, the shrinkage factors satisfy
$\cov(\kappa_{gj},\kappa_{gk}\mid\tau_g^2=t)<0$. If the group scale is integrated out, the sign of the covariance might change; when the group scale is integrated out and  $a_1=Ga_G$ and $p_gc_g=a_G$ the covariance is positive.
\end{proposition}

We now derive the unconditional joint density of the shrinkage factors
$\kappa_g = (\kappa_{g1},\ldots,\kappa_{g p_g})'$. Its relevance comes from the fact that it describes the prior probability of different shrinkage configurations within a group such as complete group shrinkage, concentrated signal, and distributed signal. To do this we first derive the joint distribution of the within group scales $\lambda_{gl}^2$ and use the change of variables formula with $\lambda_{gl}^2 = \frac{1-\kappa_{gl}}{\kappa_{gl}}$. In the following let $a_1 = G a_G$,  $u_{gl} = \lambda_{gl}^2$, $u_g$ be the vector collecting $u_{gl}$, and $ s_g(\kappa_g) = \sum_{l=1}^{p_g} \frac{1-\kappa_{gl}}{\kappa_{gl}}$. Using the fact that $\tau_g^2 = \sum_{l=1}^{p_g} u_{gl} = s_g(\kappa_g)$ and $\varphi_{gl} = \frac{u_{gl}}{\sum_{r=1}^{p_g}u_{gr}}$ and $\tau_g^2 \sim \betaprime(a_G,a_2), \varphi_g \sim \dirichlet s(c_g)$ independently we arrive at the joint distribution
\begin{equation}
p(u_g)=\frac{\Gamma(p_gc_g)}{\Gamma(c_g)^{p_g}B(a_G,a_2)}
\left(\sum_lu_{gl}\right)^{a_G-p_gc_g}
\left(1+\sum_lu_{gl}\right)^{-a_G-a_2}\prod_lu_{gl}^{c_g-1}.  
\end{equation}
Applying the component-wise transformation $u_{gl}=\frac{1-\kappa_{gl}}{\kappa_{gl}}$ we obtain the joint density of $\kappa_g$ on $(0,1)^{p_g}$:

\begin{equation}
p(\kappa_g)=\frac{\Gamma(p_gc_g)}{\Gamma(c_g)^{p_g}B(a_G,a_2)}
s_g(\kappa_g)^{a_G-p_gc_g}\{1+s_g(\kappa_g)\}^{-a_G-a_2}
\prod_l(1-\kappa_{gl})^{c_g-1}\kappa_{gl}^{-c_g-1}.
\end{equation}
The dependence structure of the joint density is expressed through the terms containing $s_g(\cdot)$. Finally, we also show the distribution of the marginal shrinkage factor $\kappa_{gl}$. Marginally, Proposition~\ref{prop:properties_lambda_gl} implies that $\lambda_{gl}^2 \sim \betaprime(c_g,a_2)$. Setting $\kappa_{gl}=\frac{1}{1+\lambda_{gl}^2}$ we have
\begin{equation}
  p(\kappa_{gl})=  \frac{1}{B(a_2,c_g)}
    \kappa_{gl}^{a_2-1}
    (1-\kappa_{gl})^{c_g-1},
    \qquad 0<\kappa_{gl}<1.
\end{equation}
Hence $\kappa_{gl}\sim \betadist(a_2,c_g).$ This representation gives a direct interpretation of the two hyperparameters. Values of $\kappa_{gl}$ close to one correspond to strong shrinkage, since $\lambda_{gl}^2$ is close to zero. The behavior near $\kappa_{gl}=1$ is governed by $c_g$. Thus, $c_g<1$ places high prior density near complete shrinkage. Values of $\kappa_{gl}$ close to zero correspond to weak shrinkage and large coefficient-specific variances. The behavior near $\kappa_{gl}=0$ is governed by $a_2$. Hence, $a_2<1$ places high prior density near no shrinkage. In particular, when $a_2=c_g=1/2$, the shrinkage factor has the classical horseshoe shape $\kappa_{gl}\sim \betadist(1/2,1/2)$ \citep{carvalho_horseshoe_2010}. 

Since our prior can exhibit horseshoe-type shrinkage (see Proposition~\ref{prop:r2d2-horseshoe}), the posterior mean based on $y_{gl}$ alone approaches $y_{gl}$ as $|y_{gl}|$ grows \citep{carvalho_horseshoe_2010}. In general, we are also interested in how much shrinkage is exerted as we observe the whole vector $y$, since the other observations can also inform the scale of $b_{gl}$. We can measure this via the quantity
\begin{equation}
\label{eq:posterior_normal_regression_expectation}
\mbe \left(  b_{gl} \mid y \right) = \left( 1- \mbe (\kappa_{gl} \mid y) \right) y_{gl} ,
\end{equation}
where we are now conditioning on the vector $y$. If several observations are large, we can quantify the probability that any of their coefficients undergoes substantial shrinkage. We show this in the following theorem using the joint posterior.

\begin{theorem}[Joint posterior tail robustness]
\label{thm:shrinkage_bounds}
Let  $\sigma>0$ be known and index the coefficients by $j=1,\ldots,p$. Consider independent observations $y_j\mid b_j,\sigma\sim\normal(b_j,\sigma^2)$. Under the Group-R2 prior, let $\tau^2\sim\betaprime(a_1,a_2)$ be independent of the Dirichlet allocations. Let $z_j=\frac{y_j^2}{2\sigma^2}$ and define 
\begin{equation}
\kappa_j=\frac{1}{1+ \tau^2 \phi_g \varphi_{gl}},\qquad
R(y)=1+\sum_{j=1}^p z_j, \qquad
\alpha=a_2+\frac p2.
\end{equation}
There is a constant $C>0$, depending only on the prior hyperparameters and the fixed group structure, such that the following statements hold. 
\begin{enumerate}
\item[\emph{(a)}] For every nonempty $A\subseteq\{1,\ldots,p\}$, every $\delta\in(0,1)$, and every $y\in\mathbb R^p$,
\begin{equation}
\label{eq:joint_shrinkage_probability}
\begin{aligned}
\Pr\!\left(\max_{j\in A}\kappa_j>\delta\mid y,\sigma\right)
&\leq C R(y)^\alpha\sum_{j\in A}e^{-\delta z_j}\\
&\leq C|A|R(y)^\alpha e^{-\delta\min_{j\in A}z_j}.
\end{aligned}
\end{equation}
\item[\emph{(b)}] For each $j$ with $y_j\ne0$,
\begin{equation}
\label{eq:full_posterior_mean_bound}
\left|\mbe(b_j\mid y,\sigma)-y_j\right|
\leq \frac{4\alpha\sigma^2}{|y_j|}
+C|y_j|R(y)^\alpha e^{-y_j^2/(4\sigma^2)}.
\end{equation}
In particular, fix $\delta\in(0,1)$, a nonempty set $A$ of indices, nonzero constants $v_j$ for $j\in A$, and observations outside $A$. When $y_j(r)=rv_j$ for $j\in A$, as $r\to\infty$,
\begin{equation}
\label{eq:joint_tail_robustness_limit}
\begin{aligned}
\Pr\!\left(\max_{j\in A}\kappa_j>\delta\mid y(r),\sigma\right)&\longrightarrow0.\\
\end{aligned}
\end{equation}
Moreover $\left\|\mbe(b_A\mid y(r),\sigma)-y_A(r)\right\|_2 =O(r^{-1})$.
\end{enumerate}
\end{theorem}


For instance, consider first a group in which only one observation, $y_j$, grows in magnitude, while all other observations remain fixed. Taking $A=\{j\}$ in Theorem~\ref{thm:shrinkage_bounds}, we see that the posterior probability of $\kappa_j>\delta$ tends to zero for every fixed $\delta\in(0,1)$. Thus, even when the other observations in the group are small, the posterior mean of $b_j$ approaches $y_j$. The conclusion also holds jointly, namely, if the observations in $A$ (they need not belong to the same group) grow and those outside $A$ remain fixed, the probability that even one coefficient in $A$ undergoes substantial shrinkage tends to zero. 

This example is related to the conditional Lindley paradox \citep{som_lindley_conditional_2016}. They consider two nested regression models and show that a large signal common to both can drive the Bayes factor for including additional predictors toward zero. Here we ask how a growing observation affects the posterior mean of its own coefficient when the coefficients share a group scale. An important question remains for the other members of the group, namely, if one observation stays small while its neighbors grow, how much shrinkage does its coefficient retain? The theorem leaves that question open.

\subsubsection{Effective number of non-zero coefficients}
\label{subsubsec:meff}

A global summary of the shrinkage induced by the prior is provided by the effective number of nonzero coefficients, $m_{\text{eff}}$, defined as \citep{piironen_sparsity_2017}
\begin{equation}
\label{eq:meff}
m_{\text{eff}} = \sum_{i=1}^p (1 - \kappa_i),
\end{equation}
where $\kappa_i$ denotes the shrinkage factor associated with the $i$th coefficient $b_i$ and by construction $m_{\text{eff}} \leq p$. This quantity captures how the combination of global and local scales, encoded through $\kappa_i$, reduces model complexity. 

Visualizing the prior distribution of $m_{\text{eff}}$ is a practical tool for understanding how the choice of hyperparameters controls shrinkage. We give two ways to set the hyperparameters from a prior guess of the number of relevant coefficients: simulating the prior of $m_{\text{eff}}$, which works for any choice of the hyperparameters, and a closed form for its prior mean, which holds under the coupling of Proposition~\ref{prop:properties_lambda_gl}. If the user has prior knowledge or expectations about the number of relevant coefficients, they can simulate the distribution of $m_{\text{eff}}$ under different hyperparameter settings and select the configuration that aligns with their prior beliefs \citep{piironen_sparsity_2017, aguilar2025dependency}. This approach is particularly useful in high-dimensional applications, where direct specification of priors on individual coefficients may be unintuitive or impractical.
We can also define a group-wise effective number of nonzero coefficients as
\begin{equation}
    m_{\text{eff}, g} = \sum_{l=1}^{p_g} (1 - \kappa_{gl}),
\end{equation}
so that the total effective number of nonzero coefficients decomposes as
\begin{equation}
    m_{\text{eff}} = \sum_{g=1}^G m_{\text{eff}, g}.
\end{equation}
Under strong global shrinkage, we expect $m_{\text{eff}}$ to concentrate near zero, while under weak global shrinkage, it may approach $p$. Similarly, for each group $g$, strong shrinkage at the group level (e.g., small values of $\tau_g^2$) will result in $m_{\text{eff}, g}$ concentrating near zero, indicating that most coefficients in the group are shrunk strongly toward zero. In contrast, when $\tau_g^2$ is large, we expect $m_{\text{eff}, g}$ to approach $p_g$, reflecting a weakly regularized group with more non-negligible coefficients. 
\begin{figure}
    \centering
    \includegraphics[width=\linewidth]{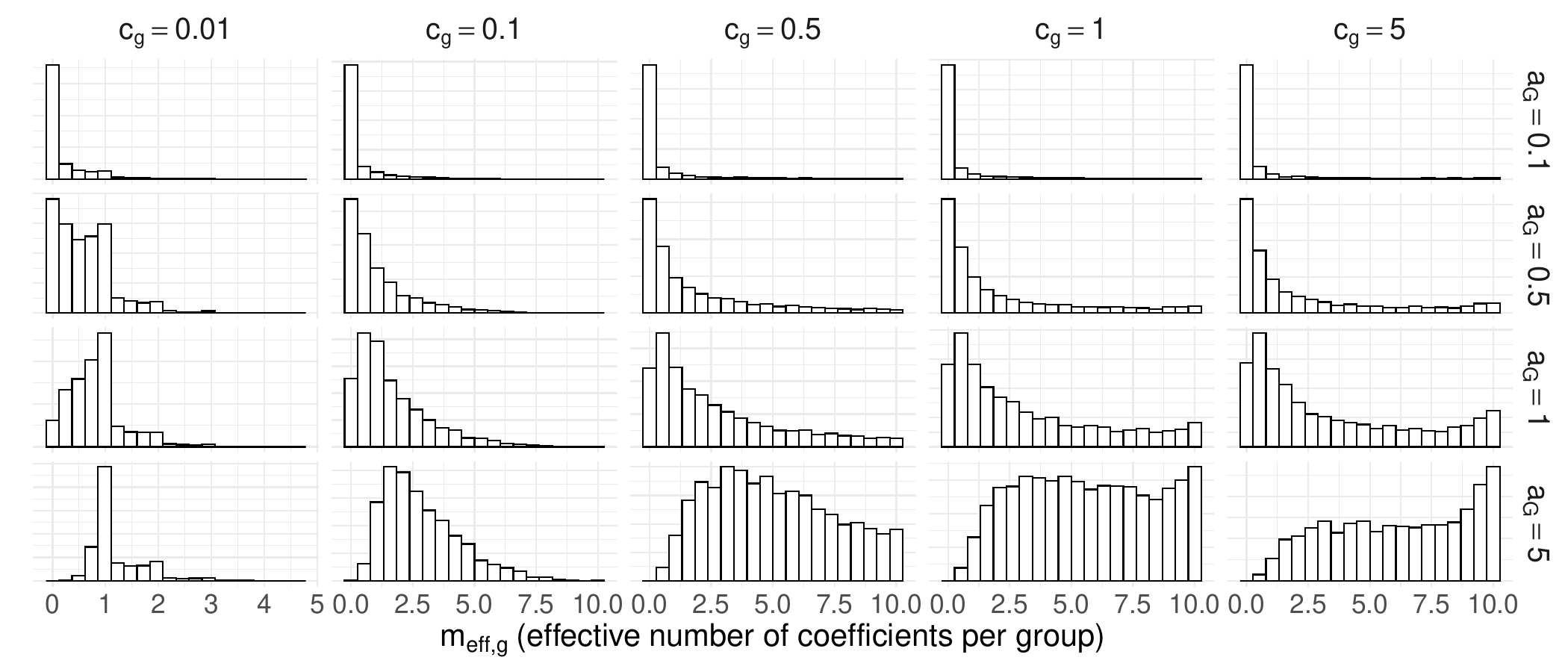}
   \caption{Prior predictive distributions of the group-wise effective number of non-zero coefficients ($m_{\mathrm{eff},g}$) for different combinations of $a_G$ and $c_g$. Here, $a_G$ is the concentration parameter of the group-level Dirichlet decomposition (controlling between-group shrinkage) and $c_g$ is the within-group concentration (controlling within-group shrinkage). Rows vary $a_G$ and columns vary $c_g$. Results are shown for $G=10$ groups with 10 covariates per group, with $a_1 = G a_G$ and $a_2 = 0.5$.}
    \label{fig:prior_group_meff}
\end{figure}
Figure \ref{fig:prior_group_meff} displays the prior predictive distribution of $m_{\text{eff}, g}$, under varying values of $a_G$ and $c_g$. Recall that $a_G$ is the parameter of the symmetric Dirichlet distribution used in the first-stage decomposition (across groups), while $c_g$ controls the second-stage allocation of variance within each group. Values of $(a_G,c_g)$ below 1 push mass toward the edges of the simplexes, inducing sparsity in $b_{gl}$. The value of $a_G$ also controls the prior mean of $R^2$, via $R^2 \sim \betadist (Ga_G,a_2)$. 

As $a_G$ increases, so does the prior mean of $R^2$, from about $0.67$ when $a_G = 0.1$ to about $0.99$ when $a_G = 5$. When $a_G < 1$, the Dirichlet distribution concentrates near the corners of the simplex, allocating most variance to a few groups. In contrast, for $a_G \geq 1$, the distribution tends to allocate variance more evenly across groups.

Given a fixed $a_G$, the total variance, $\tau_g^2$, assigned to a group is limited by the Dirichlet partition. This constraint is evident in Figure \ref{fig:prior_group_meff}: when $a_G$ is small, stronger shrinkage occurs within groups regardless of $c_g$. Within-group sparsity is influenced by $c_g$. In line with this, we see that smaller values of $c_g$ promote sparsity concentrating mass toward lower $m_{\text{eff}, g}$, while larger values of $c_g$ shift the prior mass toward less sparse configurations. These prior predictive distributions will naturally depend on the tail behavior of the prior, influenced largely by $a_2$. In the Appendix, we generate the figure for $a_2=0.1$.

As an alternative to simulation, under the coupling of Proposition~\ref{prop:properties_lambda_gl}, the prior mean of $m_{\text{eff}}$ has a closed form, which can be used to set $a_G$ from a prior guess $m_0$ of the number of relevant coefficients.

\begin{corollary}[Prior mean of the effective number of nonzero coefficients]
\label{cor:meff}
Let the assumptions of Proposition~\ref{prop:properties_lambda_gl} hold with $c_g = a_G/p_g$ for all $g$, and consider the normal means model of Section~\ref{subsubsec:shrinkage_factors}. Then
\begin{equation}
\label{eq:meff_mean}
\mbe(m_{\text{eff},g}) = \frac{p_g c_g}{c_g + a_2}, \qquad
\mbe(m_{\text{eff}}) = \sum_{g=1}^G \frac{p_g c_g}{c_g + a_2} = \sum_{g=1}^G \frac{p_g a_G}{a_G + p_g a_2}.
\end{equation}
For fixed $a_2$ and every prior guess $0 < m_0 < p$, there is exactly one $a_G > 0$ with $\mbe(m_{\text{eff}}) = m_0$. If all groups have the same size, $c_g = a_2 m_0/(p - m_0)$ and $a_G = p_g c_g$.
\end{corollary}

Since $m_{\text{eff}}$ is a sum of the terms $1 - \kappa_{gl}$, and the expectation of a sum is the sum of the expectations, Corollary~\ref{cor:meff} only requires the marginal distribution $\kappa_{gl} \sim \betadist(a_2, c_g)$. The dependence between the shrinkage factors (Proposition~\ref{prop:kappa_dependence}) changes the spread of $m_{\text{eff}}$, but not its mean. The proof is given in the Appendix. For groups of different sizes, $a_G$ may be found from a one-dimensional search, since the right-hand side of \eqref{eq:meff_mean} increases in $a_G$. Without the hyperparameter coupling, the second part of Proposition~\ref{prop:properties_lambda_gl} does not apply, and we recommend simulating the prior of $m_{\text{eff},g}$ as in Figure~\ref{fig:prior_group_meff}.

Note that under the coupling, $\mbe(m_{\text{eff}})$ in Equation~\eqref{eq:meff_mean} and $\mbe(R^2) = G a_G/(G a_G + a_2)$ depend on $(a_G, a_2)$ only through $a_G/a_2$. Hence, a prior guess $m_0$ fixes the prior mean of $R^2$, and a prior mean of $R^2$ fixes $\mbe(m_{\text{eff}})$. The remaining parameter $a_2$ sets the precision $\nu_{R^2} = G a_G + a_2$ and the tails of the marginal prior (Proposition~\ref{prop:tailprior}).

\subsubsection{Shrinkage Factors in the Normal Linear Regression Setting}
\label{subsubsec:meff_regression}
The preceding discussions were based on the normal means setting \citep{castillo_bayesian_2015}. We now shift our attention to the general normal linear regression setting. In this case shrinkage will depend on the correlation between predictors and on the entire vector of observations $y$. We now study how much the conditional posterior mean is shrunk relative to the least squares estimate, both for the full vector and for each group. 

\begin{theorem}[Conditional posterior mean in linear regression]
\label{thm:concentration_linreg}
Consider a centered regression. Let $p<n$ and assume $Q=X'X$ is positive definite. Let $W=\diag(\phi_g\varphi_{gl})$, and $\Lambda=\tau^2W$. Let $S_X=(Q+\Lambda^{-1})^{-1}Q$. The conditional posterior mean is
\begin{equation}
m=\mbe(b\mid y,\sigma^2,\tau^2,\phi,\{\varphi_g\})
=S_X\hat b^{\mathrm{OLS}},
\end{equation}
where the least squares estimate is $\hat b^{\mathrm{OLS}}=Q^{-1}X'y$. Moreover, we have
\begin{equation}
\|\hat b^{\mathrm{OLS}}-m\|_2\ge
\frac{\|\hat b^{\mathrm{OLS}}\|_2}
{1+\lambda_{\max}(Q)\tau^2\max_{g,l} \phi_g\varphi_{gl}}.
\end{equation}
Additionally, $\|m\|_2\le\tau^2\max_{g,l}w_{gl}\,\|X'y\|_2$, so $m\to0$ uniformly over allocations as $\tau^2\to0$.
\end{theorem}

Since $\max_{g,l}w_{gl}\leq1$, taking $\tau^2$ close to zero pulls the entire conditional mean toward zero for any allocation. When the columns of one group are orthogonal to those of the others, $S_X$ acts separately on that group. If this holds for every group, we obtain a bound in terms of each group's allocation and its own design matrix.

\begin{corollary}[Group shrinkage under block orthogonality]
\label{cor:group_shrinkage}
Under the assumptions of Theorem~\ref{thm:concentration_linreg}, suppose $X_g'X_h=0$ for $g\ne h$. Then
\begin{equation}
\|\hat b_g^{\mathrm{OLS}}-m_g\|_2\ge
\frac{\|\hat b_g^{\mathrm{OLS}}\|_2}
{1+\lambda_{\max}(X_g'X_g)\tau^2\phi_g\max_l\varphi_{gl}}
\ge\frac{\|\hat b_g^{\mathrm{OLS}}\|_2}{1+\lambda_{\max}(X_g'X_g)\tau^2}.
\end{equation}
\end{corollary}

The matrix $S_X$ can be used to define the corresponding $m_{\text{eff}}$. To see this, suppose first that $X'X=I$. The diagonal entries of $S_X$ are then the fractions $1-\kappa_{gl}$ that we consider in Equation~\eqref{eq:meff}. When the predictors are correlated, defining a shrinkage factor might not be straightforward since $S_X$ can have off-diagonal entries.  This has been studied by \cite{aguilar2025dependency}, who use this same construction for their dependency-aware shrinkage priors, where the conditional prior covariance can itself have off-diagonal entries. As they do, we propose to use the trace to define $m_{\text{eff}}$. Since the conditional prior covariance of $b$ is diagonal, with entries $\sigma^2\tau^2\phi_g\varphi_{gl}$, $m_{\text{eff}}$ depends on both the design and the allocations of variance across and within groups. Namely, $m_{\text{eff}}$ has the form
\begin{equation}
\label{eq:meff_trace}
m_{\text{eff}} = \operatorname{tr}(S_X) = \sum_{i=1}^{n} \frac{d_i}{1 + d_i},
\end{equation}
where $d_1,\ldots,d_n$ are the eigenvalues of $X \Lambda X'$. Even when $X$ is rank deficient, $X'X+\Lambda^{-1}$ remains invertible, so the trace is well defined for $p>n$. Moreover, the quantity $m_{\text{eff}}$ is bounded by the rank of $X$, since at most  $\operatorname{rank}(X)$ of the $d_i$ are positive. To define a per group effective number of non-zero coefficients in this case, we can consider only the diagonal entries of $S_X$ belonging to group $g$, giving $m_{\text{eff},g}$.

We now study how the design $X$ interacts with the choice of $a_G$ if we have a prior guess $m_0$ for the number of relevant coefficients. Following \citet{piironen_sparsity_2017}, we use $m_0$ as a target for the prior mean of $m_{\text{eff}}$. Consider first orthogonal columns scaled so that $X'X=nI$. Each coordinate again has its own shrinkage factor, now $\kappa_{gl}=1/(1+n\lambda_{gl}^2)$, and Equation~\eqref{eq:meff_trace} reduces to $\sum_{g,l}(1-\kappa_{gl})$. Under the full coupling $a_1=Ga_G$ and $c_g=a_G/p_g$ of Proposition~\ref{prop:properties_lambda_gl}, the expected contribution of one coefficient is
\begin{equation}
\label{eq:meff_mean_n}
\mbe(1 - \kappa_{gl}) = \frac{n c_g}{c_g + a_2}\; {}_2F_1\left(1, c_g + 1; c_g + a_2 + 1; 1 - n\right),
\end{equation}
which recovers the normal-means value $c_g/(c_g+a_2)$ when we set $n=1$. For $c_g=a_2=1/2$, we have $\sqrt{n}/(1+\sqrt{n})$, which is the same expression of \citet{piironen_sparsity_2017} for the horseshoe with global scale set to one (see Proposition~\ref{prop:r2d2-horseshoe}).

At fixed hyperparameters, the expected contribution in Equation~\eqref{eq:meff_mean_n} increases with $n$ \citep{olver_nist_2010}. Thus an $a_G$ chosen to give prior mean $m_0$ in the normal means model of Corollary~\ref{cor:meff} gives a prior mean larger than $m_0$ when $X'X=nI$ and $n>1$. To retain the same prior guess, we can fix $a_2$ and choose $a_G$ so that the contributions in Equation~\eqref{eq:meff_mean_n} sum to $m_0$ over all coefficients.

For correlated predictors we can use the same prior guess with the observed design $X$. For each candidate $a_G$, we draw the scales and allocations from their prior and compute the trace in Equation~\eqref{eq:meff_trace}, holding $X$ fixed. Averaging these traces lets us compare the implied prior mean with $m_0$ before fitting the model. When $X$ is rank deficient, the rank bound above also limits the effective model size we can elicit in this way. We use this prior check in Section~\ref{subsec:hyperparameter_specification}.

\subsection{Hyperparameter specification}
\label{subsec:hyperparameter_specification}
Here we summarize, based on the theoretical discussion above, how the hyperparameters $(a_1,a_2,a_G,c_g)$ can be selected. Recall that $R^2 \sim \betadist(a_1,a_2)$, $\phi \sim \dirichlet s(a_G)$ and $\varphi_g \sim \dirichlet s(c_g)$. The recommendations below differ in how the four hyperparameters are linked. We distinguish three levels:
\begin{itemize}
    \item \emph{Full coupling}: $a_1 = G a_G$ and $c_g = a_G/p_g$, so that only $a_G$ and $a_2$ are free. This is the setting of the second part of Proposition~\ref{prop:properties_lambda_gl}, of Propositions~\ref{prop:marginalpriors} to~\ref{prop:r2d2-horseshoe}, and of Corollary~\ref{cor:meff}. The marginal prior of $\lambda^2_{gl}$ is a Beta prime distribution, and the prior mean of $m_{\text{eff}}$ has a closed form.
    \item \emph{Group-level coupling}: $a_1 = G a_G$, and $c_g$ is chosen separately, so that $a_G$, $a_2$ and $c_g$ are free. The prior on $R^2$ is $\betadist(G a_G, a_2)$, and marginally $\tau^2_g \sim \betaprime(a_G, a_2)$ (first part of Proposition~\ref{prop:properties_lambda_gl}). Proposition~\ref{prop:uncoupled_marginals} gives the behavior of the marginal prior at the origin and in the tails, but the marginal prior of $\lambda^2_{gl}$ has no closed form, and the prior of $m_{\text{eff}}$ has to be simulated as in Figure~\ref{fig:prior_group_meff}.
    \item \emph{No coupling}: the prior on $R^2$ is set directly through $(\mu_{R^2}, \nu_{R^2})$, and $a_G$ and $c_g$ are chosen separately. This is the most flexible level.
\end{itemize}
We recommend starting with full coupling, since it has the fewest choices, and moving to other levels only when the prior information cannot be expressed otherwise.

\paragraph{Tails.} The tails of the marginal prior are set by $a_2$ alone, under full coupling (Proposition~\ref{prop:tailprior}) and under group-level coupling (Proposition~\ref{prop:uncoupled_marginals}). For $a_2 \in (0,1/2)$, the marginal prior, $p(b_{gl})$, has heavier than Cauchy tails, and the prior behaves more and more like a variable selection type prior, the smaller $a_2$ is set within this range. For $a_2 = 1/2$, the tails decay as $|b_{gl}|^{-2}$, as for the Cauchy and the horseshoe, whatever the value of $c_g$. If also $c_g = 1/2$, the marginal prior is the horseshoe, which has bounded influence (Proposition~\ref{prop:r2d2-horseshoe}). We therefore recommend $a_2 = 1/2$ as a flexible default choice. We do not recommend also fixing $c_g = 1/2$ under full coupling, since this requires $a_G = p_g/2$. For groups of equal size, this gives $\mbe(R^2) = p/(p+1)$, which is close to one and increases with the number of predictors. Instead, $a_G$ is set from prior knowledge as described next, and $c_g = a_G/p_g$ follows.

\paragraph{Prior knowledge about $R^2$ or about the number of non-zero coefficients.} Under full coupling with $a_2$ fixed, only $a_G$ is left, and it can be set from the quantity that is known best. A prior mean $\mu_{R^2}$ from past studies or expert knowledge gives $a_G = \mu_{R^2} a_2 / \{G(1-\mu_{R^2})\}$, using $\mu_{R^2} = a_1/(a_1 + a_2)$. A prior guess $m_0$ of the number of relevant coefficients gives $a_G$ from Corollary~\ref{cor:meff}, or, with $n$ observations, from \eqref{eq:meff_mean_n}. Selecting $\mu_{R^2}$ or $m_0$ therefore presents a trade-off, since a value of $\mu_{R^2}$ fixes the prior mean of $m_{\text{eff}}$, and a value of $m_0$ fixes $\mu_{R^2}$. 

\paragraph{Prior knowledge about the signal type within groups.} If the user expects concentrated or distributed signals within groups, group-level coupling allows setting $c_g$ separately from $a_G$. Figure~\ref{fig:prior_group_meff} shows that larger $a_G$ should be chosen for less shrinkage, or when many relevant signals are expected in a given group (distributed signal setting), and low $a_G$ to exert stronger shrinkage, when only few signals (concentrated signal setting) or many small signals are expected. For $c_g$, Figure~\ref{fig:log_cor_analytical} suggests that $c_g = 1/2$ allows $a_G$ to control a wide range of correlation in variances. Hence, for $a_G$ relatively small, signals in a group are shrunk jointly towards zero, while $a_G$ relatively large encourages relatively little shrinkage with little or negative correlation in variances. Since $a_1 = G a_G$ still holds, the choice of $a_G$ also sets $\mu_{R^2} = G a_G/(G a_G + a_2)$, which should be checked.

\paragraph{No prior knowledge.} When one wants to express general lack of knowledge, a uniform $R^2$ prior, $a_1 = a_2 = 1$, and uniform priors over the simplexes, $a_G = c_g = 1$, may be chosen (no coupling). This, however, does not lead to uniformity in the shrinkage coefficient space (Figure~\ref{fig:prior_group_meff}).

\paragraph{Prior check with the observed design matrix.} Corollary~\ref{cor:meff} and \eqref{eq:meff_mean_n} assume an orthogonal design. For any of the choices above, we therefore recommend checking the implied priors before fitting. That is, we draw from the prior and compute $m_{\text{eff}}$ and $m_{\text{eff},g}$ with the observed $X$ using \eqref{eq:meff_trace}, together with the implied $R^2$. If the priors on $R^2$ and on $m_{\text{eff}}$ cannot both be matched under full coupling, we recommend moving to group-level coupling or no coupling, and choosing $a_G$ and $c_g$ with this check. In that case, we give priority to the prior on $R^2$, since the simulations in Section~\ref{sec:experiments} show that predictive performance depends strongly on how well it is centered.

Finally, \cite{boss_group_2023} propose using Marginal Maximum Likelihood Estimation (MMLE), an empirical Bayes approach, to automatically estimate hyperparameters. They implement MMLE iteratively within their Gibbs sampler for posterior inference. The benefit of automatic selection is, however, offset by the additional computational complexity it introduces and is contrary to the focus of this paper to set hyperparameters based on theoretical principles or understanding of the model.

\subsection{Posterior consistency}
\label{subsec:posterior_consistency}

We now study posterior consistency of the Group-R2 prior when the number of coefficients and the group sizes may increase with the sample size. We are interested in whether the posterior concentrates around the true coefficient vector and residual variance. We also consider prediction when $p_n>n$, where different coefficient vectors can produce the same fitted mean. In this case, we establish concentration around the true mean vector on the observed design. 

Consider the centered sequence of Gaussian regressions,
\begin{equation}
y_n=X_nb_n^\star+\varepsilon_n,\qquad
\varepsilon_n\sim\normal_n(0,\sigma^{2\star}I_n),\quad \sigma^\star>0,
\end{equation}
where $X_n$ is a fixed design and $b_n^\star$ is the true coefficient vector. We denote the true residual standard deviation by $\sigma^\star$. We write $b$ and $\sigma^2$ for the coefficients and residual variance under the fitted model. The coefficient dimension $p_n$, the number of groups $G_n$, and the group sizes $p_{g,n}$ may depend on $n$ \citep{ghosal_fundamentals_2017}. Denote by $\Pi_n$ the prior induced by the Group-R2 hierarchy and by $\Pi_n(\cdot\mid y_n)$ its posterior. The statements below refer to the true data-generating process.

Theorem~\ref{thm:consistency_growing_groups} establishes consistency for the joint posterior of $b$ and $\sigma^2$. We assume that the eigenvalues of $X_n'X_n/n$ remain bounded above and away from zero, so that changes in the coefficients correspond to changes in the mean vector. We also require $p_n\log n=o(n)$, which allows the dimension to grow while retaining enough prior probability near the true coefficients. For the within-group allocations, the individual shapes $c_{g,n}$ may decrease as groups grow, provided their total concentration $p_{g,n}c_{g,n}$ stays bounded away from zero and each shape satisfies the stated upper bound.

\begin{theorem}[Consistency with growing groups and unknown variance]
\label{thm:consistency_growing_groups}
Suppose $p_n\log n=o(n)$, $\sup_n\|b_n^\star\|_\infty<\infty$, and, for constants $0<d_-\le d_+<\infty$,
\begin{equation}
d_-\le\lambda_{\min}(X_n'X_n/n)
\le\lambda_{\max}(X_n'X_n/n)\le d_+.
\end{equation}
Additionally, consider the following priors 
\begin{equation}
\begin{aligned}
\phi_n&\sim\dirichlet(a_{G,n},\ldots,a_{G,n}),\\
\varphi_{g,n}&\sim\dirichlet(c_{g,n},\ldots,c_{g,n}),\\
\tau_n^2&\sim\betaprime(a_{1,n},a_2),
\end{aligned}
\end{equation}
and conditionally independent, zero-mean normal coefficients with variances $\sigma^2\tau_n^2\phi_{g,n}\varphi_{gl,n}$. Assume, for fixed positive constants,
\begin{equation}
\begin{aligned}
a_-\le a_{G,n}\le a_+,\qquad
c_-/p_{g,n}\le c_{g,n}\le c_+, \qquad
a_0\le a_{1,n}\le A G_n.
\end{aligned}
\end{equation}
where $a_2>0$ is fixed.  Let $\sigma^2$ have a proper prior that is fixed as $n$ varies, independently, with a continuous, positive density in a neighborhood of $\sigma^{2\star}$. Then, for every $\epsilon>0$,
\begin{equation}
\Pi_n\bigl(\|b-b_n^\star\|_2>\epsilon\ \text{or}\ |\sigma^2-\sigma^{2\star}|>\epsilon
\mid y_n\bigr)\longrightarrow0
\quad\text{almost surely}.
\end{equation}
\end{theorem}

Theorem~\ref{thm:consistency_growing_groups} shows that the posterior probability outside any fixed neighborhood of the true coefficient vector and residual variance tends to zero. Both the number of groups and their sizes may increase. Our result does not impose sparsity of the true coefficient vector. For instance, the coupled specification $a_{1,n}=G_na_{G,n}$ and $c_{g,n}=a_{G,n}/p_{g,n}$ satisfies both the allocation and scale conditions when $a_{G,n}$ remains bounded above and away from zero. The result also includes specifications without these couplings and a fixed prior on $R^2$ insofar as assumptions of the theorem are satisfied.

\citet{boss_group_2023} also prove almost-sure $\ell_2$ consistency for the regression coefficients under the GIGG prior. They assume $p_n=o(n)$, restrict the number of nonzero coefficients to $o(n/\log n)$, and condition on the noise variance and a prescribed global scale that decreases with $n$. We impose the stronger condition $p_n\log n=o(n)$, which allows every coefficient to be nonzero, and infer the residual variance jointly with the coefficients. The result also applies if the value of $\sigma^\star$ is known.

\citet{boss_group_2023} build on the result of \citet{armagan_posterior_2013}, where it is required that there is enough prior probability near the true coefficient vector, while our proof uses the testing approach of \citet{ghosal_convergence_2000} directly, which allows the residual variance to be unknown. In the GIGG prior, coefficients in the same group share a scale and conditioning on these scales, \citet{boss_group_2023} factor the coefficient probabilities and bound the contributions of active and inactive coordinates separately. In the Group-R2, however, we must account for the fact that the proportions of variance sum to one across and within each group, so assigning a larger share to one predictor leaves less for the others. The proof uses normalized gamma variables to bound the probability of a \emph{joint} allocation event at both levels. We then integrate the conditional normal probabilities over suitable intervals for $\tau_n^2$ and $\sigma^2$. This gives a lower bound of $\exp(-Cp_n\log n)$ for the prior probability near the true coefficients and variance. The requirement on the dimension condition becomes clear when noticing that the proof considers errors bounded by $\exp(-cn)$. 

Whenever $p_n>n$, the problem requires a different approach, since in this case, $X$ can not have full column rank. We study the concentration of the fitted mean. To this end, define the prediction distance
\begin{equation}
\label{eq:consistency_prediction_distance}
d_n(b,b_n^\star)=\frac{\|X_n(b-b_n^\star)\|_2}{\sqrt n}.
\end{equation}
This distance measures the discrepancy between two mean vectors on the observed design. It can be zero for distinct coefficient vectors, so concentration in $d_n$ does not by itself imply recovery of $b_n^\star$.  We now assume that the coefficient vector is sparse and that the residual variance is known. We also use a different allocation specification, with fixed total concentration at each level of the decomposition.

\begin{theorem}[Prediction contraction, including $p_n>n$]
\label{thm:consistency_prediction}
Suppose that $\sigma^\star$ is known, $\sup_n\|b_n^\star\|_\infty<\infty$, and
\begin{equation}
\max_{1\le j\le p_n}\|X_{n,j}\|_2/\sqrt n\le K<\infty.
\end{equation}
For fixed positive $\alpha,\zeta,a_1,a_2$, consider 
\begin{equation}
\begin{aligned}
\phi_n&\sim\dirichlet(\alpha/G_n,\ldots,\alpha/G_n),\\
\varphi_{g,n}&\sim\dirichlet(\zeta/p_{g,n},\ldots,\zeta/p_{g,n}),\qquad
\tau_n^2\sim\betaprime(a_1,a_2).
\end{aligned}
\end{equation}
Define
\begin{equation}
k_n=\max\{1,\|b_n^\star\|_0\},\qquad
q_n=k_n\log(e p_n n),\qquad
\epsilon_n=(q_n/n)^{1/4}.
\end{equation}
If $q_n=o(n)$, there is a finite constant $L$, depending only on the fixed bounds and hyperparameters, such that
\begin{equation}
\Pi_n\left(\frac{\|X_n(b-b_n^\star)\|_2}{\sqrt n}>L\epsilon_n
\,\middle|\,y_n\right)\longrightarrow0
\quad\text{almost surely}.
\end{equation}
The result holds for any partition into nonempty groups  whatever the rank of $X_n$.
\end{theorem}

Theorem~\ref{thm:consistency_prediction} allows both the number of groups and their sizes to grow, including settings with more coefficients than observations. For example, $p_n$ may grow polynomially in $n$ if $k_n\log n=o(n)$ where $k_n >0$ is the number of nonzero coefficients. Here the allocation shapes are $a_{G,n}=\alpha/G_n$ and $c_{g,n}=\zeta/p_{g,n}$, which need not satisfy either coupling used to obtain closed-form marginals. These choices allow the prior to assign most of the variance to groups and coordinates containing nonzero coefficients. In the proof, we use this property to obtain the joint prior bound
\begin{equation}
\Pi_n(\|b-b_n^\star\|_1\le\delta)
\ge\exp\{-C k_n\log(e p_n/\delta)\},\qquad 0<\delta\le1,
\end{equation}
uniformly over group partitions. The exponent in this lower bound is proportional to $k_n$, while the full dimension $p_n$ enters only through the logarithm. Using the column-norm condition, we then relate this coefficient neighborhood to one in prediction distance in the proof. We obtain this bound by keeping the total concentrations $\alpha$ and $\zeta$ fixed as the dimensions grow.

The fourth-root rate comes from the test used in the proof. If a coefficient vector is at prediction distance $r$ from the truth, then the expected squared residual norm per observation increases by $r^2$. We use this idea to construct a test with errors bounded by $\exp(-c n r^4)$. Comparing this bound with the prior probability above gives $\epsilon_n=(q_n/n)^{1/4}$. The exponent is determined by the test we use, however using a sharper test could lead to a faster rate.

As a noteworthy comparison, the spike-and-slab group lasso of \citet{bai_group_lassos_2022} obtains contraction rates governed by the number of active groups. Among their assumptions, the number of active groups is $o(n/\log G_n)$, the largest true coefficient is $O(\log G_n)$ (in absolute value), and the design satisfies eigenvalue bounds on sparse collections of groups. In Theorem~\ref{thm:consistency_prediction}, $k_n$ counts individual nonzero coefficients. For example, a handful of nonzero coefficients contribute the same $k_n$ whether they occupy group $i$ or $j$. Our proof allocates variance at both levels of the hierarchy, but the resulting prediction bound holds uniformly over those partitions.

\section{Experiments}
\label{sec:experiments}

We have implemented the models with the probabilistic programming language and framework Stan \citep{carpenter_stan_2017, stan_development_team_stan_2024}, which employs an adaptive Hamiltonian Monte Carlo (HMC) sampler known as the No-U-Turn Sampler (NUTS) \citep{neal_mcmc_2011, brooks_handbook_2011, hoffman_no-u-turn_2014} to sample draws from posterior distributions. The associated data and code are available at \mygithub.

\subsection{Evaluation Metrics}  
\label{sec:eval_metrics}

Our goal is to evaluate whether incorporating grouping structures into the prior improves the performance of $R^2$ decomposition priors, and under what conditions. To do so, we compare a baseline model $M$ with its grouped counterpart $M_G$. For a given quantity of interest $\mathcal{Q}$, we define the performance difference as $\Delta \mathcal{Q} = \mathcal{Q}(M_G) - \mathcal{Q}(M)$. When the distribution of $\Delta \mathcal{Q}$ is centered around zero, both models perform similarly. The interpretation of $\Delta \mathcal{Q}$ depends on the specific quantity: higher values may indicate improvement or deterioration.

We evaluate four aspects of model performance: predictive performance, parameter recovery, coverage, and convergence. Results on predictive performance and parameter recovery are reported in the main text while coverage and convergence diagnostics appear in the Appendix.

\textbf{Out-of-sample Predictive Performance:} We assess predictive performance using the expected log-pointwise predictive density ($\elpd$) \citep{vehtari_survey_2012, vehtari_practical_2016}, computed as
\begin{equation}
\elpd = \sum_{i=1}^{N_{\rm new}}\ln \left( \frac{1}{S} \sum_{s=1}^S p(y_{\text{new},i} \mid \theta^{(s)}) \right),    
\end{equation}
where $\theta^{(s)}$ represents the $s$th draw from the posterior distribution $p( \theta \mid y)$ for $s = 1, \dots, S$. The $\elpd$ quantifies predictive performance across a set of $N_{\rm new}$ observations unseen during model training. A value of $\Delta \elpd > 0$ indicates that the grouped version outperforms the nongrouped. To account for variation in the scale of $\Delta \elpd$ across different models and simulation settings, we report asinh-transformed $\Delta \elpd$ values in all figures. This transformation is given by $\operatorname{asinh}(x) = \log(x + \sqrt{x^2 + 1})$ and behaves like a logarithm for large values while remaining linear around zero, facilitating interpretation of differences.

\textbf{Parameter Recovery:} We assess accuracy of parameter recovery by the posterior root mean squared error (RMSE) for each regression coefficient:
\begin{equation}
\rmse = \frac{1}{p} \sum_{k=1}^p \sqrt{\frac{1}{S} \sum_{s=1}^S \left(b_k^{(s)} - b_k\right)^2},    
\end{equation}
where $b_k^{(s)}$ is the $s$th posterior draw for coefficient $b_k$, and $b_k$ is the true value. This global metric captures both systematic bias and estimation variance \citep{robert_bayesian_2007}. To refine this assessment, we also report three types of RMSE: 1) across all coefficients, 2) for true zeros (evaluating shrinkage of irrelevant predictors), and 3) for true nonzeros (measuring signal recovery). Lower RMSE values indicate better recovery; thus, $\Delta \rmse < 0$ favors the grouped model. We highlight that this is not the typical RMSE based on the posterior mean, but we are using the whole posterior distribution to calculate it. 

\textbf{Coverage:}  We assess $95\%$ marginal credible interval coverage by reporting average coverage rates, interval widths, specificity, sensitivity (power), and nonzero coverage \citep{benjamini_controlling_1995}. We also examine how these metrics vary with interval width via Receiver Operating Characteristic (ROC) curves.

\textbf{Convergence Diagnostics:} We monitor the quality of posterior samples using two convergence indicators: 1) The potential scale reduction factor $\widehat{R}$ \citep{vehtari_rank_normalization_2021}, which should be close to 1 if chains are well mixed. 2) The effective sample size (ESS), reflecting the effect of dependency compared to a sample of independent draws. High ESS values signal efficient sampling and low autocorrelation \citep{brooks_handbook_2011}.

\subsection{Simulations}

\subsubsection{Data generating procedure}

We base our simulation design on a modified version of the setup proposed by \citet{boss_group_2023}. The design matrix $X$ is drawn from a multivariate normal distribution with zero mean and covariance matrix $\Sigma_X$. The structure of $\Sigma_X$ reflects a block exchangeable correlation pattern: variables within the same group have pairwise correlation $\rho_{\text{in}} = 0.8$, while variables across groups have $\rho_{\text{out}} = 0.2$. All covariates are standardized to have unit variance. The error variance $\sigma^2$ is set to achieve a prespecified $R^2 \in \{0.25, 0.80\}$. We fix the sample size to $n = 200$ for all experiments. Each group contains $p_g =10$ covariates, so the total number of groups is $G = p / 10$, with $p \in \{100, 500\}$.

We evaluate five scenarios for generating the regression coefficients $b$, each reflecting different assumptions about sparsity and signal distribution within groups. 1) \textbf{Concentrated}: The signal is sparse and localized within each group. The first coefficient of each group is nonzero and set to $b_{g1} = 2$ while all others are zero. 2) \textbf{Random concentrated}: A random variant of the previous one in which the first coefficient in each group is sampled independently from $\mathcal{N}(0, 3^2)$, with all other coefficients in the group set to zero.  3) \textbf{Distributed}: The signal is spread across the multiple covariates that belong to a group. In the deterministic version, only the first group is active, leading to the first 10 coefficients being nonzero. We set $b_i = 0.5$ for $i = 1,\dots,5$ and $b_i = 1$ for $i = 6,\dots,10$. 4) \textbf{Random distributed}: Similar to the distributed, but the first 10 coefficients of the active group are independently drawn from $\mathcal{N}(0, 3^2)$. In both cases, the remaining coefficients are set to zero. 5) \textbf{Random Coefficients}: Following \citet{boss_group_2023}, the first group is randomly assigned either a concentrated or distributed signal with equal probability, ensuring the presence of at least one active group. For each of the remaining groups, a concentrated signal is assigned with probability 0.2, a distributed signal with probability 0.2, and no signal with probability 0.6.

\subsubsection{Models}


We compare the Group-R2 prior with three sets of models. First, we compare each Group-R2 prior with its nongrouped counterpart, the R2D2 prior with the same prior on $R^2$. This isolates the effect of the group decomposition. Second, we compare with the GIGG prior \citep{boss_group_2023}, calibrated to the same prior on $R^2$ (Section~\ref{subsec:gigg_comparison}). Third, we compare with five grouped priors from the literature (Appendix~\ref{subsec:external_comparators}). Three of them can set coefficients exactly to zero. These are the Bayesian group lasso with spike and slab (BGL-SS) and the Bayesian sparse group selection with spike and slab (BSGS-SS) of \citet{xu_bayesian_2015}, and the group-informed $g$-prior of \citet{agarwal_bayesian_2025}. The other two are continuous, the group horseshoe (GHS) of \citet{xu_bayesian_2016} and GRASP \citep{graspgroupedregressionadaptive2025}. We fit each of the five priors with the code of its authors.

The grouped and nongrouped versions of the R2 priors differ only in their hierarchical structure. The grouped models include the $\phi$ and $\varphi_g$ decompositions, while the nongrouped ones do not. For Group-R2 priors, we specify symmetric Dirichlet distributions for both $\phi$ and $\varphi_g$, with $a_G \in \{0.1, 0.5, 1 \}$ and $c_g = 0.5$, respectively. For the nongrouped (standard) R2D2 model, we also use a symmetric Dirichlet with $a_\pi = a_G$.

To set the hyperparameters, we use group-level coupling (Section~\ref{subsec:hyperparameter_specification}) where $a_1 = G a_G$, and fix $a_2 = 0.5$, so that by Proposition~\ref{prop:uncoupled_marginals} the marginal prior has Cauchy-type tails. Since $p_g c_g \neq a_G$, Propositions~\ref{prop:marginalpriors} to~\ref{prop:r2d2-horseshoe} and Corollary~\ref{cor:meff} do not apply, and the prior of $m_{\text{eff},g}$ is as simulated in Figure~\ref{fig:prior_group_meff}. 
The value $c_g = 0.5$ was chosen based on the expected effective number of nonzero coefficients $m_\textrm{eff}$. As discussed in Section~\ref{subsubsec:meff}, it allows for a variety of shapes for its distribution, enforcing sparsity when $a_G$ is small and a uniform like behavior that spreads signal across group elements when $a_G$ increases. Additionally, Figure~\ref{fig:log_cor_analytical} shows that the chosen values for $a_G$ when $c_g = 0.5$, allow for a wide range of correlation in log-marginal variances.   A priori, these behaviors should accommodate for the different mechanisms by which we generate the coefficients. Since these models are fully characterized by the choice of $a_G$, we denote them as \texttt{R2-a}$_\texttt{G}$ in the results.

\begin{figure}[t]%
\includegraphics[width=\linewidth]{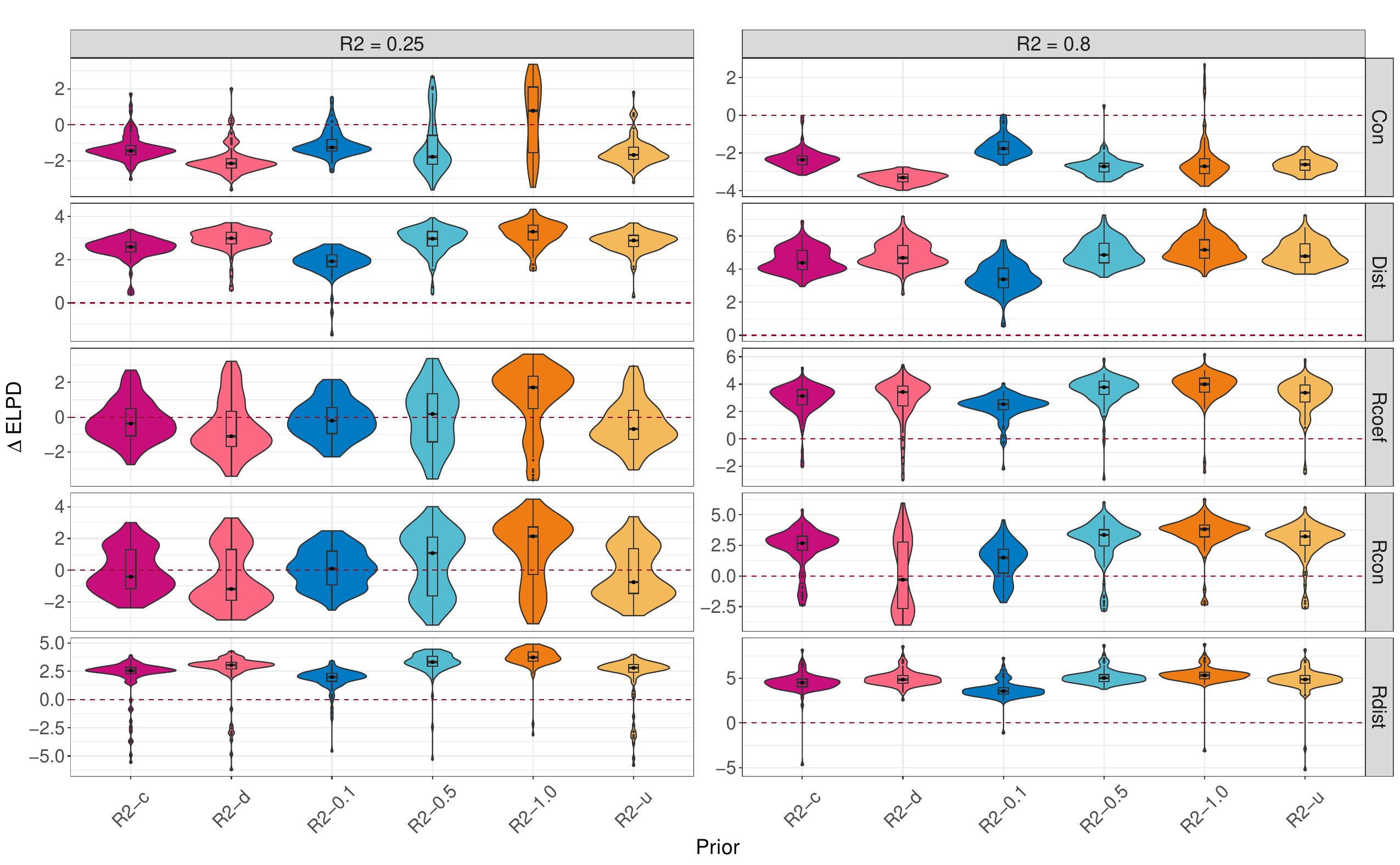}
  \caption{ \textbf{$\Delta \elpd$ (asinh-transformed) for the lower-dimensional scenario ($p = 100$, $n = 200$)}. Values above the line indicate improvement in predictive performance when considering groups. ``Con” denotes the concentrated signal, ``Dist” the distributed signal, ``Rcon” and ``Rdist” their random counterparts, and ``Rcoef” the random coefficients setting. Group-R2 priors improve predictive performance in distributed and mixed signal settings. Gains increase with $a_G$, especially when $R^2$ is high.}
  \label{fig:elpd_low_dimension}
\end{figure}%

We also evaluate two additional prior configurations and their grouped counterparts:
1) Uniform R2 prior \texttt{R2-u}: This baseline sets $a_1 = a_2 = 1$ for the $R^2$ prior, and uses symmetric Dirichlet distributions with $a_G = c_g = 1$ for the grouped version, or $a_\pi = 1$ for the nongrouped. This corresponds to uniform priors over both the total $R^2$ and the simplex partitions. 2) Concentrated and Distributed R2 Priors: These models share the same prior mean and variance for $R^2$, with $(\mu_{R^2}, \nu_{R^2}) = (1/3, 3)$. They differ in how signal sparsity is structured. The concentrated version \texttt{R2-c} sets $a_G = 1$ and $c_g = 0.5$, promoting sparsity within each group. This setup should be better suited for concentrated signals, where only a few predictors per group are truly active. In contrast, the distributed version \texttt{R2-d} uses $a_G = 0.5$ and $c_g = 1$, placing more uniform prior mass across coefficients within a group and thereby supporting distributed signals, where multiple predictors share the signal. These configurations allow us to test whether tuning the decomposition hyperparameters to match the underlying sparsity pattern improves model performance.

\begin{figure}[t!]%
\includegraphics[width=\linewidth]{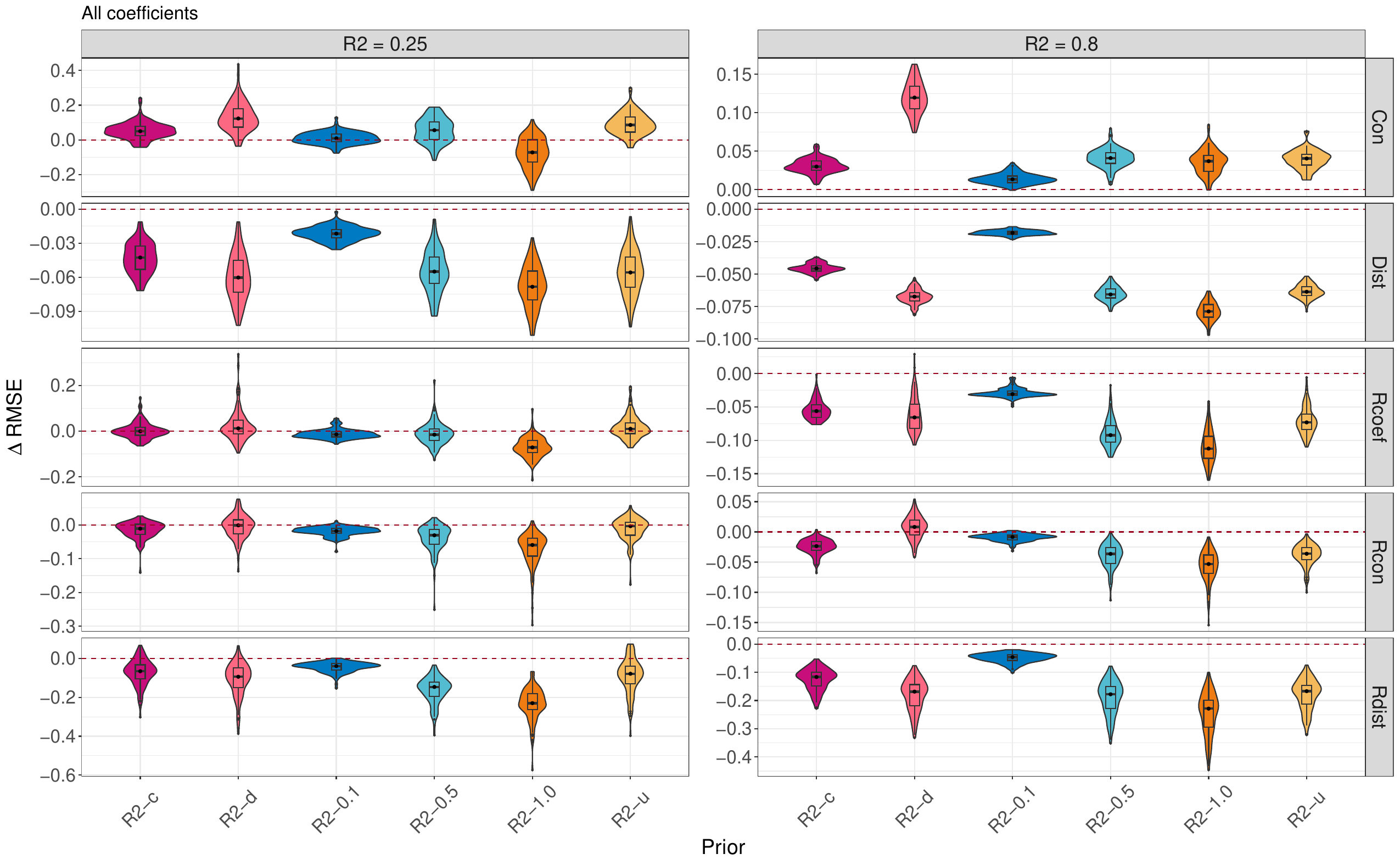}
  \caption{ \textbf{$\Delta \rmse$ for the lower-dimensional scenario ($p = 100$, $n = 200$).} Values below the line indicate improvement in parameter recovery when considering groups. ``Con” denotes the concentrated signal, ``Dist” the distributed signal, ``Rcon” and ``Rdist” their random counterparts, and ``Rcoef” the random coefficients setting. Group-R2 priors consistently improve parameter recovery in distributed and mixed signal settings. Improvements grow with $a_G$; the uniform prior performs robustly across different scenarios.
 }
  \label{fig:rmse_all_low_dimension}
\end{figure}%

\subsubsection{Results - Comparison to non-grouped R2 priors }

\paragraph{Lower-dimensional results}

We first evaluate the performance of Group-R2 priors in the lower-dimensional setting ($p = 100$, $n = 200$). Figure \ref{fig:elpd_low_dimension} reports the difference in out-of-sample predictive performance $\Delta \elpd$, while Figure \ref{fig:rmse_all_low_dimension} reports the difference in parameter recovery $\Delta \rmse$. Overall, grouping improves predictive performance in scenarios where signal is distributed across covariates or randomly assigned (Distributed, Random Distributed, Random Coefficients), particularly when $R^2 = 0.8$. Gains in predictive performance tend to increase with the group-level concentration parameter $a_G$, as seen in the progression from models \texttt{R2-}$0.1$ to \texttt{R2-}$1.0$. In contrast, when signals are highly concentrated within groups (Concentrated), grouping provides little to no benefit in predictive performance. Moreover, priors assuming distributed signals such as \texttt{R2-d} perform worse in the concentrated scenario than priors assuming concentrated scenario like \texttt{R2-c}.

\begin{figure}[t]%
\includegraphics[width=\linewidth]{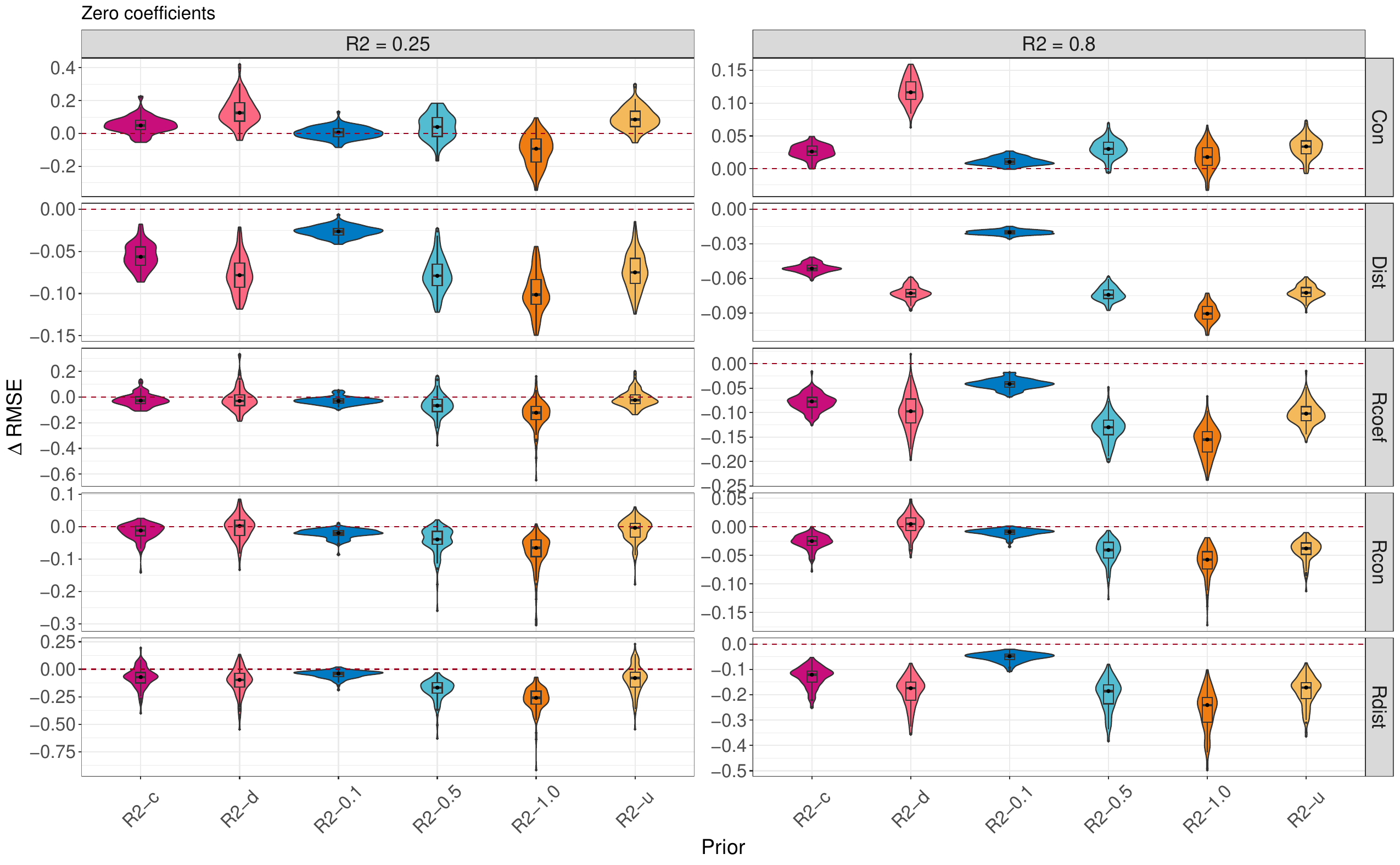}
  \caption{ \textbf{$\Delta \rmse$ for truly zero coefficients for the lower-dimensional scenario ($p = 100$, $n = 200$).} Values below the line indicate improvement in parameter recovery when considering groups. ``Con” denotes the concentrated signal, ``Dist” the distributed signal, ``Rcon” and ``Rdist” their random counterparts, and ``Rcoef” the random coefficients setting. 
 }
\label{fig:rmse_true_zero_low_dimension}
\end{figure}

\begin{figure}[t]%
\includegraphics[width=\linewidth]
{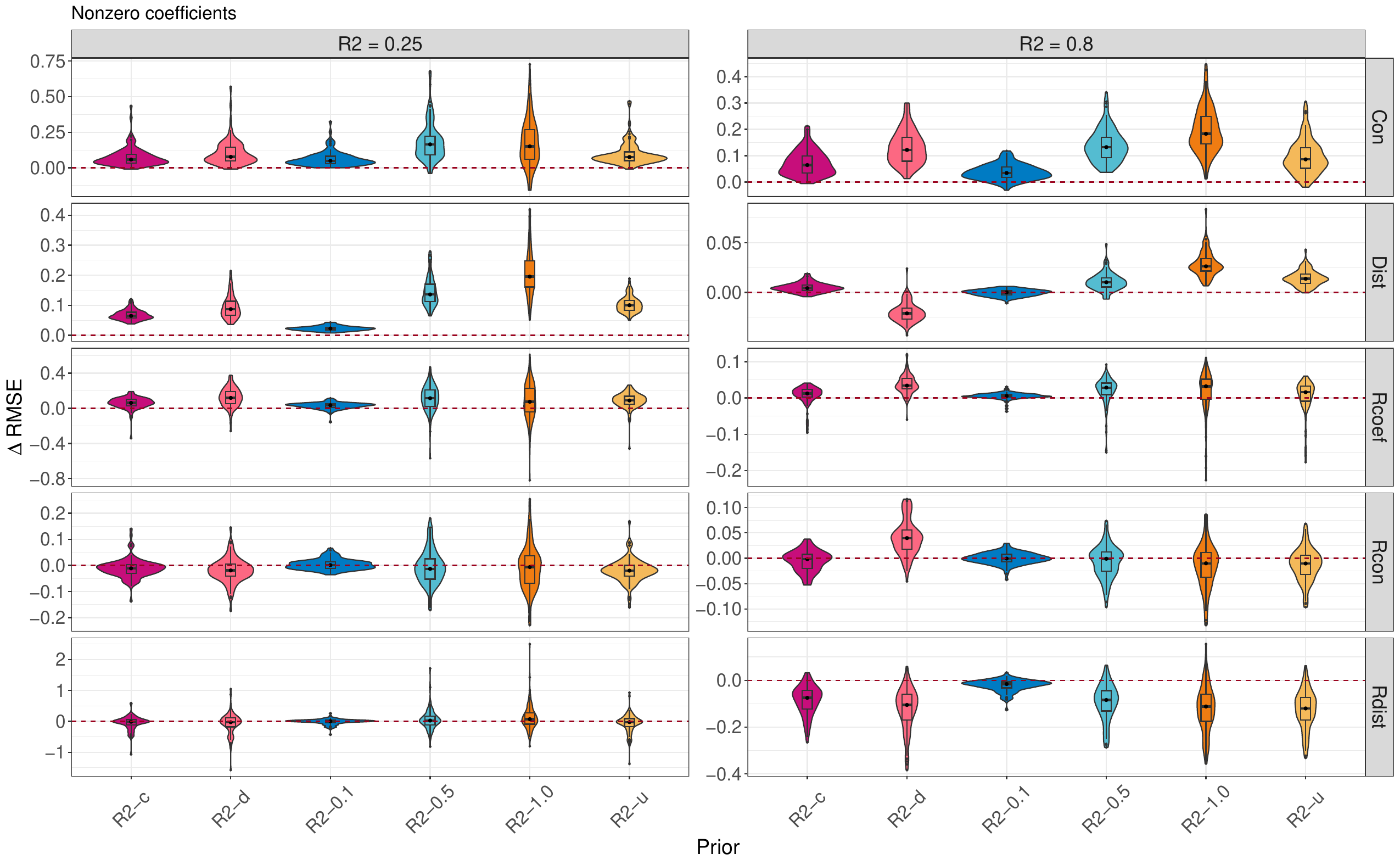}
  \caption{ \textbf{$\Delta \rmse$ for truly nonzero coefficients for the lower-dimensional scenario ($p = 100$, $n = 200$).} Values below the line indicate improvement in parameter recovery when considering groups. ``Con” denotes the concentrated signal, ``Dist” the distributed signal, ``Rcon” and ``Rdist” their random counterparts, and ``Rcoef” the random coefficients setting. 
 }
\label{fig:rmse_true_nonzero_low_dimension}
\end{figure}

Figure \ref{fig:rmse_all_low_dimension} shows that, except for the concentrated ``Con" case, parameter recovery is consistently improved when grouping is taken into account, especially when $R^2$ is high. Once again, concentrated signals represent the most challenging scenario, where non-grouped models can sometimes match or outperform grouped ones. As with predictive performance, RMSE improves monotonically with increasing $a_G$ in the Group-R2 priors. The uniform prior \texttt{R2-u}, which does not favor any particular sparsity structure, emerges as a robust default. It delivers competitive performance across different signal patterns when no prior knowledge about sparsity is available. The improvement in the cases in which the grouped priors outperform, as can be seen from Figures~\ref{fig:rmse_true_zero_low_dimension} and~\ref{fig:rmse_true_nonzero_low_dimension}, results from improved shrinkage on the truly zero coefficients groups. This holds true for low and high $R^2$. Relatively weak performance of the grouped priors in the concentrated case comes from the fact that the truly non-zero coefficients in each group are over-shrunk, as indicated by higher $\Delta \rmse$ seen in the first row of Figure~\ref{fig:rmse_true_nonzero_low_dimension}. Taken together, these results echo findings found in the group-prior literature such as \citet{boss_group_2023}, in which the benefit to using group-priors comes from imposing correlation in shrinkage for groups of covariates for which the signal is weak.

\paragraph{High-dimensional results}
\begin{figure}[t]%
\includegraphics[width=\linewidth]{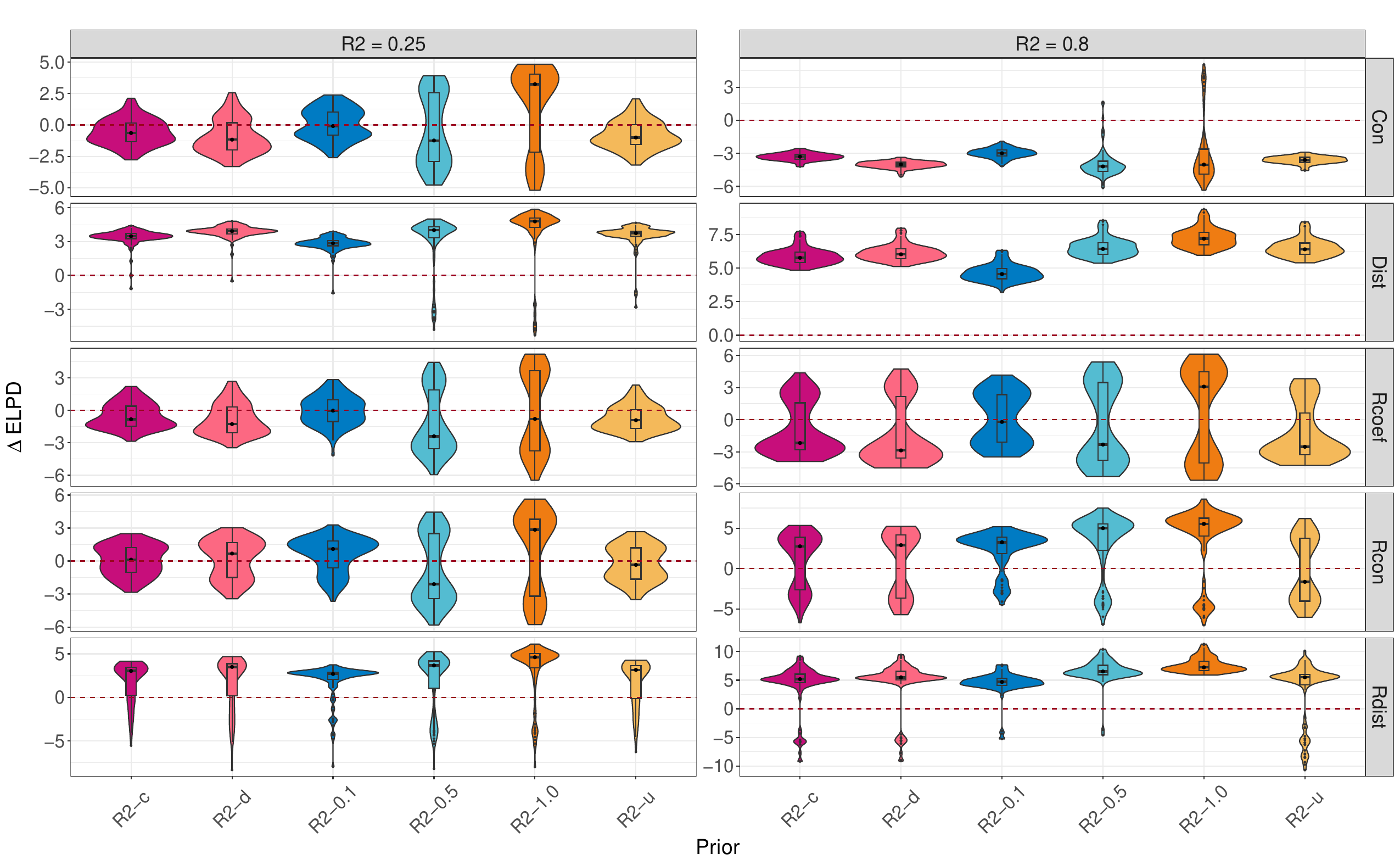}
  \caption{\textbf{ $\Delta \elpd$ (asinh-transformed) for the high-dimensional scenario ($p = 500$, $n = 200$)}. Values above the line indicate improvement in predictive performance when considering groups. ``Con” denotes the concentrated signal, ``Dist” the distributed signal, ``Rcon” and ``Rdist” their random counterparts, and ``Rcoef” the random coefficients setting. Group-R2 priors improve predictive performance in distributed signal settings. Performance in Random Coefficients (Rcoef) is mixed, with limited gains except when $a_G$ is sufficiently high and high $R^2$ is present. Concentrated signals remain challenging.
 }
\label{fig:elpd_high_dimension}
\end{figure}%

We now consider the high-dimensional setting with $p = 500$ and $n = 200$. Figures~\ref{fig:elpd_high_dimension} and~\ref{fig:rmse_all_high_dimension} show $\Delta \elpd$ and $\Delta \rmse$, respectively\footnote{The break down of the $\Delta \rmse$ in terms of true zero and true non-zero is relegated to the Appendix due to similarity to the lower-dimensional DGP results.}.

The interpretation of the results remains the same. The results show similar patterns to the lower-dimensional case, but the benefits of incorporating group structure are amplified in high dimensions. In distributed signal settings, grouping leads to large gains in both predictive performance and parameter recovery, particularly when $R^2 = 0.8$. For the Random Coefficients case, improvement only comes when considering sufficiently high values for $a_G = 1$  and under high $R^2$. Concentrated signals are still challenging. Inclusion of grouping structures does not outperform the baseline.

\subsection{Comparison with the GIGG prior}
\label{subsec:gigg_comparison}

Among the grouped shrinkage priors discussed in the introduction, the GIGG prior of \citet{boss_group_2023} is the closest competitor to ours. It targets the same setting of block correlated regressors, it belongs to the same class of continuous shrinkage priors, and its three scales play the role that the two stage decomposition plays in the Group-R2 prior. Additionally it compares favorably against competing continuous as well as discrete mixture priors, extensively discussed in \citet{boss_group_2023}. 

In particular, the GIGG prior employs a three tiered shrinkage prior $b_{gl}\sim \normal(0,\tau^2\gamma_g^2\lambda_{gl}^2)$, with a global scale $\tau^2$, a group scale $\gamma_g^2$ and a local scale $\lambda_{gl}^2$, where $\tau \sim \mathrm{C}_+\left(0,1\right)$,  $\gamma_g^2\sim\gammadist(a_g,1)$ with shape $a_g$ and rate $1$, and $\lambda_{gl}^2\sim\invgammadist(b_g,1)$ such that $\gamma_g^2\lambda_{gl}^2 \sim \betaprime(a_g,b_g)$ and thus compares under the coupled hyperparameters suggested above to the Group-R2 in that $\tau_g^2 = \phi_g\tau^2 \sim \betaprime(a_G,a_2)$, albeit the Group-R2 adds further decomposition by $\varphi_{gl}$. We write the GIGG shapes as $(a_g, b_g)$ which occupy distinct roles. The group shape $a_g$ governs the mass that the marginal prior of $b_{gl}$ places at the origin and therefore how strongly a group is thresholded as a whole, with smaller values shrinking entire groups more aggressively. The local shape $b_g$ governs the tail of that same marginal and the degree to which coefficients within a group share their shrinkage, with larger values tying them more closely to the group scale and smaller values allowing individual coefficients to escape it. Since a major conceptual difference is direct control of the Group-R2 over the model's $R^2$, we suggest below a calibration method to find $a_g$ and $b_g$ to match the implied $R^2$ to the one set by our prior. Figure~\ref{fig:gigg_implied_r2} in the Appendix shows how much the implied prior on $R^2$ changes with $(a_g, b_g)$ before calibration.


\subsubsection{Calibrating the GIGG to the Group-R2 prior}

Under this hierarchy, and taking $\sigma^2 = 1$ and $X'X/n = I$ for the purpose of the calibration, the total prior signal variance under the GIGG is
\begin{equation}
  V = \tau^2 \sum_{g=1}^{G} \gamma_g^2 \sum_{l=1}^{p_g} \lambda_{gl}^2,
  \qquad
  R^2_{\text{GIGG}} = \frac{V}{V+1},
  \label{eq:gigg_implied_r2}
\end{equation}
whose induced distribution depends on $(a_g, b_g)$ but has no closed form. The Group-R2 prior with concentration $a_G$ places $R^2 \sim \betadist(G a_G, a_2)$ directly, with $a_2 = 0.5$ as above. For each $a_G \in \{0.1, 0.5, 1\}$ we therefore choose
\begin{equation}
  (a_g, b_g)
  = \argmin_{(a_g, b_g)}\;
    W_1\!\left( R^2_{\text{GIGG}}(a_g, b_g),\; \betadist(G a_G, a_2) \right),
  \label{eq:gigg_calibration}
\end{equation}
where $W_1$ is the $1$-Wasserstein distance between the two distributions of $R^2$. We denote the resulting models \texttt{GIGG-}$a_G$ and pair each with the Group-R2 model \texttt{R2-a}$_\texttt{G}$ that shares its target. Both sides of Equation~\eqref{eq:gigg_calibration} move with the number of groups, so the calibration is carried out separately for $G = 10$ and $G = 50$. The optimization, its constraints and its diagnostics are described in Section~\ref{subsec:gigg_calibration} of the Appendix.

Table~\ref{tab:gigg_calibration} reports the calibrated values. The reported values of $W_1$ are small throughout, but two qualifications apply. At $G = 10$ the constraint $a_g \leq 0.5$, which keeps the pole of the marginal prior at the origin \citep{boss_group_2023}, binds for \texttt{GIGG-}$0.5$ and \texttt{GIGG-}$1.0$, so these two are matched only as closely as this bound allows. At $G = 50$ no constraint binds, but the targets are themselves close to degenerate, since $\betadist(50, 0.5)$ places almost all of its mass above $0.98$, and the small $W_1$ reflects both distributions sitting near the boundary rather than agreement in shape. Figure~\ref{fig:gigg_implied_r2_calibration} in the Appendix shows the two distributions of $R^2$ side by side for each design and makes both qualifications visible.

\begin{table}[t]
\small
\centering
\caption{GIGG shapes calibrated to the $R^2$ prior implied by each Group-R2 model, with the out of sample $1$-Wasserstein distance to the target. Values marked $^{\dagger}$ sit at the boundary of the feasible region.}
\label{tab:gigg_calibration}
\begin{tabular}{lcccccc}
\toprule
 & \multicolumn{3}{c}{$G = 10$ $(p = 100)$} & \multicolumn{3}{c}{$G = 50$ $(p = 500)$} \\
\cmidrule(lr){2-4}\cmidrule(lr){5-7}
Model & $a_g$ & $b_g$ & $W_1$ & $a_g$ & $b_g$ & $W_1$ \\
\midrule
\texttt{GIGG-}$0.1$ & $0.363$ & $10.000^{\dagger}$ & $0.038$ & $0.488$ & $3.826$ & $0.058$ \\
\texttt{GIGG-}$0.5$ & $0.500^{\dagger}$ & $1.394$ & $0.058$ & $0.051$ & $0.393$ & $0.020$ \\
\texttt{GIGG-}$1.0$ & $0.500^{\dagger}$ & $0.714$ & $0.042$ & $0.029$ & $0.201$ & $0.010$ \\
\bottomrule
\end{tabular}
\end{table}

\subsubsection{Results}

\begin{figure}[t]%
\includegraphics[width=\linewidth]{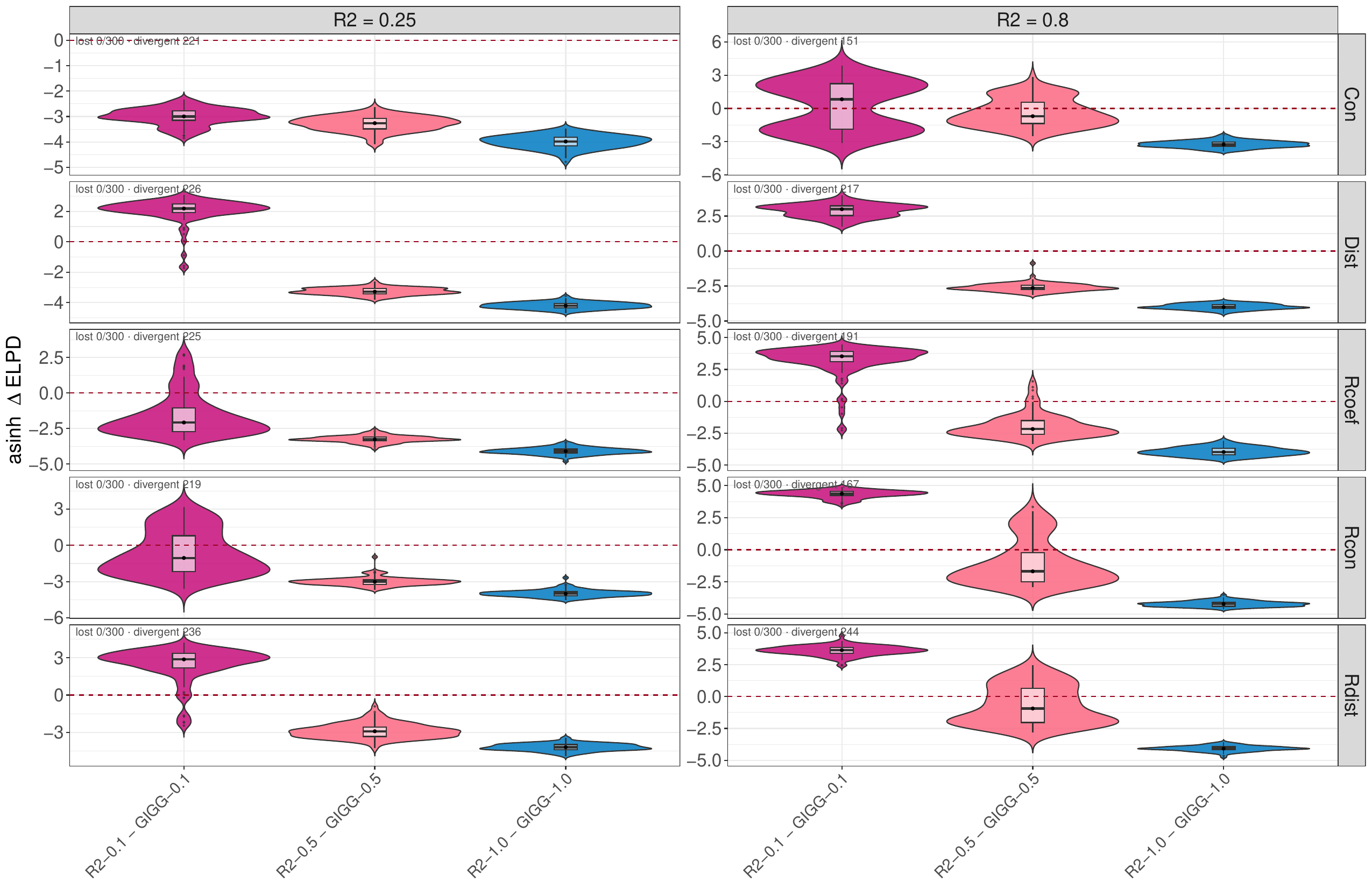}
  \caption{ \textbf{$\Delta \elpd$ (asinh-transformed) against the calibrated GIGG for the lower-dimensional scenario ($p = 100$, $n = 200$)}. Values above the line indicate better predictive performance of the Group-R2 prior. Each contrast pairs a Group-R2 model with the GIGG calibrated to the $R^2$ prior it implies, so that the two differ only in how the explained variance is decomposed. ``Con” denotes the concentrated signal, ``Dist” the distributed signal, ``Rcon” and ``Rdist” their random counterparts, and ``Rcoef” the random coefficients setting. The panel annotations give the number of contrasts lost to failed fits and the number of retained contrasts in which at least one of the two fits had divergent transitions.}
  \label{fig:gigg_elpd_low_dimension}
\end{figure}

Figure~\ref{fig:gigg_elpd_low_dimension} reports $\Delta \elpd = \elpd(\texttt{R2-a}_\texttt{G}) - \elpd(\texttt{GIGG-}a_G)$ for the lower-dimensional setting, so that positive values favor the Group-R2 prior. The corresponding high-dimensional figure and the $\Delta \rmse$ figures, both over all coefficients and split by coefficient type, are relegated to Section~\ref{subsec:gigg_calibration} of the Appendix.

The comparison depends strongly on $a_G$. At $a_G = 0.1$ the Group-R2 prior predicts better, with a median $\Delta \elpd$ of $5.4$ and an advantage in $69\%$ of replications. At $a_G = 0.5$ and $a_G = 1$ the ordering reverses, with medians of $-7.8$ and $-28.6$. The same ordering holds in the high-dimensional setting ($p = 500$, Figure~\ref{fig:gigg_elpd_high_dimension} in the Appendix). Parameter recovery follows the same pattern, shown together with the high-dimensional case in Figures~\ref{fig:gigg_rmse_low_dimension} and~\ref{fig:gigg_rmse_high_dimension} of the Appendix.

One possible contributor to this ordering is the rule $a_1 = G a_G$ used to set the $R^2$ prior. Since $a_2 = 0.5$ is fixed, the prior mean of $R^2$ is $G a_G / (G a_G + 0.5)$, which is $0.67$, $0.91$ and $0.95$ at $G = 10$ and $0.91$, $0.98$ and $0.99$ at $G = 50$. The data are generated with $R^2 \in \{0.25, 0.80\}$, so for the larger values of $a_G$ the prior places almost all of its mass above the truth, and increasingly so as $G$ grows. For $a_G = 0.1$ the median $\Delta \elpd$ is $14.6$ when $R^2 = 0.80$ and $-0.6$ when $R^2 = 0.25$ at $p = 100$, and $39.8$ against $-20.4$ at $p = 500$, so the Group-R2 prior is favored precisely where its implied prior mean is close to the data generating value. The calibrated GIGG shares this target. For $a_G \in \{0.5, 1\}$, however, its prior on $R^2$ is even more concentrated near one than the target (Figure~\ref{fig:gigg_implied_r2_calibration} in the Appendix). The location of the prior on $R^2$ therefore does not explain why the GIGG predicts better in these cases. A more likely cause is how the two priors allocate the variance across groups. At $G = 50$, the calibrated GIGG has very small group shapes, $a_g = 0.051$ and $a_g = 0.029$ (Table~\ref{tab:gigg_calibration}). It therefore gives most of the variance to a few groups and shrinks the other groups strongly toward zero. In contrast, the Group-R2 prior with $a_G \in \{0.5, 1\}$ spreads the variance more evenly across the $50$ groups.

The difference in recovery is concentrated in the coefficients that are truly zero. Splitting $\Delta \rmse$ by coefficient type, the median difference on the truly nonzero coefficients stays close to zero in every configuration, whereas on the truly zero coefficients it grows with $a_G$ and with the dimension and accounts for most of the aggregate difference. 
The split figures are given in Section~\ref{subsec:gigg_calibration} of the Appendix.

The quality of the posterior sampling qualifies these numbers, and convergence diagnostics for all six models, split by coefficient type and by scenario, are reported in Tables~\ref{tab:gigg_diag_low} and~\ref{tab:gigg_diag_high} of the Appendix. The posterior geometry induced by the GIGG is the more difficult of the two, and increasingly so as $a_G$ grows and as the dimension increases. In the high-dimensional setting the Group-R2 models attain higher effective sample sizes and lower $\widehat{R}$ than their calibrated counterparts at $a_G \in \{0.5, 1\}$, and for \texttt{GIGG-}$1.0$ the mixing is poor enough that the posterior cannot be regarded as adequately explored. 

Taken together, the Group-R2 compares favorably to the GIGG when the R2 prior is well calibrated to the true $R^2$ and offers the clear advantage of interpretability of the global scale being matched to a prior on the models' $R^2$ and control over the decomposition. 

\section{Discussion}

In this work, we introduced the Group-R2 prior, analyzed its mathematical properties, and evaluated its empirical performance through simulations. The Group-R2 prior is an extension of the R2-based shrinkage priors that incorporates known grouping structures through a two-stage hierarchical decomposition of the prior coefficient of determination $R^2$. We first distribute the total variance across groups and then within each group across individual coefficients. We have shown that this construction preserves desirable shrinkage properties, such as heavy tails and strong concentration near the origin, both at a marginal and group level. We have further proved posterior consistency when the number and the size of the groups grow with the sample size, without assuming sparsity of the true coefficients, and contraction of the fitted mean when there are more predictors than observations. We have also shown how the choice of hyperparameters of the decomposition governs shrinkage behavior at both levels of the hierarchy. We have also provided a practical framework for hyperparameter calibration using interpretable diagnostics such as the effective number of non-zero coefficients, with recommendations on how to further tune hyperparameters based on knowledge of $R^2$ or the type of signal expected from a given group.

Our experiments complement our theoretical results and illustrate the practical value of incorporating grouping structures into R2-based priors. In comparison to non-grouped R2 priors in both lower- and high-dimensional settings, Group-R2 priors consistently improve parameter recovery when the signal is distributed across predictors or randomly mixed. In line with our discussion, increasing $a_G$ enhances recovery by encouraging more even allocations of variance across groups. Gains in predictive accuracy follow similar trends, especially in high dimensions. When the signal was concentrated in a single coefficient within each group, grouping brought little or no benefit and could make recovery worse, depending on how well the prior matched the data generating process. However, we conjecture that different hyperparameter specification might lead to improvements in this scenario.

Beyond the comparison with non-grouped R2-based priors, we compared the Group-R2 prior with a GIGG prior calibrated to the same implied distribution of $R^2$, and with several grouped priors from the literature, including spike-and-slab formulations, the group horseshoe, GRASP and a grouped $g$-prior. The Group-R2 prior was competitive in these comparisons across the different settings considered, particularly when the prior on $R^2$ was close to the explained variance in the data. Its posterior was also sampled more reliably than those of all competing priors. When the prior placed most of its mass near the true value, the Group-R2 prior showed excellent performance. This showcases that R2 elicitation is a crucial step in R2-based priors. In applications we therefore recommend choosing the location of the $R^2$ prior from domain knowledge.

A key limitation of the current framework is its reliance on pre-specified group structures, which are assumed to be known and fixed in advance. In practice, groups may be misspecified due to noisy prior knowledge or arbitrary definitions, leading to misallocation of variance and suboptimal shrinkage. For example, assigning signal covariates to low-variance groups can suppress important effects, while allocating variance to purely noisy groups may inflate uncertainty. This makes the method vulnerable to structural assumptions, particularly in settings where group definitions are uncertain or high-dimensional. Future work should focus on extending the Group-R2 framework to settings where group structure is unknown or partially observed, for instance by coupling with latent group discovery methods or using external similarity measures to inform grouping.

\begin{appendix}

\newpage

\label{sec:supplementary_results}
\section{Proofs}

\subsection*{Proof of Proposition \ref{prop:properties_r2g}}

Recall that $R_g^2=\phi_gR^2$, where $\phi$ and $R^2$ are independent. Since the beta-distributed variable $R^2$ is positive almost surely, division by $R^2$ gives
\begin{equation}
\left(\frac{R_1^2}{R^2},\ldots,\frac{R_G^2}{R^2}\right) =(\phi_1,\ldots,\phi_G)\sim\dirichlet(\alpha_1,\ldots,\alpha_G).
\end{equation}
Write $\alpha_0=\sum_g\alpha_g$. To obtain the distribution of a  single coordinate, represent the Dirichlet vector as $\phi_g=Y_g/\sum_hY_h$, using independent gamma variables $Y_h\sim\gammadist(\alpha_h,1)$, with shape and rate parameterization. Then $\sum_{h\ne g}Y_h$ is gamma with shape $\alpha_0-\alpha_g$, independently of $Y_g$. The ratio of these two gamma components to their sum has the beta distribution, so
\begin{equation}
\frac{R_g^2}{R^2}=\phi_g\sim\betadist(\alpha_g,\alpha_0-\alpha_g).
\end{equation}
Independence ensures that conditioning on $R^2=r^2$ leaves the law of $\phi$ unchanged. The conditional group contributions are therefore a Dirichlet vector multiplied by $r^2$. For distinct $g,h$, the Dirichlet moments give \citep{lin_dirichlet_2016}
\begin{equation}
\begin{aligned}
\mbe(\phi_g)&=\frac{\alpha_g}{\alpha_0},\qquad
\mbe(\phi_g\phi_h)=\frac{\alpha_g\alpha_h}{\alpha_0(\alpha_0+1)},\\
\cov(R_g^2,R_h^2\mid R^2=r^2)
&=(r^2)^2\left\{\frac{\alpha_g\alpha_h}{\alpha_0(\alpha_0+1)}
-\frac{\alpha_g\alpha_h}{\alpha_0^2}\right\} =-\frac{(r^2)^2\alpha_g\alpha_h}{\alpha_0^2(\alpha_0+1)}.
\end{aligned}
\end{equation}
This proves the three assertions of the proposition. \qed

\subsection*{Proof of Proposition \ref{prop:properties_lambda_gl}} 
To prove the proposition we will use the following auxiliary propositions. Their proofs can be found in \cite{zhang_bayesian_2020}.

\begin{proposition}
\label{prop:proof_aux_1}
If $\tau^2 \mid \xi \sim Ga(a_1 , \xi)$ and $\xi \sim Ga(a_2, 1)$, then $\tau^2 \sim \betaprime(a_1,a_2)$. 
\end{proposition}

\begin{proposition}
\label{prop:proof_aux_2}
If $\tau^2 \mid \xi \sim Ga(a_1 , \xi)$ , $\phi \sim \dirichlet s(a_\pi)$ of dimension $k$ and $a_1= k a_\pi$, then $ \phi_j \tau^2  \mid \xi \sim Ga(a_\pi, \xi), j = 1,..., k$ independently.
\end{proposition}

\begin{enumerate}
    \item By hypothesis we have that $\tau^2 \sim \betaprime(G a_G, a_2)$. Applying Proposition~\ref{prop:proof_aux_1} yields

    $$\tau^2 \mid \xi \sim Ga(G a_G , \xi) , \xi \sim Ga(a_2, 1).$$
    
    Using Proposition~\ref{prop:proof_aux_2} with $a_1 =G a_G$ shows that $\tau_g^2 \mid \xi \sim Ga(a_G, \xi)$. Therefore we have

    $$\tau_g^2 \mid \xi \sim Ga(a_G , \xi) , \xi \sim Ga(a_2, 1).$$ 

    Applying Proposition~\ref{prop:proof_aux_1} again gives 
    $\tau_g^2 \sim \betaprime(a_G , a_2)$. $\qed$

    \item The result follows by applying the same procedure that we have used to show the previous result but at a group level. $\qed$
    
\end{enumerate}

\subsection*{Proof of Proposition~\ref{prop:flat_reduction}}

We construct both allocation levels from one family of independent gamma variables. Let $Z_{gl}\sim\gammadist(c_{gl},1)$ independently, and define
\begin{equation}
H_g=\sum_{l=1}^{p_g}Z_{gl},\qquad
H_+=\sum_{g=1}^GH_g,\qquad
a_g^*=\sum_{l=1}^{p_g}c_{gl}.
\end{equation}
Each $H_g$ is gamma with shape $a_g^*$. The different group sums are independent. Within group $g$, gamma normalization gives
\begin{equation}
\left(\frac{Z_{g1}}{H_g},\ldots,\frac{Z_{gp_g}}{H_g}\right)
\sim\dirichlet(c_{g1},\ldots,c_{gp_g}),
\end{equation}
independently of $H_g$. Since different groups use disjoint independent gamma variables, the within-group vectors are mutually independent and independent of the entire vector of group sums. Therefore we may set
\begin{equation}
\phi_g=H_g/H_+,\qquad \varphi_{gl}=Z_{gl}/H_g.
\end{equation}
This construction has the group Dirichlet distribution and all the independence properties in the proposition. The product of the two allocations simplifies to
\begin{equation}
\phi_g\varphi_{gl}=\frac{Z_{gl}}{H_+} =\frac{Z_{gl}}{\sum_{h,r}Z_{hr}}.
\end{equation}
The concatenated product vector is the normalization of independent gamma variables with shapes $(c_{gl})_{g,l}$. Its distribution is the stated Dirichlet distribution. In the case of singleton group, their within-group proportion equals one. \qed

\subsection*{Proof of Proposition \ref{prop:covariance_lambdas}} 

\begin{enumerate}
    \item  We are interested in the quantity
\begin{equation}
\operatorname{Corr} \left( \log( \phi_g \varphi_{gj} \right), \log(\phi_g \varphi_{gk}) )   = \frac{\operatorname{Cov} \left( \log( \phi_g \varphi_{gj} \right), \log(\phi_g \varphi_{gk}) )  }{ \sqrt{  \var( \log( \phi_g \varphi_{gj} ) \var( \log( \phi_g \varphi_{gk} ) } }  , \ \ j \neq k
\end{equation}
Applying properties of the covariance operator and using the fact that $\phi_g$ and $\varphi_g$ are independent we find that:
\begin{equation}
\begin{aligned}
    \operatorname{Cov}( \log(\phi_g \varphi_{gj}), \log(\phi_g \varphi_{gk}) ) &= \operatorname{Cov} \left( \log(\phi_g) + \log( \varphi_{gj}), \log(\phi_g) + \log( \varphi_{gk})    \right)  \\ 
&= \operatorname{Var}(\log \phi_g)+ \operatorname{Cov}(\log \varphi_{gj}, \log \varphi_{gk}).
\end{aligned}
\end{equation}
The main quantity in the denominator is $ \var( \log( \phi_g \varphi_{gj} )$, which can be written as
\begin{equation}
\begin{aligned}
    \var( \log( \phi_g \varphi_{gj} ) & = \var \left( \log(\phi_g) +  \log(\varphi_{gj}) \right) \\
&= \var ( \log(\phi_g) ) +  \var (\log(\varphi_{gj}).
\end{aligned}
\end{equation}
Since $\varphi_{g} \sim \dirichlet s(c_g)$, both $\varphi_{gj}$ and $\varphi_{gk}$ have the same marginal distributions and we have $\var( \log( \phi_g \varphi_{gj} ) = \var ( \log(\phi_g ) +  \var (\log(\varphi_{g\cdot})$. Therefore 
\begin{equation}
\label{eq:corr_proof}
\operatorname{Corr} \left( \log( \phi_g \varphi_{gj} \right), \log(\phi_g \varphi_{gk}) )   = \frac{ \operatorname{Var}(\log \phi_g)+ \operatorname{Cov}(\log \varphi_{gj}, \log \varphi_{gk})  }{ \var ( \log(\phi_g) +  \var (\log(\varphi_{g\cdot}) }. \qed
\end{equation}
The result can be generalized to the case in which the proportions of explained variance do not follow a symmetric Dirichlet, but we do not show that here. 

\item The result follows by substituting the following quantities in Equation~\eqref{eq:corr_proof}:
\begin{equation}
\begin{aligned}
\var(\log \phi_g) &= \psi_1(a_G) - \psi_1(G a_G) \\
\var(\log \varphi_{gl}) &= \psi_1(c_g) - \psi_1(p_g c_g), \ \ l \in \{ 1,..., p_g \} \\
\cov(\log \varphi_{gj}, \log \varphi_{gk}) &= - \psi_1(p_g c_g),  \ \ j \neq k\\
\end{aligned}
\end{equation}
where $\psi_1$ is the trigamma function \citep{olver_nist_2010}. \qed

\item We have 
\begin{equation}
\label{eq:cov_proof}
\begin{aligned}
\cov(\phi_g \varphi_{gj},\ \phi_g \varphi_{gk}) &= \mbe[\phi_g^2 \varphi_{gj} \varphi_{gk}] - \mbe[\phi_g \varphi_{gj}]\mbe[\phi_g \varphi_{gk}]   \\
&= \mathbb{E}[\phi_g^2] \cdot \mathbb{E}[\varphi_{gj} \varphi_{gk}] - \mbe[\phi_g]^2 \mbe[\varphi_{gj} ] \mbe[\varphi_{gk} ], \ \ j \neq k,
\end{aligned}
\end{equation}
where we have used the independence of $\phi_g, \varphi_g$ to expand the expectations. The expressions in the last line have closed analytical forms given by:
\begin{equation}
\begin{aligned}
    \mbe[\phi_{g} ] &= \frac{1}{G}, \ \ \mbe[\varphi_{gl} ] = \frac{1}{p_g}, \ \ l \in \{ 1,..., p_g \},  \\  \mathbb{E}[\phi_g^2] &=\frac{a_g(a_g + 1)}{(G a_g)(G a_g + 1)}  \ \ \mathbb{E}[\varphi_{gj} \varphi_{gk}] = \frac{c_g}{p_g (c_g p_g + 1)}, \ \ j \neq k
\end{aligned}
\end{equation}
The result follows by substituting the expressions in Equation~\eqref{eq:cov_proof}. $\qed$

\end{enumerate}

\subsection*{Proof of Proposition \ref{prop:marginalpriors}} 

Under the assumptions considered, the Group-R2 prior has the following alternative representation
\begin{equation}
b_{gl} \mid \lambda_{gl}^2 \sim \normal \left( 0, \lambda_{gl}^2 \right) \ \ \lambda_{gl}^2 \sim \betaprime(c_g, a_2).
\end{equation}	
The prior marginal density of $b_i$ is given by
\begin{equation}
\begin{aligned}
        p( b_{gl} ) &= \int_0^\infty \frac{1}{ \sqrt{2\pi \lambda_{gl}}^2 }\exp \left\lbrace  -\frac{ b_{gl}^2  }{ 2 \lambda_{gl}^2}    \right\rbrace \frac{1}{\text{B}(c_g, a_2) } (\lambda_{gl}^2)^{c_g-1} \left( \lambda_{
gl}^2+1
    \right)^{-c_g -a_2} d\lambda_{gl}^2 \\
    &= \frac{1}{\sqrt{2\pi } \operatorname{B}( c_g, a_2) } \int_0^\infty \exp \left\lbrace  -\frac{ b_{gl}^2  }{ 2 } t_{gl}    \right\rbrace t_{gl}^{\eta-1} (t_{gl}+1)^{ \nu-\eta-1  } dt_{gl}\\
    &= \frac{1}{ \sqrt{2\pi} \operatorname{B}(c_g, a_2) } \Gamma(\eta) U( \eta, \nu, z_{gl} ), 
\end{aligned}
\end{equation}
where $t_{gl}=\frac{1}{\lambda_i^2}$, $\eta=a_2+1/2$ and $\nu= 3/2-c_g$. $U(\eta, \nu, z_{gl})$ represents the confluent hypergeometric function of the second kind \citep{Zwillinger}) and $z_{gl}=\frac{|b_{gl}|^2}{2}$. The function $U(\eta,\nu, z_{gl})$ is defined as long as the real part of $\eta$ and $z_{gl}$ are positive, which is always the case since $a_2>0$ and $|b_{gl}|>0$. $\qed$ 

\subsection*{Proof of Proposition \ref{prop:originprior}} 

In the following we drop the index of $z_{gl}$, since all cases are handled equally. The confluent hypergeometric function of the second kind $U(\eta, \nu, z)$ satisfies the following identity \cite{nisthandbook}).
\begin{equation}
\label{eq:Utransformation}
   U(\eta,\nu, z)=z^{1-\nu} U(\eta-\nu+1, 2-\nu, z). 
\end{equation}
Substituting $\eta=a_2+1/2, \nu= 3/2-c_g$ and $z= \frac{t^2}{2}$ results in
\begin{equation}
U\left(a_2+1/2, 3/2-c_g, \frac{t^2}{2} \right)= 2^{1/2-c_g} |t|^{2 c_g-1} U\left(a_2+ c_g , c_g +1/2,  \frac{t^2}{2}\right).
\end{equation}
The right hand side has a singularity at $t=0$ when $c_g <1/2$. We consider the limiting form $U$ has as $|z|\to 0$ to study the order of concentration near the origin for different values of $c_g$. Equation 13.2.18 in \cite{nisthandbook} states
\begin{equation}
U(\eta, \nu, z)= \frac{\Gamma(\nu-1)}{\Gamma(\eta)} z^{1-
\nu}+\frac{\Gamma(1-\nu)}{\Gamma(\eta-\nu+1)} +\mathcal{O}\left( z^{2-\nu}\right) , \ \ 1< \nu <2 .
\end{equation}
Substituting $\eta=a_2+1/2, \nu= 3/2-c_g$ and $z= \frac{t^2}{2}$ results in
\begin{equation}
U\left( a_2+1/2, 3/2-c_g, \frac{t^2}{2}\right)= C t^{2c_g-1} + \frac{\Gamma(c_g+1/2)}{\Gamma(a_2+1/2)} +\mathcal{O}\left( |t|^{1+2c_g} \right) , \ \ 0< c_g <1/2 ,    
\end{equation}
where $C= 2^{1/2-c_g} \frac{\Gamma(1/2-c_g)}{\Gamma(a_2+c_g)}$. The last expression is unbounded at $t=0$ when $c_g<1/2$. The case when $c_g=1/2$ follows equation 13.2.19 in \cite{nisthandbook} given by 
\begin{equation}
 U(\eta, \nu, z)= -\frac{1}{\Gamma(\eta)}\left( \ln(z)+\psi(\eta)+2\gamma \right)  +\mathcal{O}\left(  z \ln (z)  \right) , \ \ \nu = 1 ,
\end{equation}
where $\psi(\cdot), \gamma$ represent the digamma function and the Euler-Mascheroni constant respectively. Substituting $\eta=a_2+1/2, \nu=3/2-c_g$ and $z=\frac{t^2}{2}$ gives
\begin{equation}
  \begin{aligned}
      U\left( a_2+1/2, 3/2-c_g, \frac{t^2}{2}\right)&= -\frac{1}{\Gamma(a_2+1/2)}\left( \ln\left( \frac{t^2}{2} \right)+\psi(a_2+1/2)+2\gamma \right) \\&\qquad +\mathcal{O}\left(  t^2 \ln \left(\frac{t^2}{2} \right) \right), \ \  c_g = 1/2.  
  \end{aligned}
\end{equation}
The last expression is not defined at $t=0$ and goes to infinity as $|t| \to 0$ . To see the marginal priors are bounded when $c_g>1/2$ it is sufficient to consider the cases given by equations 13.2.20-13.2.22 in  \cite{nisthandbook} and making the proper substitutions of $\eta=a_2+1/2, \nu= 3/2-c_g$ and $z=\frac{t^2}{2} $. In terms of the values  $c_g$ takes we have
\begin{equation}
U(\eta, \nu, t)\sim
\begin{cases}
	 \mathcal{O} \left( t^{2c_g-1}   \right), \ \ \  &1/2<c_g<3/2 \\ 
	 \mathcal{O} \left( t^2 \ln(t^2/2)   \right), \ \ \  &c_g=3/2 \\ 
	 \mathcal{O} \left( t^2  \right), \ \ \  &c_g>3/2 \\ 
\end{cases}
\end{equation}
The derivative of $U(\eta, \nu, z)$ is given by equation 13.3.22 in \cite{nisthandbook} as
\begin{equation}
\frac{d}{dz}  U(\eta, \nu, z)= - \eta U(\eta+1, \nu+1, z).      
\end{equation}
Combining this with Equation \eqref{eq:Utransformation} gives
\begin{equation}
\frac{d}{dz}  U(\eta, \nu, z) = -\eta z^{-\nu} U(\eta-\nu  +1, 1- \nu, z ). 
\end{equation}
Substituting $\eta=a_2+1/2, \nu= 3/2-c_g$ and $z= \frac{t^2}{2}$ results in
\begin{equation}
\label{eq:dudt}
\frac{d}{dt}  U\left( a_2+1/2 , 3/2-c_g , \frac{t^2}{2} \right) = C t^{2c_g-2} U\left( a_2+c_g, c_g-1/2, \frac{t^2}{2} \right),     
\end{equation}
where $C=-(a_2+1/2)2^{3/2-c_g}$. Equation \eqref{eq:dudt} is undefined at $t=0$ when $c_g<1$. $\qed$

\subsection*{Proof of Proposition \ref{prop:tailprior}} 

To prove the proposition we make use of Watson's lemma \citep{miller_applied_2006}.

\begin{proposition}[Watson's lemma]
Let $0\leq T \leq \infty$ be fixed. Assume $f(t)= t^{\lambda} g(t)$, where $g(t)$ has an infinite number of derivatives in the neighborhood of $t=0$, with $g(0)\neq 0$, and $\lambda > -1$. Suppose, in addition, either that $ |f(t)| < K e^{ct}$ for any $t>0$, where $K$ and $c$ are independent of $t$. Then it is true that for all positive $x$ that 
\begin{equation}
\left| \int_0^T e^{-xt} f(t) dt \right| <\infty     
\end{equation}
and that the following asymptotic equivalence holds:
\begin{equation}
\int_0^T e^{-xt} f(t) dt  \sim \sum_{n=0}^\infty \frac{ g^{(n)}(0) \Gamma \left( \lambda+n+1 \right) }{n! x^{\lambda+n+1 }}	,   
\end{equation}
 for $x>0$ as $x\to \infty$. \\
\end{proposition}

The marginal distribution of a coefficient $b$ is given by
\begin{equation}
   \begin{aligned}
        p(b)&= \frac{1}{\sqrt{2\pi} \operatorname{B}(c_g, a_2) } \int_{0}^\infty \exp \left\lbrace  -\frac{|b|^2}{2} t \right\rbrace t^{\eta-1} (t+1)^{\nu-\eta-1}  dt \\   &= \int_0^\infty \exp \left\lbrace  -z t  \right\rbrace f(t) dt,
   \end{aligned} 
\end{equation}
where $ z=\frac{|b|^2}{2},f(t)= C \, t^{\eta-1} (t+1)^{\nu-\eta-1} = t^{\eta-1} g(t), C=  \left(\sqrt{2\pi } \text{B}(c_g, a_2) \right)^{-1},$ and $g(t)=C \,(t+1)^{\nu-\eta-1}$.  Setting $\lambda=\eta-1$, the hypothesis $\lambda >-1$ is satisfied since $a_2-1/2 > -1$ for $a_2>0$. Moreover, $g(t)$ is infinitely differentiable around $t=0$ and $g(0) \neq 0$. By Watson's Lemma, since $|f(t)|< K e^{ct}$ for all $t>0$ where $K$ and $c$ are independent of $t$, then as $|b| \to \infty,$
\begin{equation}
p(z \mid \sigma)= \sum_{n=0}^\infty  \frac{g^{(n)}(0) \Gamma(\lambda+n+1)}{n! z^{\lambda+n+1}}.        
\end{equation}
Truncating the sum at $n=2$ gives
\begin{equation}
\begin{aligned}
        p(z )&= C\left\lbrace \frac{ \Gamma(a_2+1/2)}{z^{a_2+1/2}} -
    \frac{ (c_g+a_2) \Gamma(a_2+3/2)}{z^{a_2+3/2}}+
    \frac{ (c_g+a_2) (c_g+a_2+1) \Gamma(a_2+5/2)}{z^{a_2+5/2}}  \right\rbrace  \\ 
    &\qquad + \mathcal{O}\left(  \frac{1}{ z^{a_2+7/2}}\right) \\
    &\sim \mathcal{O} \left(  \frac{1}{ z^{a_2+1/2}}\right).
\end{aligned}
\end{equation}
Therefore $p\left( |b| \right)\sim \mathcal{O} \left(  \frac{1}{ |b|^{2a_2+1}}\right)$. When $a_2< 1/2$ as $|b|\to \infty$ and comparing with a Cauchy distribution represented by $\frac{1}{b^2}$, we have that 
\begin{equation}
    \frac{p(b )}{ \frac{1}{b^2} } \sim \mathcal{O} \left( \frac{ 1 }{ b^{2a_2-1}  } \right)      \to \infty  .
\end{equation}

\subsection*{Proof of Proposition \ref{prop:r2d2-horseshoe}} 

Since we have shown that  $\lambda_{gl}^2 \sim \betaprime( c_g, a_2 )$, we can write the Group-R2 prior as
\begin{equation}
    b_{gl} \mid \lambda_{gl}^2 \sim \normal(0, \lambda_{gl}^2), \ \ 
    \lambda_{gl}^2 \sim \betaprime \left( c_g, a_2 \right).
\end{equation}
When $c_g= a_2 = 1/2$, then $\lambda_{gl} \sim \text{Cauchy}^+(0,1)$ and we have

\begin{equation}
 b_{gl} \mid \lambda_{gl}^2 \sim \normal(0, \lambda_{gl}^2), \ \  \lambda_{gl} \sim \text{Cauchy}^+(0,1).
\end{equation}

The horseshoe prior has the following representation \citep{carvalho_horseshoe_2010}:
\begin{equation}
    \begin{aligned}
        b_{gl} \mid \lambda_{gl}, \omega &\sim \normal(0,  \lambda_{gl}^2 \omega^2 ),\ \
    \lambda_{gl} \sim \text{Cauchy}^+(0,1),
    \end{aligned}
\end{equation}
where $\lambda_{gl}$ are per coefficient shrinkage parameters and $\omega$ is the global shrinkage parameter. Therefore the Group-R2 prior with $c_g = a_2 =\tfrac{1}{2}$ corresponds to a horseshoe prior with global scale $\omega= 1$. This proves the first part of the proposition.

To see that bounded influence holds, consider the scalar normal-means model $y\mid b\sim\normal(b,1)$. \citet{carvalho_horseshoe_2010} shows the following posterior-mean identity
\begin{equation}
\label{eq:postmeannormalmeans}
\mbe(b\mid y)-y=\frac{m'(y)}{m(y)}.
\end{equation}
We call the right hand part of the equation the score. Since $y\mid\lambda\sim\normal(0, 1+\lambda^2)$, integrating over the half-Cauchy density gives
\begin{equation}
\begin{aligned}
m(y)&=\frac{2}{\pi\sqrt{2\pi}}\int_0^\infty
\exp\left\{-\frac{y^2}{2(1+\lambda^2)}\right\}
(1+\lambda^2)^{-3/2}\,d\lambda. \\
&= \frac{1}{\pi\sqrt{2\pi}}\int_0^1
e^{-y^2z/2}(1-z)^{-1/2}\,dz.
\end{aligned}
\end{equation}
where we have used the change of variables $z=(1+\lambda^2)^{-1}$. To evaluate this integral, set $u=\sqrt{1-z}$. Define the imaginary error function by 
\begin{equation}
\operatorname{erfi}(x)=\frac{2}{\sqrt\pi}\int_0^x e^{t^2}\,dt.
\end{equation}
Then, for $y\ne0$,
\begin{equation}
\begin{aligned}
m(y) &=\frac{2e^{-y^2/2}}{\pi\sqrt{2\pi}} \int_0^1 e^{y^2u^2/2}\,du\\
&=\frac{e^{-y^2/2}}{\pi y} \operatorname{erfi}\left(\frac{y}{\sqrt2}\right).
\end{aligned}
\end{equation}
At $y=0$, the preceding integral gives $m(0)=2/(\pi\sqrt{2\pi})$. Differentiating for $y\ne0$, using $\operatorname{erfi}'(x)=2e^{x^2}/\sqrt\pi$, gives
\begin{equation}
\label{eq:horseshoe_score}
\frac{m'(y)}{m(y)}
=-y-\frac1y+
\frac{\sqrt{2/\pi}\,e^{y^2/2}}
{\operatorname{erfi}(y/\sqrt2)}.
\end{equation}

We now evaluate the tail limit of this expression. The error-function asymptotic expansion \citep{olver_nist_2010}, applied to $\operatorname{erfi}(x)=-i\operatorname{erf}(ix)$, gives
\begin{equation}
\operatorname{erfi}\left(\frac{y}{\sqrt2}\right)
=\frac{\sqrt{2/\pi}\,e^{y^2/2}}{y}
\left\{1+\frac1{y^2}+O(y^{-4})\right\},
\qquad y\to+\infty.
\end{equation}
Hence, the last term of Equation~\eqref{eq:horseshoe_score} satisfies
\begin{equation}
\begin{aligned}
\frac{\sqrt{2/\pi}\,e^{y^2/2}}
{\operatorname{erfi}(y/\sqrt2)}
&=y\left\{1+\frac{1}{y^2}+O(y^{-4})\right\}^{-1}=y-\frac{1}{y}+O(y^{-3}).
\end{aligned}
\end{equation}
Substituting into Equation~\eqref{eq:horseshoe_score} gives
\begin{equation}
\frac{m'(y)}{m(y)}
= -\frac{2}{y}+  O(y^{-3}).
\end{equation}
Since $m$ is even, $\frac{m'(y)}{m(y)}$ is odd and the expansion holds in both tails. Therefore
\begin{equation}
\mbe(b\mid y)-y=\frac{m'(y)}{m(y)}
=-\frac2y+O(|y|^{-3})\longrightarrow0,
\qquad |y|\to\infty.
\end{equation}

It remains to check boundedness near zero. The Taylor expansion at zero gives
\begin{equation}
\begin{aligned}
\operatorname{erfi}(y/\sqrt2)
&=\sqrt{\frac{2}{\pi}}\left\{y+\frac{y^3}{6}+O(y^5)\right\}.
\end{aligned}
\end{equation}
Therefore
\begin{equation}
\frac{m'(y)}{m(y)} =-\frac{2y}{3}+O(y^3),\qquad y\to 0.    
\end{equation}

Thus the score extends continuously to zero with value zero. Since $m(y)>0$, it is continuous and bounded elsewhere. We have shown that its limit is zero in both tails. Together this shows it is bounded on $\mathbb R$. \qed

\subsection*{Proof of Proposition~\ref{prop:uncoupled_marginals}}

In the following we suppress the coefficient subscripts. Write $T=\tau_g^2$ and $V=\varphi_{gl}$. Then
\begin{equation}
\begin{aligned}
T&\sim\betaprime(a_G,a_2),\qquad b\mid T,V\sim\normal(0,TV),\\
V&\sim\betadist(c_g,(p_g-1)c_g)\quad\text{if }p_g>1,
\end{aligned}
\end{equation}
with $T$ independent of $V$. For $p_g=1$, we let $V=1$.

We start with the behavior near zero. The mixture density is
\begin{equation}
p(b)=\int k_b(t,v)\,d(P_T\otimes P_V)(t,v),
\end{equation}
where $k_b(t,v)= e^{-b^2/(2tv)}/ \sqrt{2\pi tv}$. For any sequence $r_m\downarrow0$ with $r_m>0$, the kernels satisfy $k_{r_m}(t,v)\uparrow k_0(t,v)$ for every $t,v>0$. Monotone convergence, followed by Tonelli's theorem and independence, gives
\begin{equation}
\begin{aligned}
\lim_{m\to\infty}p(r_m)
&=\int\lim_{m\to\infty}k_{r_m}(t,v)\,d(P_T\otimes P_V)(t,v)\\
&=\int\frac{1}{\sqrt{2\pi tv}}\,d(P_T\otimes P_V)(t,v)\\
&=\frac{1}{\sqrt{2\pi}}\mbe(T^{-1/2})\mbe(V^{-1/2})=p(0).
\end{aligned}
\end{equation}
Since $p(b)$ is even and nonincreasing in $|b|$, this also establishes the limit as $|b|\to0$.

The first negative half moment is
\begin{equation}
\begin{aligned}
\mbe(T^{-1/2})
&=\frac{1}{B(a_G,a_2)}
\int_0^\infty t^{a_G-3/2}(1+t)^{-a_G-a_2}\,dt\\
&=\begin{cases}
\dfrac{B(a_G-1/2,a_2+1/2)}{B(a_G,a_2)},&a_G>1/2,\\
+\infty,&a_G\le1/2.
\end{cases}
\end{aligned}
\end{equation}
When $p_g>1$, the second moment is
\begin{equation}
\begin{aligned}
\mbe(V^{-1/2})
&=\frac{1}{B(c_g,(p_g-1)c_g)}
\int_0^1v^{c_g-3/2}(1-v)^{(p_g-1)c_g-1}\,dv\\
&=\begin{cases}
\dfrac{B(c_g-1/2,(p_g-1)c_g)}{B(c_g,(p_g-1)c_g)},&c_g>1/2,\\
+\infty,&c_g\le1/2.
\end{cases}
\end{aligned}
\end{equation}
The finiteness conditions follow from the powers $t^{a_G-3/2}$ and $v^{c_g-3/2}$ at zero. Hence,  $p(0)<\infty$ exactly when $a_G>1/2$ and $c_g>1/2$, and otherwise $p(b)\to\infty$ as $|b|\to0$. For $p_g=1$, $\mbe(V^{-1/2})=1$, so only $a_G>1/2$ is needed.

We now show the tail behavior. For fixed $v>0$, integrating over $T$ gives
\begin{equation}
p(b\mid v)=\frac{1}{\sqrt{2\pi v}\,B(a_G,a_2)}
\int_0^\infty t^{a_G-3/2}(1+t)^{-a_G-a_2}e^{-b^2/(2vt)}\,dt.
\end{equation}
For $b\ne0$, substitute $u=b^2/(2vt)$ and then integrate over $V$. This combined with Tonelli yields
\begin{equation}
\begin{aligned}
|b|^{2a_2+1}p(b)
&=K\int_{(0,1]}\int_0^\infty
v^{a_2}e^{-u}u^{a_2-1/2}
\left(1+\frac{2vu}{b^2}\right)^{-a_G-a_2}\,du\,dP_V(v),
\end{aligned}
\end{equation}
with $K=\frac{2^{a_2}}{\sqrt{\pi}\,B(a_G,a_2)}$. 

Now, let $(b_m)$ be any sequence of nonzero numbers with $|b_m|\to\infty$. Define
\begin{equation}
\begin{aligned}
F_m(u,v)&=v^{a_2}e^{-u}u^{a_2-1/2}
\left(1+\frac{2vu}{b_m^2}\right)^{-a_G-a_2},\quad
D(u,v)&=v^{a_2}e^{-u}u^{a_2-1/2}.
\end{aligned}
\end{equation}
For every $u>0$ and $0<v\le1$, $F_m(u,v)\to D(u,v)$ and $0\le F_m(u,v)\le D(u,v)$. The dominating function is integrable with respect to $du\,dP_V(v)$, since
\begin{equation}
\int_{(0,1]}\int_0^\infty D(u,v)\,du\,dP_V(v)
=\Gamma(a_2+1/2)\mbe(V^{a_2})
\le\Gamma(a_2+1/2)<\infty.
\end{equation}
Dominated convergence therefore gives
\begin{equation}
\begin{aligned}
\lim_{m\to\infty}|b_m|^{2a_2+1}p(b_m)
&=K\int_{(0,1]}\int_0^\infty
\lim_{m\to\infty}F_m(u,v)\,du\,dP_V(v)\\
&=\frac{2^{a_2}\Gamma(a_2+1/2)}{\sqrt{\pi}\,B(a_G,a_2)}
\mbe(V^{a_2}),
\end{aligned}
\end{equation}
where
\begin{equation}
\mbe(V^{a_2})=
\begin{cases}
\dfrac{B(c_g+a_2,(p_g-1)c_g)}{B(c_g,(p_g-1)c_g)},&p_g>1,\\
1,&p_g=1.
\end{cases}
\end{equation}
Since the sequence was arbitrary, this proves the tail limit as $|b|\to\infty$. \qed

\subsection*{Proof of Proposition \ref{prop:jointdist_group}}

For fixed group $g$, write $\Sigma_g = \tau_g^2 \operatorname{diag}(\varphi_{g1},\ldots,\varphi_{g p_g}).$ Then
\begin{equation}
b_g \mid \tau_g^2,\varphi_g \sim \normal_{p_g}(0,\Sigma_g). 
\end{equation}
We are interested in the marginal density $p(b_g\mid \varphi_g)$ obtained by integrating out $\tau_g^2$. For notational convenience, set $t=\tau_g^2$. Since $\tau_g^2\sim \betaprime(a_G,a_2)$, we have
\begin{equation}
 p(t) = \frac{1}{B(a_G,a_2)} t^{a_G-1}(1+t)^{-(a_G+a_2)}.
\end{equation}
Therefore,
\begin{equation}
  \begin{aligned}
    p(b_g\mid \varphi_g)
    &=
    \int_0^\infty
    p(b_g\mid t,\varphi_g)p(t)\,dt \\
    &= \int_0^\infty (2\pi)^{-p_g/2} t^{-p_g/2}
    \left(\prod_{l=1}^{p_g}\varphi_{gl}\right)^{-1/2}
    \exp\left\{ -\frac{1}{2t}\sum_{l=1}^{p_g}\frac{b_{gl}^2}{\varphi_{gl}}
    \right\}
    \\
    &\quad \times
    \frac{1}{B(a_G,a_2)}
    t^{a_G-1}
    (1+t)^{-(a_G+a_2)}
    dt.
\end{aligned}  
\end{equation}
Collecting terms gives

\begin{equation}
 p(b_g\mid \varphi_g) = K_g \int_0^\infty t^{a_G-p_g/2-1} (1+t)^{-(a_G+a_2)} \exp\left\{ -\frac{z_g}{t} \right\}
    dt,
    \end{equation}
where $ K_g = (2\pi)^{-p_g/2} / B(a_G,a_2) \left(\prod_{l=1}^{p_g}\varphi_{gl}\right)^{-1/2}$ and $z_g= 1/2 \sum_{l=1}^{p_g} b_{gl}^2/ \varphi_{gl}$. Now use the change of variables $u=1/t$, we have
\begin{equation}
 p(b_g\mid \varphi_g) = K_g \int_0^\infty u^{a_2+p_g/2-1} (1+u)^{-(a_G+a_2)} \exp\{-z_g u\} du.   
\end{equation}
Recall the integral representation of the confluent hypergeometric function of the second kind \citep{olver_nist_2010},
\begin{equation}
  U(\eta,\nu,z) = \frac{1}{\Gamma(\eta)} \int_0^\infty e^{-zu} u^{\eta-1} (1+u)^{\nu-\eta-1} du.
\end{equation}
Matching terms gives $\eta=a_2+ p_g/2$, $\nu = 1+ p_g/2-a_G$. Hence,
\begin{equation}
  p(b_g \mid \varphi_g) = \frac{(2\pi)^{-p_g/2}}{B(a_G,a_2)} \left(\prod_{l=1}^{p_g}\varphi_{gl}\right)^{-1/2} \Gamma\left(a_2+\frac{p_g}{2}\right) U\left( a_2+\frac{p_g}{2}, 1+\frac{p_g}{2}-a_G, z_g \right).
\end{equation}

We now study the behavior at the group origin. Conditional on fixed $\varphi_g$ in the interior of the simplex, $b_g\to 0$ implies $z_g\to 0$. The small-argument behavior of $U(\eta,\nu,z)$ \citep{olver_nist_2010} implies that, when $\nu>1$,
\begin{equation}
   U(\eta,\nu,z)
    \sim
    \frac{\Gamma(\nu-1)}{\Gamma(\eta)} z^{1-\nu},
    \qquad z\to 0. 
\end{equation}
Here $\nu>1$ is equivalent to $1+\frac{p_g}{2}-a_G>1$ or $a_G < p_g/2 $. Therefore, if $a_G<p_g/2$, the joint density diverges as $b_g\to 0$. When $a_G=p_g/2$, we have $\nu=1$ and the divergence is logarithmic. When $a_G>p_g/2$, the density is bounded at the group origin \citep{olver_nist_2010}. This proves the result. \qed

\subsection*{Proof of Proposition~\ref{prop:kappa_dependence}}

Fix a group $g$ with $p_g\ge2$ and two distinct coordinates $j,k$. We first study the covariance at a conditional level. Condition on $\tau_g^2=t>0$, and write $X=\varphi_{gj}$ and $Y=\varphi_{gk}$. Then
\begin{equation}
\kappa_{gj}=f_t(X),\qquad \kappa_{gk}=f_t(Y),\qquad
f_t(x)=\frac{1}{1+tx}.
\end{equation}
Since the allocation vector is independent of the group scale, we can condition on $t$ while leaving its Dirichlet law unchanged. Given $X=x$, the remaining allocations sum to $1-x$, and the conditional Dirichlet distribution gives
\begin{equation}
Y\mid X=x\ \stackrel{d}{=}\ (1-x)U,
\end{equation}
where $U\sim\betadist(c_g,(p_g-2)c_g)$ if $p_g>2$, and $U=1$ if $p_g=2$. The law of $U$ does not depend on $x$, therefore
\begin{equation}
h_t(x):=\mbe(\kappa_{gk}\mid X=x,\tau_g^2=t) =\mbe\left\{\frac{1}{1+t(1-x)U}\right\}.
\end{equation}
For $x<x'$ and $u>0$,
\begin{equation}
\frac{1}{1+t(1-x)u}<\frac{1}{1+t(1-x')u}.
\end{equation}
Since $U>0$, $h_t$ is strictly increasing, whereas $f_t$ is strictly decreasing.

Let $X'$ be an independent copy of $X\sim\betadist(c_g,(p_g-1)c_g)$. Conditioning first on $X$ gives
\begin{equation}
\begin{aligned}
\cov(\kappa_{gj},\kappa_{gk}\mid\tau_g^2=t)
&=\mbe\{f_t(X)h_t(X)\}-\mbe\{f_t(X)\}\mbe\{h_t(X)\}\\
&=\frac12\mbe\left\{ [f_t(X)-f_t(X')][h_t(X)-h_t(X')]
\right\}.
\end{aligned}
\end{equation}
The second equality follows by expanding the product and using independence for its cross terms. Since the functions are bounded, all expectations are finite. Whenever $X\ne X'$ strict monotonicity makes the product negative. Since $\Pr(X\ne X')=1$, this proves the strict negative conditional covariance.

We now consider the unconditional covariance case. Assume $a_1=Ga_G$ and $p_gc_g=a_G$. Proposition~\ref{prop:properties_lambda_gl} gives the representation
\begin{equation}
\begin{aligned}
\xi&\sim\gammadist(a_2,1),\qquad
Z_{gj},Z_{gk}\stackrel{\mathrm{iid}}{\sim}\gammadist(c_g,1),\\
\lambda_{g\ell}^2&=\frac{Z_{g\ell}}{\xi},\qquad
\kappa_{g\ell}=\frac{\xi}{\xi+Z_{g\ell}},
\quad \ell\in\{j,k\},
\end{aligned}
\end{equation}
where the gamma is parametrized in terms of shape and rate, and $\xi,Z_{gj},Z_{gk}$ are mutually independent.

Given $\xi$, the two shrinkage factors are independent with the same conditional mean. Writing $Z\sim\gammadist(c_g,1)$, define
\begin{equation}
m_g(x)=\mbe\left(\frac{x}{x+Z}\right),\qquad x>0.
\end{equation}
Thus $\mbe(\kappa_{gj}\mid\xi)=\mbe(\kappa_{gk}\mid\xi)=m_g(\xi)$. Therefore, the law of total covariance yields
\begin{equation}
\begin{aligned}
\cov(\kappa_{gj},\kappa_{gk})
&=\mbe\{\cov(\kappa_{gj},\kappa_{gk}\mid\xi)\} + \cov\{\mbe(\kappa_{gj}\mid\xi),\mbe(\kappa_{gk}\mid\xi)\}\\
&=\var\{m_g(\xi)\}.
\end{aligned}
\end{equation}
To establish strict positivity, take $0<x<x'$. Since $Z>0$ almost surely,
\begin{equation}
m_g(x')-m_g(x)
=\mbe\left\{\frac{Z(x'-x)}{(x+Z)(x'+Z)}\right\}>0.
\end{equation}
Thus $m_g$ is strictly increasing. Since $\xi$ is nondegenerate, $m_g(\xi)$ is nonconstant. Moreover it is bounded between zero and one. Consequently,
\begin{equation}
\cov(\kappa_{gj},\kappa_{gk})=\var\{m_g(\xi)\}>0.
\end{equation}
\qed

\subsection*{Shrinkage Factors moments}
\label{subsec:proof_kappa_moments}

For $p_g\ge2$, the beta integral for $\varphi_{gl}$ gives
\begin{equation}
\mbe(\kappa_{gl}^m\mid t)=\frac1{B(c_g,(p_g-1)c_g)}
\int_0^1\frac{v^{c_g-1}(1-v)^{(p_g-1)c_g-1}}{(1+tv)^m}\,dv,
\end{equation}
which is the Euler representation of ${}_2F_1(m,c_g;p_gc_g;-t)$ \citep{olver_nist_2010}. \qed

\subsection*{Proof of Theorem~\ref{thm:shrinkage_bounds}}
We first consider the probability in part~(a). For notational convenience, let $T=\tau^2$ and let $W_j=\phi_g\varphi_{gl}$ denote the proportion of total variance assigned to coefficient $j=(g,l)$. The allocation vector $W=(W_1,\ldots,W_p)$ satisfies $W_j>0$ almost surely and $\sum_jW_j=1$. For given values $T=t$ and $W=w$, the shrinkage factor is $\kappa_j(t,w)=(1+tw_j)^{-1}$. Integrating over the conditional normal prior of $b_j$ gives
\begin{equation}
y_j\mid T=t,W=w,\sigma
\sim\normal\!\left(0,\frac{\sigma^2}{\kappa_j(t,w)}\right).
\end{equation}
The observations are independent conditional on $T,W$ and $\sigma$, so their joint density is the product of these normal densities. Denote this product, after removing the common factor $(2\pi\sigma^2)^{-p/2}$, by $L_y(t,w)$. Using $z_j=y_j^2/(2\sigma^2)$ as in the theorem, we obtain
\begin{equation}
\label{eq:tail_robustness_likelihood}
L_y(t,w)=\prod_{j=1}^p\kappa_j(t,w)^{1/2}
\exp\!\left\{-\sum_{j=1}^p z_j\kappa_j(t,w)\right\}.
\end{equation}
We can now express the posterior probability of an event involving $T$ and $W$ by averaging this likelihood under their joint prior. Let $\mbe_\pi$ denote expectation under that prior, and let $D(y)=\mbe_\pi\{L_y(T,W)\}$ be the normalizing constant. Since $0<L_y(t,w)\leq1$, we have $0<D(y)\leq1$, and Bayes' formula gives, for any event $E$ involving the scale and allocations,
\begin{equation}
\label{eq:tail_robustness_bayes}
\Pr(E\mid y,\sigma) =\frac{\mbe_\pi\{\mathbf{1}_E L_y(T,W)\}}{D(y)}.
\end{equation}
For the event $\{\kappa_j>\delta\}$, the factor $e^{-z_j\kappa_j}$ bounds the numerator by $e^{-\delta z_j}$. Hence, we need a lower bound for $D(y)$ to determine how this likelihood bound carries over to the posterior.

Recall that $R(y)=1+\sum_jz_j$ and denote it by $R$ for the remainder of the calculation. We obtain the lower bound by considering allocations for which every coefficient receives at least a fixed proportion $\eta$ of the total variance, and scales $t$ between $R$ and $2R$. Such allocations have positive prior probability for sufficiently small $\eta>0$. Indeed, since $p$ is fixed and every $W_j$ is positive almost surely, the events $\{\min_jW_j\geq1/m\}$ increase to $\{\min_jW_j>0\}$ as $m\to\infty$. Continuity of probability from below gives
\begin{equation}
\lim_{m\to\infty}\Pr\!\left(\min_jW_j\geq1/m\right) =\Pr\!\left(\min_jW_j>0\right)=1.
\end{equation}
Choose $\eta>0$ for which $q_\eta=\Pr(\min_jW_j\geq\eta)>0$. For an allocation in this event, $w_j\leq1$ and $R\leq t\leq2R$ imply
\begin{equation}
\frac{1}{3R}\leq\frac{1}{1+2R} \leq\kappa_j(t,w)\leq\frac{1}{1+\eta R}.
\end{equation}
The lower bound controls the product of the factors $\kappa_j^{1/2}$ in $L_y$, while the upper bound controls its exponential term. In particular,
\begin{equation}
\sum_{j=1}^p z_j\kappa_j(t,w)
\leq\frac{R-1}{1+\eta R}\leq\frac1\eta,
\end{equation}
so substituting both bounds into Equation~\eqref{eq:tail_robustness_likelihood} gives
\begin{equation}
L_y(t,w)\geq3^{-p/2}R^{-p/2}e^{-1/\eta}.
\end{equation}
It remains to account for the prior probability of $R\leq T\leq2R$. The beta prime density is $\pi_T(t)=t^{a_1-1}(1+t)^{-a_1-a_2}/B(a_1,a_2)$ for $t>0$. Since $1+t\leq2t$ on this interval, integration yields
\begin{equation}
\begin{aligned}
\Pr(R\leq T\leq2R)
&\geq\frac{2^{-(a_1+a_2)}}{B(a_1,a_2)}
\int_R^{2R}t^{-a_2-1}\,dt =\frac{2^{-(a_1+a_2)}(1-2^{-a_2})}{a_2B(a_1,a_2)}
R^{-a_2}.
\end{aligned}
\end{equation}
Denote the positive coefficient of $R^{-a_2}$ by $d_0$. Restricting the prior expectation in $D(y)$ to these scales and allocations, and using the independence of $T$ and $W$, gives
\begin{equation}
\label{eq:tail_robustness_denominator}
\begin{aligned}
D(y)
&\geq q_\eta\,3^{-p/2}R^{-p/2}e^{-1/\eta}
\Pr(R\leq T\leq2R)\\
&\geq c_0R^{-\alpha},
\end{aligned}
\end{equation}
where $c_0=q_\eta d_0 3^{-p/2}e^{-1/\eta}>0$ and $\alpha=a_2+p/2$. These constants depend only on the fixed prior hyperparameters and group structure.

We now return to the probability of $\kappa_j>\delta$. Each factor in $L_y$ is at most one, and on this event $e^{-z_j\kappa_j}\leq e^{-\delta z_j}$. Thus, for any $j$ and $\delta\in(0,1)$,
\begin{equation}
\begin{aligned}
\Pr(\kappa_j>\delta\mid y,\sigma)
&=\frac{\mbe_\pi\{\mathbf{1}_{\{\kappa_j>\delta\}}L_y(T,W)\}}{D(y)}\\
&\leq c_0^{-1}R^\alpha e^{-\delta z_j}.
\end{aligned}
\end{equation}
Let $C=c_0^{-1}$. Since $\max_{j\in A}\kappa_j>\delta$ means that at least one of the events $\{\kappa_j>\delta\}$ occurs, the union bound gives
\begin{equation}
\begin{aligned}
\Pr\!\left(\max_{j\in A}\kappa_j>\delta\mid y,\sigma\right)
&\leq\sum_{j\in A}\Pr(\kappa_j>\delta\mid y,\sigma)\\
&\leq CR^\alpha\sum_{j\in A}e^{-\delta z_j}\\
&\leq C|A|R^\alpha e^{-\delta\min_{j\in A}z_j}.
\end{aligned}
\end{equation}
This proves part~(a) for the joint posterior of the shrinkage factors.

For part~(b), we are interested in the difference between the posterior mean and the observation. Combining the normal likelihood and prior gives
\begin{equation}
b_j\mid T,W,y,\sigma
\sim\normal\!\left((1-\kappa_j)y_j,\sigma^2(1-\kappa_j)\right).
\end{equation}
The conditional mean has absolute value at most $|y_j|$ and the conditional standard deviation is at most $\sigma$, so the conditional absolute first moment is bounded by $|y_j|+\sigma\sqrt{2/\pi}$. The posterior mean therefore exists, and averaging over the posterior of $(T,W)$ gives
\begin{equation}
\label{eq:full_posterior_mean_identity}
\mbe(b_j\mid y,\sigma)-y_j
=-y_j\mbe(\kappa_j\mid y,\sigma).
\end{equation}
To control this difference as $|y_j|$ increases, we need a bound on how quickly $\mbe(\kappa_j\mid y,\sigma)$ decreases. We first bound $\mbe\{\kappa_j(1-\kappa_j)\mid y,\sigma\}$, since this product appears when differentiating the likelihood with respect to $t$.

For a fixed allocation $w$, let $F_y(t,w)=t\pi_T(t)L_y(t,w)$. We will integrate its derivative over $t$, using the fact that $F_y$ tends to zero at both ends of the integration interval. In the following calculation, $\kappa_j$ denotes $\kappa_j(t,w)$. Differentiating the beta prime density and the likelihood, with $t\,\partial_t\kappa_j=-\kappa_j(1-\kappa_j)$, yields
\begin{equation}
\label{eq:tail_robustness_derivative}
\begin{aligned}
\partial_tF_y(t,w) =\pi_T(t)L_y(t,w)\bigg\{
&a_1-(a_1+a_2)\frac{t}{1+t} -\frac12\sum_{j=1}^p(1-\kappa_j)+\sum_{j=1}^p z_j\kappa_j(1-\kappa_j)\bigg\}.
\end{aligned}
\end{equation}
Before integrating this expression, we check its boundary values and integrability. Since $L_y(t,w)\leq1$ and $a_1,a_2>0$,
\begin{equation}
\begin{aligned}
0\leq F_y(t,w)&\leq t\pi_T(t)
=\frac{t^{a_1}(1+t)^{-a_1-a_2}}{B(a_1,a_2)},\\
\lim_{t\downarrow0}F_y(t,w)&=\lim_{t\to\infty}F_y(t,w)=0.
\end{aligned}
\end{equation}
Also, $0<\kappa_j<1$ and $\kappa_j(1-\kappa_j)\leq1/4$. For fixed $y$, let $M_y=2a_1+a_2+p/2+\frac14\sum_jz_j$, which is finite. Bounding each term in Equation~\eqref{eq:tail_robustness_derivative} gives
\begin{equation}
|\partial_tF_y(t,w)|\leq M_y\pi_T(t),
\qquad
\int_0^\infty M_y\pi_T(t)\,dt=M_y<\infty.
\end{equation}
We can now justify integration over $(0,\infty)$ by first working on finite intervals. For $m\geq2$, let $g_m(t,w)=\mathbf{1}_{[1/m,m]}(t)\partial_tF_y(t,w)$. For every $t>0$ and fixed $w$, $g_m(t,w)\to\partial_tF_y(t,w)$ and $|g_m(t,w)|\leq M_y\pi_T(t)$. This bound is integrable with respect to $dt$, so dominated convergence, followed by the fundamental theorem of calculus on $[1/m,m]$, gives
\begin{equation}
\begin{aligned}
\int_0^\infty\partial_tF_y(t,w)\,dt
&=\int_0^\infty\lim_{m\to\infty}g_m(t,w)\,dt\\
&=\lim_{m\to\infty}\int_0^\infty g_m(t,w)\,dt\\
&=\lim_{m\to\infty}\{F_y(m,w)-F_y(1/m,w)\}=0.
\end{aligned}
\end{equation}
Let $P_W$ denote the prior distribution of $W$. Since the bound $M_y\pi_T(t)$ is independent of $w$ and has integral $M_y$ under $dt\,dP_W(w)$, Fubini's theorem allows us to average the last identity over $W$. Substituting Equation~\eqref{eq:tail_robustness_derivative}, dividing by $D(y)$, and rearranging the terms gives
\begin{equation}
\label{eq:tail_robustness_moment_identity}
\begin{aligned}
\sum_{j=1}^p z_j\mbe\{\kappa_j(1-\kappa_j)\mid y,\sigma\}
&=(a_1+a_2)\mbe\!\left(\frac{T}{1+T}\,\middle|\,y,\sigma\right)-a_1+\frac12\sum_{j=1}^p\mbe(1-\kappa_j\mid y,\sigma)\\
&\leq a_2+\frac p2=\alpha.
\end{aligned}
\end{equation}
Every term on the left is nonnegative. Thus, for $z_j>0$, we have $\mbe\{\kappa_j(1-\kappa_j)\mid y,\sigma\}\leq\alpha/z_j$. To obtain a bound for $\mbe(\kappa_j\mid y,\sigma)$ itself, consider separately $\kappa_j\leq1/2$ and $\kappa_j>1/2$. On the first event, $\kappa_j\leq2\kappa_j(1-\kappa_j)$, while part~(a) bounds the posterior probability of the second event. Combining the two gives
\begin{equation}
\begin{aligned}
\mbe(\kappa_j\mid y,\sigma)
&=\mbe\{\kappa_j\mathbf{1}_{\{\kappa_j\leq1/2\}}\mid y,\sigma\}
+\mbe\{\kappa_j\mathbf{1}_{\{\kappa_j>1/2\}}\mid y,\sigma\}\\
&\leq2\mbe\{\kappa_j(1-\kappa_j)\mid y,\sigma\}
+\Pr(\kappa_j>1/2\mid y,\sigma)\\
&\leq\frac{2\alpha}{z_j}+CR^\alpha e^{-z_j/2}.
\end{aligned}
\end{equation}
Multiplying this bound by $|y_j|$ in Equation~\eqref{eq:full_posterior_mean_identity} and substituting $z_j=y_j^2/(2\sigma^2)$ gives Equation~\eqref{eq:full_posterior_mean_bound}.

Finally, consider the observations $y_j(r)=rv_j$ for $j\in A$, with every $v_j\ne0$, and fixed observations outside $A$. Along this sequence we have
\begin{equation}
R(y(r))=O(r^2),\qquad
\min_{j\in A}z_j(r)=\frac{r^2}{2\sigma^2}\min_{j\in A}v_j^2.
\end{equation}
For fixed $\delta\in(0,1)$, the exponential term in part~(a) therefore decreases faster than the polynomial factor grows, proving that the posterior probability of $\max_{j\in A}\kappa_j>\delta$ tends to zero. In part~(b), the first term is $O(r^{-1})$ and the second is $o(r^{-1})$ for each $j\in A$. Since $A$ is fixed and finite, the Euclidean norm of the differences is also $O(r^{-1})$. This proves the result.

\subsection*{Proof of Corollary~\ref{cor:meff}}

By the second part of Proposition~\ref{prop:properties_lambda_gl}, $\lambda_{gl}^2 \sim \betaprime(c_g, a_2)$, and hence $\kappa_{gl} = 1/(1+\lambda_{gl}^2) \sim \betadist(a_2, c_g)$. Thus $\mbe(1 - \kappa_{gl}) = c_g/(c_g + a_2)$. By linearity of expectation, $\mbe(m_{\text{eff},g}) = p_g c_g/(c_g + a_2)$, and summing over $g$ gives $\mbe(m_{\text{eff}})$. Inserting $c_g = a_G/p_g$ gives the second form in \eqref{eq:meff_mean}. Each term $p_g a_G/(a_G + p_g a_2)$ is continuous and strictly increasing in $a_G$, and tends to $0$ as $a_G \to 0$ and to $p_g$ as $a_G \to \infty$. Hence $\mbe(m_{\text{eff}})$ takes every value in $(0, p)$ exactly once. If $p_g = p/G$ for all $g$, then $p\, c_g/(c_g + a_2) = m_0$ gives $c_g = a_2 m_0/(p - m_0)$. $\qed$

\subsection*{Proof of Theorem~\ref{thm:concentration_linreg}}
Fix $\tau^2$ and let $\phi_g, \varphi_{gl} > 0$ and form the matrix the diagonal matrix $D=(\tau^2W)^{-1}$, which is clearly positive definite. We denote the OLS estimator by $\hat b=\hat b^{\mathrm{OLS}}=Q^{-1}X'y$. Conditional on the scales and $\sigma^2$, we have that the posterior density is proportional to 
\begin{equation}
\exp\left[-\frac1{2\sigma^2}
\{(y-Xb)'(y-Xb)+b'Db\}\right].
\end{equation}
Completing the multivariate square leads to
\begin{equation}
\begin{aligned}
b'(Q+D)b-2b'X'y &=(b-m)'(Q+D)(b-m)-m'(Q+D)m,\\ 
m&=(Q+D)^{-1}X'y=(Q+D)^{-1}Q\hat b.
\end{aligned}
\end{equation}
Thus the conditional posterior is normal with mean $m$ and covariance $\sigma^2(Q+D)^{-1}$. Note that this explains why  $\sigma^2$ does not appear in the mean. 

We are interested in bounding the difference  between the posterior and least-squares means. To do so consider
\begin{equation}
\hat b-m=(Q+D)^{-1}D\hat b=(I+D^{-1}Q)^{-1}\hat b,
\end{equation}
where the second equality follows from $Q+D=D(I+D^{-1}Q)$. For any invertible matrix $M$ and vector $v$, the induced Euclidean operator norm satisfies
\begin{equation}
\|v\|_2=\|M(M^{-1}v)\|_2\le\|M\|_2\|M^{-1}v\|_2.
\end{equation}
Applying this inequality with $M=I+D^{-1}Q$ gives
\begin{equation}
\|\hat b-m\|_2\ge\frac{\|\hat b\|_2}{\|I+D^{-1}Q\|_2}.
\end{equation}
The triangle inequality and submultiplicativity of the operator norm bound the denominator by
\begin{equation}
\begin{aligned}
\|I+D^{-1}Q\|_2
&\le1+\|D^{-1}\|_2\|Q\|_2\\
&=1+\tau^2\max_iw_i\,\lambda_{\max}(Q).
\end{aligned}
\end{equation}
Here $Q$ is clearly symmetric positive definite. Substitution proves the claimed lower bound. 

For the upper bound on the posterior mean, let $m=(Q+D)^{-1}X'y$. Since $Q$ is positive semidefinite, for every unit vector $v$ we have $v'(Q+D)v\ge v'Dv\ge\lambda_{\min}(D)$. Hence
\begin{equation}
\begin{aligned}
\|m\|_2 &\le\|(Q+D)^{-1}\|_2\|X'y\|_2\\
&=\frac{\|X'y\|_2}{\lambda_{\min}(Q+D)} \le\frac{\|X'y\|_2}{\lambda_{\min}(D)} =\tau^2\max_iw_i\,\|X'y\|_2.
\end{aligned}
\end{equation}
Since the proportions of variance are nonnegative and sum to one, we have $\max_iw_i\le1$. Whenever the data is fixed the bound $\tau^2\|X'y\|_2$ tends to zero independently of $\phi_g$ and $\varphi_{gl}$. This proves uniform convergence as $\tau^2\to 0$.  \qed

\subsection*{Proof of Corollary~\ref{cor:group_shrinkage}}

The assumption $X_g'X_h=0$ for distinct groups makes $Q=X'X$ block diagonal, with blocks $Q_g=X_g'X_g$. The matrix $D$ which , represents the prior precision, is also block diagonal. Its $g$th block and the corresponding posterior and least-squares means are given by
\begin{equation}
\begin{aligned}
D_g&=\{\tau^2\phi_g\diag(\varphi_g)\}^{-1},\\
m_g&=(Q_g+D_g)^{-1}X_g'y,\\
\hat b_g^{\mathrm{OLS}}&=Q_g^{-1}X_g'y.
\end{aligned}
\end{equation}
Since $Q$ is positive definite the inverses are well defined.
Now,
\begin{equation}
\hat b_g^{\mathrm{OLS}}-m_g
=(I+D_g^{-1}Q_g)^{-1}\hat b_g^{\mathrm{OLS}}.
\end{equation}
Applying norm properties yields
\begin{equation}
\begin{aligned}
\|\hat b_g^{\mathrm{OLS}}-m_g\|_2
&\ge\frac{\|\hat b_g^{\mathrm{OLS}}\|_2}
{\|I+D_g^{-1}Q_g\|_2}\\
&\ge\frac{\|\hat b_g^{\mathrm{OLS}}\|_2}
{1+\lambda_{\max}(Q_g)\tau^2\phi_g\max_l\varphi_{gl}}.
\end{aligned}
\end{equation}
Since $\phi_g\le1$ and $\max_l\varphi_{gl}\le1$ we can enlarge the denominator, weakening the lower bound to
\begin{equation}
\|\hat b_g^{\mathrm{OLS}}-m_g\|_2
\ge\frac{\|\hat b_g^{\mathrm{OLS}}\|_2}
{1+\lambda_{\max}(X_g'X_g)\tau^2}.
\end{equation}
This proves both inequalities.

\subsection*{Proof of Equations~\eqref{eq:meff_trace} and~\eqref{eq:meff_mean_n}}

Conditional on the scales, the prior of $b$ is $b \sim \normal(0, \sigma^2 \Lambda)$ with conditional posterior mean $\bar{b} = (X'X + \Lambda^{-1})^{-1} X'y$. Let $M = \Lambda^{1/2} X'X \Lambda^{1/2}$, then
\begin{equation}
(X'X + \Lambda^{-1})^{-1} X'X = \Lambda^{1/2} (I + M)^{-1} M \Lambda^{-1/2}.    
\end{equation}
The trace and the diagonal elements do not change under this similarity transform with the diagonal matrix $\Lambda^{1/2}$. Hence $m_{\text{eff}} = \sum_i d_i/(1+d_i)$, where $d_i$ are the eigenvalues of $M$, whose nonzero values are those of $X\Lambda X'$. 

Each term lies in $[0,1)$ and at most $\operatorname{rank}(X)$ terms are nonzero. The matrix $(I+M)^{-1}M$ is symmetric with eigenvalues in $[0,1)$, so its diagonal elements also lie in $[0,1)$, and $m_{\text{eff},g}$ is well defined. 

Finally, if $X'X = nI$, then $M = n\Lambda$ and $m_{\text{eff}} = \sum_{g,l} n\lambda_{gl}^2/(1 + n\lambda_{gl}^2) = \sum_{g,l}(1 - \kappa_{gl})$ as claimed in the main text.

To prove Equation~\eqref{eq:meff_mean_n}, we can proceed similarly to the moments of $\kappa_{gl}$. Let $u = \lambda_{gl}^2/(1+\lambda_{gl}^2) \sim \betadist(c_g, a_2)$, then $1 - \kappa_{gl} = nu/\{1 + (n-1)u\}$, and Euler's integral representation of ${}_2F_1$ gives \citep{olver_nist_2010}
\begin{equation}
\begin{aligned}
\mbe(1-\kappa_{gl}) &<= \frac{n}{\operatorname{B}(c_g,a_2)} \int_0^1 u^{c_g}(1-u)^{a_2-1}\{1+(n-1)u\}^{-1}\,du \\ 
&= \frac{n c_g}{c_g + a_2}\, {}_2F_1(1, c_g+1; c_g+a_2+1; 1-n).    
\end{aligned}
\end{equation}
The function $nu/\{1+(n-1)u\}$ increases in $n$ for every $u \in (0,1)$, so \eqref{eq:meff_mean_n} increases with $n$. $\qed$

\subsection{Proofs of posterior consistency}
\label{subsec:proof_consistency}
We are interested in the posterior probability assigned to coefficients (or coefficients and variance) outside a neighborhood of the truth. For both theorems we use Bayes' formula to compare prior mass near the truth with the errors of tests outside the target neighborhood, following the testing approach to posterior concentration \citep{ghosal_convergence_2000}. 

In the following, let $\theta=(b,\sigma^2)$ in Theorem~\ref{thm:consistency_growing_groups}, and $\theta=b$ with $\sigma^2=\sigma^{2\star}$ fixed in Theorem~\ref{thm:consistency_prediction}. Denote by $f_n(y_n\mid\theta)$ the Gaussian sampling density and by $\theta_n^\star$ the true parameter. We use $E_\theta$ for expectation under parameter $\theta$, and a subscript $\star$ for expectation or probability under the truth, suppressing dependence on $n$. The prior $\Pi_n$ is the distribution of $\theta$ after integrating out the latent scales and allocations.

For a parameter set $B_n$, define
\begin{equation}
\begin{aligned}
L_n(\theta)&=\frac{f_n(y_n\mid\theta)}{f_n(y_n\mid\theta_n^\star)},\\
N_n(B_n)&=\int_{B_n}L_n(\theta)\,d\Pi_n(\theta),\\
D_n&=\int L_n(\theta)\,d\Pi_n(\theta).
\end{aligned}
\label{eq:consistency_likelihood_notation}
\end{equation}
These integrals are just averages with respect to the prior. The likelihoods we consider are all positive. An application of Tonelli's theorem gives $E_{\star}D_n=\int E_{\star}L_n(\theta)\,d\Pi_n(\theta)=1$. This implies that $0<D_n<\infty$ almost surely. Bayes' formula and any test $\psi_n\in\{0,1\}$ give
\begin{equation}
\Pi_n(B_n\mid y_n)=\frac{N_n(B_n)}{D_n}
\leq\psi_n+(1-\psi_n)\frac{N_n(B_n)}{D_n}.
\label{eq:consistency_posterior_decomposition}
\end{equation}
Here $\psi_n=1$ means that the test rejects the true parameter. The idea is to choose tests for which rejection under the truth and acceptance throughout $B_n$ both have small probability. The latter bound controls $(1-\psi_n)N_n(B_n)$ through the likelihood-ratio identity, as shown below. Together with a lower bound on $D_n$, this gives the required posterior limit. These tests are used to establish consistency and may therefore depend on the true parameter \citep{ghosal_convergence_2000}.

Constants denoted by $C$ or $c$ below are positive and depend only on the fixed bounds and hyperparameters in the relevant theorem, and on a fixed target radius when specified, but not on $n$ or the group partition.

\subsubsection{Proof of Theorem~\ref{thm:consistency_growing_groups}}

Fix $\epsilon>0$. We want to bound the posterior probability of the set
\begin{equation}
B_{n,\epsilon}=\{(b,\sigma^2):\|b-b_n^\star\|_2>\epsilon
\ \text{or}\ |\sigma^2-\sigma^{2\star}|>\epsilon\}.
\end{equation}
We will first lower-bound $D_n$ by $\exp\{-o(n)\}$ and construct tests of $B_{n,\epsilon}$ with errors of order $\exp(-cn)$. The testing bound then decreases faster than the denominator can decrease. To obtain the denominator bound, we use the smaller set
\begin{equation}
K_n=\{(b,\sigma^2):\|b-b_n^\star\|_2\le n^{-1/2},\ |\sigma^2-\sigma^{2\star}|\le n^{-1/2}\}.
\end{equation}
We use the shrinking set $K_n$  to lower-bound $D_n$. 

\paragraph{Prior probability near the true coefficients and variance.}
For notational convenience, we suppress $n$ and write $p=\sum_gp_g$, $m=\max_gp_g$, and $H=Gm$. We first choose a set of allocations and scales on which every conditional normal density admits a lower bound near its true coefficient. Each coefficient receives the product of a group share and a share within that group, which we write as $\phi_g\varphi_{gl}$. We therefore bound their joint contribution before multiplying the conditional coefficient probabilities.

Using independent gamma variables in the shape--rate parameterization, write
\begin{equation}
\begin{aligned}
Y_g&\sim\gammadist(a_G,1),\qquad Z_{gl}\sim\gammadist(c_g,1),\\
\phi_g&=\frac{Y_g}{\sum_hY_h},\qquad
\varphi_{gl}=\frac{Z_{gl}}{\sum_r Z_{gr}}.
\end{aligned}
\end{equation}
Let $E$ be the event that every $Y_g$ and $Z_{gl}$ lies in $[1,2]$. Our idea is that this keeps the group allocations comparable to $1/G$ and the within-group allocations comparable to $1/p_g$. Since $a_G\in[a_-,a_+]$, continuity and positivity of the gamma density give $\Pr(1 \leq Y_g \leq 2)\geq q>0$ uniformly in $g,n$. For the within-group variables, the shapes $c_g$ may approach zero, so we need a bound that retains their dependence on $c_g$. To this end we set $M_\Gamma=\max_{0\leq c\leq c_+}\Gamma(c+1)<\infty$. Since $z^{c-1}\geq 1/2$ for $0<c\leq c_+$ and $1\leq z\leq 2$, we have
\begin{equation}
\Pr\{1\le\operatorname{Gamma}(c,1)\le2\} =\frac{c}{\Gamma(c+1)}\int_1^2z^{c-1}e^{-z}\,dz \ge\frac{e^{-2}}{2M_\Gamma}c=:k_0c.
\end{equation}
Independence and $c_g\ge c_-/p_g$ now imply
\begin{equation}
\Pr(E)\ge q^G\prod_g(k_0c_-/p_g)^{p_g}
\ge\exp\{-Cp\log(ep)\}.
\end{equation}
For the last inequality, we can use $G\le p$ and $\sum_gp_g\log p_g\le p\log p$, decreasing $k_0$ if necessary. On $E$, the numerator and denominator of each normalized gamma coordinate give
\begin{equation}
\frac1{2G}\le\phi_g\le\frac2G,\qquad
\frac1{2p_g}\le\varphi_{gl}\le\frac2{p_g}.
\end{equation}
The same bounds hold for singleton allocations which equal one.

On $E$, the smallest product $\phi_g\varphi_{gl}$ is of order $1/(Gm)$. We therefore restrict the independent total scale $T=\tau^2$ to $[H,2H]$, where $H=Gm$, to keep the coefficient variances away from zero. We only need a lower bound for the probability of this event. Rewriting the beta-prime density gives
\begin{equation}
f_T(t)=\frac{t^{-a_2-1}(1+1/t)^{-a_1-a_2}}{B(a_1,a_2)}.
\end{equation}
For $t\ge H$, the inequalities $\log(1+1/t)\le1/t$, $a_1\le AG$ and $H\ge G$ imply
\begin{equation}
(1+1/t)^{-a_1-a_2}\ge e^{-(a_1+a_2)/H}\ge e^{-A-a_2}.
\end{equation}
Also $B(a_1,a_2)\le B(a_0,a_2)$, as is seen from the beta integral on $(0,1)$ and $a_1\ge a_0$. Integrating the remaining power over $[H,2H]$ yields
\begin{equation}
\Pr(H\le T\le2H) \ge C\int_H^{2H}t^{-a_2-1}\,dt =C\frac{1-2^{-a_2}}{a_2}H^{-a_2}
\ge e^{-C\log(ep)},
\end{equation}
because $1\le H\le p^2$. On this scale event and $E$, it can be shown that the coefficient variance multiplier satisfies
\begin{equation}
\frac14\le\frac{m}{4p_g}\le T\phi_g\varphi_{gl} \le\frac{8m}{p_g}\le8p.
\end{equation}

By assumption, the fixed prior density of $\sigma^2$ is continuous and positive near $\sigma^{2\star}$, so for sufficiently large $n$ its probability on $|\sigma^2-\sigma^{2\star}|\le n^{-1/2}$ is at least $C n^{-1/2}$. On that interval $\sigma^2\in[\sigma^{2\star}/2,2\sigma^{2\star}]$. 

Therefore, every conditional coefficient variance $v_j=\sigma^2T\phi_g\varphi_{gl}$ lies in $[\sigma^{2\star}/8,16\sigma^{2\star}p]$. Let $M=\sup_n\|b_n^\star\|_\infty$. On the coordinate interval $|b_j-b_j^\star|\le(np)^{-1/2}$, we have $|b_j|\le M+1$ and hence
\begin{equation}
\frac{e^{-b_j^2/(2v_j)}}{\sqrt{2\pi v_j}} \ge\frac{\exp\{-4(M+1)^2/\sigma^{2\star}\}}{\sqrt{32\pi \sigma^{2\star}p}} \ge C p^{-1/2}.
\end{equation}
Each interval has length $2/\sqrt{np}$. Conditional on the latent variables and $\sigma^2$, the coefficients are independent, so the probability of their product box is at least $\{C/(p\sqrt n)\}^{p}$. The choice of interval width ensures
\begin{equation}
\sum_{j=1}^p(b_j-b_j^\star)^2\leq p\frac1{np}=\frac1n.
\end{equation}
Thus this box, together with the chosen variance interval, is contained in $K_n$. The conditional lower bound holds uniformly over the latent event, allowing us to integrate it against the independent scale, allocation and variance priors:
\begin{equation}
\begin{aligned}
\Pi_n(K_n)&\ge C n^{-1/2}e^{-Cp\log(ep)} \{C/(p\sqrt n)\}^{p}
\ge e^{-Cp_n\log n}.
\end{aligned}
\label{eq:consistency_prior_mass}
\end{equation}
The last inequality follows from $p_n\log n=o(n)$, which implies $1\le p_n<n$ for all sufficiently large $n$. Equation~\eqref{eq:consistency_prior_mass} gives the required probability for a joint neighborhood of all coefficients and the residual variance.

\paragraph{Integrated likelihood.}
We next bound the likelihood ratio on $K_n$. For $L_n(b,\sigma^2)$ defined in Equation~\eqref{eq:consistency_likelihood_notation}, set $\Delta=X_n(b-b_n^\star)$. Under the true model, $y_n-X_nb=\varepsilon_n-\Delta$. Expanding the squared norm in the Gaussian likelihood gives
\begin{equation}
\begin{aligned}
\log L_n(b,\sigma^2)
={}&-\frac n2\log(\sigma^2/\sigma^{2\star})
+\left(\frac1{2\sigma^{2\star}}-\frac1{2\sigma^2}\right)\|\varepsilon_n\|_2^2\\
&+\frac{\varepsilon_n'\Delta}{\sigma^2}-\frac{\|\Delta\|_2^2}{2\sigma^2}.
\end{aligned}
\end{equation}
On $K_n$, the upper eigenvalue bound gives $\|\Delta\|_2\le\sqrt{d_+}$. Moreover, $\|\varepsilon_n\|_2^2/\sigma^{2\star}\sim\chi_n^2$. A chi-squared upper-tail bound gives $\Pr_{\star}(\|\varepsilon_n\|_2^2>2\sigma^{2\star}n)\le e^{-cn}$. Summability and the Borel--Cantelli lemma imply $\|\varepsilon_n\|_2\le C\sqrt n$ eventually almost surely. 

For $\sigma^2\in[\sigma^{2\star}/2,2\sigma^{2\star}]$, the logarithm and its reciprocal are Lipschitz functions of the variance. Thus each term in the preceding expansion is controlled uniformly on $K_n$:
\begin{equation}
\begin{aligned}
n|\log(\sigma^2/\sigma^{2\star})|&\le C\sqrt n,\\
\left|\frac{1}{\sigma^2}-\frac{1}{\sigma^{2\star}}\right|\,\|\varepsilon_n\|_2^2&\le C\sqrt n,\\
|\varepsilon_n'\Delta|/\sigma^2&\le C\sqrt n,
\qquad \|\Delta\|_2^2/(2\sigma^2)\le C.
\end{aligned}
\end{equation}
Consequently $\inf_{K_n}\log L_n\ge-C\sqrt n$ eventually almost surely. Restricting the integrated likelihood to $K_n$ and using Equation~\eqref{eq:consistency_prior_mass}, we obtain
\begin{equation}
D_n:=\int L_n\,d\Pi_n
\ge\Pi_n(K_n)\inf_{K_n}L_n
\ge e^{-Cp_n\log n-C\sqrt n}=e^{-o(n)}
\label{eq:proof_growing_denominator}
\end{equation}
eventually almost surely. We now compare this lower bound with the error probabilities of tests for $B_{n,\epsilon}$.

\paragraph{Tests for the coefficients and residual variance.}
We now show how to use the regression residuals to test the variance and the ordinary least-squares estimator to test the coefficients. The distribution of the residuals does not depend on $b$, allowing us to control variance values far from $\sigma^{2\star}$ first. For the remaining variance values, the least-squares estimator gives a uniform bound for coefficient error.

Let $Q_n=X_n'X_n$, let $P_n$ be the orthogonal projection onto the columns of $X_n$, and set $\nu_n=n-p_n$. Since $X$ is of full column rank, the estimators
\begin{equation}
\widehat b_n=Q_n^{-1}X_n'y_n,\qquad
\widehat{\sigma}_n^2=\frac{\|(I-P_n)y_n\|_2^2}{\nu_n}
\end{equation}
are well defined for large $n$. Under $(b,\sigma^2)$ they satisfy
\begin{equation}
\widehat b_n-b\sim\normal_{p_n}(0,\sigma^2Q_n^{-1}),\qquad
\nu_n\widehat{\sigma}_n^2/\sigma^2\sim\chi^2_{\nu_n}.
\end{equation}
Consider the following test: we reject the true parameter if either estimator is farther than $\epsilon/2$ from its true value:
\begin{equation}
\psi_n=\mathbf1\{\|\widehat b_n-b_n^\star\|_2>\epsilon/2
\ \text{or}\ |\widehat{\sigma}_n^2-\sigma^{2\star}|>\epsilon/2\}.
\end{equation}
We now calculate the error estimates explicitly. For $V\sim\chi^2_\nu$, the moment generating function is $E(e^{tV})=(1-2t)^{-\nu/2}$ for $t<1/2$. Markov's inequality applied to $e^{tV}$ with $t=(1-1/a)/2$ gives the upper-tail bound when $a>1$ and the lower-tail bound when $0<a<1$ (for which $t<0$). In both cases, substitution yields
\begin{equation}
\begin{aligned}
\Pr(V\ge a\nu)&\le e^{-\nu(a-1-\log a)/2},&&a>1,\\
\Pr(V\le a\nu)&\le e^{-\nu(a-1-\log a)/2},&&0<a<1.
\end{aligned}
\label{eq:proof_chisq_bounds}
\end{equation}
Also, for $Z\sim\normal_{p_n}(0,I)$, the lower design eigenvalue bound allows a representation with $\|\widehat b_n-b\|_2^2\le \sigma^2\|Z\|_2^2/(nd_-)$. If $\sigma^2\le V_{\max}$ for a fixed $V_{\max}>0$, Markov's inequality applied to $e^{\|Z\|^2/4}$ gives
\begin{equation}
\begin{aligned}
\Pr_{b,\sigma^2}(\|\widehat b_n-b\|_2>\epsilon/2)
&\le\Pr\left(\chi^2_{p_n}>\frac{nd_-\epsilon^2}{4V_{\max}}\right)\\
&\le 2^{p_n/2}\exp\left\{-\frac{nd_-\epsilon^2}{16V_{\max}}\right\}
\le e^{-c n}
\end{aligned}
\end{equation}
for large $n$, since $p_n=o(n)$. The constant $c$ may depend on $\epsilon$ and $V_{\max}$.

Under the true parameter, use $V_{\max}=\sigma^{2\star}$ for the coefficient error. For the variance error, Equation~\eqref{eq:proof_chisq_bounds} applies at the fixed ratios $1\pm\epsilon/(2\sigma^{2\star})$, omitting a lower-tail event if its threshold is nonpositive. Since $\nu_n/n\to1$, both errors are bounded by $Ce^{-cn}$, so $E_{\star}\psi_n\le Ce^{-cn}$.

To bound the probability of acceptance under alternative hypothesis, first suppose $\sigma^2>\sigma^{2\star}+\epsilon$. Acceptance implies
\begin{equation}
\frac{\widehat{\sigma}_n^2}{\sigma^2}\le\frac{\sigma^{2\star}+\epsilon/2}{\sigma^2}
\le\frac{\sigma^{2\star}+\epsilon/2}{\sigma^{2\star}+\epsilon}<1.
\end{equation}
The lower-tail inequality in Equation~\eqref{eq:proof_chisq_bounds} therefore gives a uniform exponentially small acceptance probability. If $0<\sigma^2<\sigma^{2\star}-\epsilon$, a case that exists only when $\epsilon<\sigma^{2\star}$, acceptance instead implies
\begin{equation}
\frac{\widehat{\sigma}_n^2}{\sigma^2}\ge\frac{\sigma^{2\star}-\epsilon/2}{\sigma^2}
\ge\frac{\sigma^{2\star}-\epsilon/2}{\sigma^{2\star}-\epsilon}>1,
\end{equation}
and the upper-tail inequality gives the same conclusion. The remaining alternatives have $|\sigma^2-\sigma^{2\star}|\le\epsilon$ but $\|b-b_n^\star\|_2>\epsilon$. If the test accepts, the triangle inequality gives
\begin{equation}
\|\widehat b_n-b\|_2
\geq\|b-b_n^\star\|_2-\|\widehat b_n-b_n^\star\|_2
>\epsilon-\frac\epsilon2=\frac\epsilon2.
\end{equation}
The coefficient error bound applies uniformly with $V_{\max}=\sigma^{2\star}+\epsilon$. These cases exhaust $B_{n,\epsilon}$, so there are constants $C,c>0$ such that
\begin{equation}
\begin{aligned}
E_{\star}\psi_n &\le Ce^{-cn},\\    
\sup_{(b,\sigma^2)\in B_{n,\epsilon}}E_{b,\sigma^2}(1-\psi_n) &\le Ce^{-cn}.
\end{aligned}
\end{equation}

\paragraph{Posterior probability outside the neighborhood.}
We can now combine the testing bounds with Equation~\eqref{eq:proof_growing_denominator}. Write $N_n=N_n(B_{n,\epsilon})$. By the definition of the likelihood ratio,
\begin{equation}
\begin{aligned}
E_{\star}\{(1-\psi_n)L_n(b,\sigma^2)\}
&=\int(1-\psi_n(y)) \frac{f_n(y\mid b,\sigma^2)}{f_n(y\mid b_n^\star,\sigma^{2\star})} f_n(y\mid b_n^\star,\sigma^{2\star})\,dy\\
&=\int(1-\psi_n(y))f_n(y\mid b,\sigma^2)\,dy =E_{b,\sigma^2}(1-\psi_n).
\end{aligned}
\end{equation}
Since $(1-\psi_n)L_n(b,\sigma^2)\geq0$, by Tonelli's theorem we can integrate this identity with respect to the prior and interchange the two integrals. Applying the bound on the acceptance probability gives
\begin{equation}
\begin{aligned}
E_{\star}\{(1-\psi_n)N_n\} &=\int_{B_{n,\epsilon}}E_{\star}\{(1-\psi_n)L_n(b,\sigma^2)\}\,d\Pi_n(b,\sigma^2)\\
&=\int_{B_{n,\epsilon}}E_{b,\sigma^2}(1-\psi_n)\,d\Pi_n(b,\sigma^2) \le Ce^{-cn}.
\end{aligned}
\end{equation}
Markov's inequality yields
\begin{equation}
\Pr_{\star}\{(1-\psi_n)N_n>e^{-cn/2}\}\le Ce^{-cn/2}.
\end{equation}
Both this bound and $\Pr_{\star}(\psi_n=1)$ are summable. By the first Borel--Cantelli lemma, the corresponding events occur only finitely often with probability one. Thus $\psi_n=0$ and $(1-\psi_n)N_n\le e^{-cn/2}$ eventually almost surely. Substituting these bounds and Equation~\eqref{eq:proof_growing_denominator} into the posterior probability gives
\begin{equation}
\Pi_n(B_{n,\epsilon}\mid y_n)=\frac{N_n}{D_n}
\le\psi_n+\frac{(1-\psi_n)N_n}{D_n}
\le e^{-cn/2+o(n)}\longrightarrow0
\end{equation}
eventually almost surely, since $p_n\log n=o(n)$ and $\sqrt n=o(n)$. This proves the result for the chosen $\epsilon$. Taking the intersection of the probability-one events over all positive rational radii gives a common event on which the result holds for every $\epsilon>0$, by inclusion of the corresponding neighborhood complements. \qed

\subsubsection{Proof of Theorem~\ref{thm:consistency_prediction}}

\paragraph{Dirichlet bound for the prediction proof}

We now proceed with the prediction proof. For the prediction result, we use sparsity to bound the prior probability near the truth in terms of the number of nonzero coefficients. We first establish a property of the Dirichlet distribution, namely, a selected set of coordinates can receive almost all the allocation, while each selected coordinate retains a positive share. We will use this property for the groups containing nonzero coefficients and for the relevant coordinates within each group.

Fix $A>0$, let $U\sim\dirichlet(A/d,\ldots,A/d)$, and choose a nonempty set $J$ of $m$ coordinates. Let $R=\sum_{j\in J}U_j$. We claim that there is $c_A\in(0,1]$, depending only on $A$, such that for $0<\eta\le1/2$,
\begin{equation}
\Pr\left(1-R\le\eta,\ \frac1{2m}\le\frac{U_j}{R}\le\frac2m
\text{ for all }j\in J\right)
\ge c_A^{m+1}\eta^A d^{-(m+1)}.
\label{eq:dirichlet_face_bound}
\end{equation}
Here $1-R$ is the total allocation outside $J$, and $U_j/R$ is the proportion of the selected allocation assigned to coordinate $j$. To calculate their probabilities, first suppose $m<d$ and set $a=Am/d$, $b=A(d-m)/d$. Using the gamma representation and aggregating over the selected and unselected coordinates we have
\begin{equation}
1-R\sim\betadist(b,a),\qquad
(U_j/R)_{j\in J}\sim\dirichlet(A/d,\ldots,A/d).
\end{equation}
These two objects are independent, which follows because the selected gamma proportions are independent of their sum and of every unselected gamma variable.

Define $H_A=\max_{0\le x\le A}\Gamma(x+1)<\infty$ and $h_A=\min\{1,2^{1-A}\}>0$. For $0\le x\le\eta\le1/2$, we have $(1-x)^{a-1}\ge h_A$. Integrating the beta density and using $a+b=A$ gives
\begin{equation}
\begin{aligned}
\Pr(1-R\le\eta)
&=\frac1{B(b,a)}\int_0^\eta x^{b-1}(1-x)^{a-1}\,dx\\
&\ge\frac{h_A\eta^b}{bB(b,a)}
=\frac{h_A\Gamma(A)a\eta^b}{\Gamma(a+1)\Gamma(b+1)}\\
&\ge\frac{h_A\Gamma(A)A}{H_A^2}\frac{\eta^A}{d}.
\end{aligned}
\end{equation}
The last step uses $a\ge A/d$, $b\le A$, and $0<\eta<1$, which implies $\eta^b\ge\eta^A$.

For the selected proportions, let $W_j\sim\gammadist(A/d,1)$, $j\in J$, be independent. Normalizing these variables gives the required Dirichlet distribution. For each $j\in J$,
\begin{equation}
\begin{aligned}
\Pr(1\le W_j\le2)
&=\frac{A/d}{\Gamma(1+A/d)}
\int_1^2w^{A/d-1}e^{-w}\,dw\\
&\ge\frac{e^{-2}A}{2H_A d}.
\end{aligned}
\end{equation}
When all $W_j$ lie in $[1,2]$, their sum lies in $[m,2m]$ and every normalized coordinate lies in $[1/(2m),2/m]$. Independence gives probability at least $\left(e^{-2}A/(2H_A d)\right)^m$ for this event. Multiplying by the preceding bound for $1-R$, using the independence of the total and normalized selected proportions, proves Equation~\eqref{eq:dirichlet_face_bound} after choosing $c_A$ no larger than one and the two fixed constants in these bounds.

If $m=d$, then $R=1$ deterministically. The gamma event alone shows the bound, since the additional factor $c_A\eta^A/d$ is at most one. This also covers $d=m=1$, when the allocation is deterministic. \qed

\paragraph{Joint prior mass for Theorem~\ref{thm:consistency_prediction}}
We now use Equation~\eqref{eq:dirichlet_face_bound} to bound the probability of a coefficient neighborhood. For notational convenience, suppress $n$ and let $k=\max\{1,\|b^\star\|_0\}$. We will show that
\begin{equation}
\Pi(\|b-b^\star\|_1\le\delta)
\ge\exp\{-Ck\log(ep/\delta)\},\qquad 0<\delta\le1.
\label{eq:prediction_prior_mass}
\end{equation}
where $C$ depends only on the fixed hyperparameters, $\sigma^{2\star}$, and the uniform bound on $\|b^\star\|_\infty$. In particular, it does not depend on the group partition. We allow error at most $\delta/2$ on a selected set containing the nonzero coefficients and at most $\delta/2$ on the remaining coordinates. To achieve this, the selected coefficients need enough conditional variance to approximate their true values, while the remaining coefficients need small total conditional variance.

\paragraph{Group selection and coordinates.}

If $b^\star\ne0$, let $S$ be its support. If $b^\star=0$, choose any singleton $S$. In either case $|S|=k$ and $b_j^\star=0$ for $j\notin S$. Let $\mathcal H$ be the set of groups meeting $S$, let $h=|\mathcal H|$ and let $k_g=|S\cap g|$ for $g\in\mathcal H$. Then $1\le h\le k$ and $\sum_{g\in\mathcal H}k_g=k$. Set
\begin{equation}
\eta=\frac{\delta^2}{128(1+\sigma^{2\star})k^2p},
\end{equation}
which lies in $(0,1/2]$. With this choice, we will bound the combined error of the unselected coefficients by $\delta/2$ when the total scale is of order $k^2$.

Apply Equation~\eqref{eq:dirichlet_face_bound} first to the group vector with $(A,d,m)=(\alpha,G,h)$, and then within each selected group with $(A,d,m)=(\zeta,p_g,k_g)$. Independence of the allocation vectors gives an event $E$ with probability
\begin{equation}
\Pr(E)\ge c^{k+2h+1}\eta^{\alpha+h\zeta}
G^{-(h+1)}\prod_{g\in\mathcal H}p_g^{-(k_g+1)},
\label{eq:prediction_allocation_mass}
\end{equation}
where $ c=\min(c_\alpha,c_\zeta)$. Let $R=\sum_{g\in\mathcal H}\phi_g$ and $R_g=\sum_{j\in S\cap g}\varphi_{gj}$. On $E$, we have $R,R_g\ge1-\eta\ge1/2$, $\phi_g/R\ge1/(2h)$, and $\varphi_{gj}/R_g\ge1/(2k_g)$ for the selected coordinates. Therefore their variance proportions $w_j=\phi_g\varphi_{gj}$ satisfy
\begin{equation}
w_j\ge\frac1{16hk_g}\ge\frac1{16k^2},\qquad j\in S.
\end{equation}
The total unselected weight has the complementary bound
\begin{equation}
\sum_{j\notin S}w_j
=\sum_{g\notin\mathcal H}\phi_g
+\sum_{g\in\mathcal H}\phi_g(1-R_g)
\le\eta+\eta\sum_{g\in\mathcal H}\phi_g\le2\eta.
\end{equation}

\paragraph{Selected coefficients.}
Since $w_j\geq1/(16k^2)$ for $j\in S$, we restrict the independent total scale $T=\tau^2$ to $[k^2,2k^2]$. This keeps the conditional variance of each selected coefficient bounded away from zero. For fixed $a_1,a_2$ and $t\ge1$, the beta-prime density satisfies
\begin{equation}
f_T(t)=\frac{t^{-a_2-1}(1+1/t)^{-a_1-a_2}}{B(a_1,a_2)}
\ge\frac{2^{-a_1-a_2}}{B(a_1,a_2)}t^{-a_2-1}.
\end{equation}
Integrating over $[k^2,2k^2]$ gives $\Pr(k^2\le T\le2k^2)\ge C_Tk^{-2a_2}$, with fixed $C_T>0$. For every fixed latent configuration in $E$ with $k^2\le T\le2k^2$, each selected coefficient has variance $v_j=\sigma^{2\star}Tw_j$ in $[\sigma^{2\star}/16,2\sigma^{2\star}k^2]$. If $M$ bounds $\|b^\star\|_\infty$, then on $|b_j-b_j^\star|\le\delta/(2k)$ we have $|b_j|\le M+1$, and
\begin{equation}
\frac{e^{-b_j^2/(2v_j)}}{\sqrt{2\pi v_j}}
\ge\frac{e^{-8(M+1)^2/\sigma^{2\star}}}{2\sqrt{\pi \sigma^{2\star}}\,k}
\ge\frac{C}{k}.
\end{equation}
Each interval has length $\delta/k$. Conditional independence therefore gives probability at least $(c_N\delta/k^2)^k$, for a fixed $c_N\in(0,1]$, that all selected coefficients lie in their intervals. On that event their total $\ell_1$ error is at most $\delta/2$.

\paragraph{Remaining coefficients.}
For the unselected coordinates, we bound their total error using the conditional expectation of $\|b_{S^c}\|_1$. Their total conditional variance is at most $\sigma^{2\star}(2k^2)(2\eta)=4\sigma^{2\star}k^2\eta$. Applying Cauchy--Schwarz to each coordinate and then to the sum gives
\begin{equation}
\begin{aligned}
\mbe(\|b_{S^c}\|_1\mid T,\phi,\{\varphi_g\})
&\le\sum_{j\notin S}\sqrt{\sigma^{2\star}Tw_j}\\
&\le\sqrt{p\sum_{j\notin S}\sigma^{2\star}Tw_j}
\le2\sigma^\star k\sqrt{p\eta}
=\delta\sqrt{\frac{\sigma^{2\star}}{32(1+\sigma^{2\star})}}<\frac\delta4.
\end{aligned}
\end{equation}
The choice of $\eta$ therefore makes Markov's inequality give
\begin{equation}
\Pr\!\left(\|b_{S^c}\|_1>\delta/2
\,\middle|\,T,\phi,\{\varphi_g\}\right)
\leq\frac{2}{\delta}
\mbe(\|b_{S^c}\|_1\mid T,\phi,\{\varphi_g\})<\frac12.
\end{equation}
Thus the event $\|b_{S^c}\|_1\le\delta/2$ has conditional probability at least $1/2$, irrespective of the number of unselected coordinates. If $S^c$ is empty, its probability is one.

\paragraph{Joint probability.}
The selected and unselected coefficients are independent conditional on the latent variables. We can therefore multiply the two conditional probability bounds. On the intersection of these events,
\begin{equation}
\|b-b^\star\|_1
=\sum_{j\in S}|b_j-b_j^\star|+\|b_{S^c}\|_1
\leq\frac\delta2+\frac\delta2=\delta.
\end{equation}
Both conditional probability bounds hold uniformly over $E$ and $k^2\leq T\leq2k^2$. Integrating over these independent latent events gives
\begin{equation}
\begin{aligned}
\Pi(\|b-b^\star\|_1\le\delta)\ge{}&
\tfrac12 C_Tk^{-2a_2}c^{k+2h+1}\eta^{\alpha+h\zeta}
G^{-(h+1)}\\
&\times\prod_{g\in\mathcal H}p_g^{-(k_g+1)}
(c_N\delta/k^2)^k.
\end{aligned}
\end{equation}
To verify the claimed order, take the negative logarithm of this lower bound. Apart from a fixed constant, its terms are bounded by
\begin{equation}
\begin{aligned}
&2a_2\log k+(k+2h+1)|\log c|+(\alpha+h\zeta)\log(1/\eta)\\
&\quad +(h+1)\log G+\sum_{g\in\mathcal H}(k_g+1)\log p_g
+k\{|\log c_N|+\log(1/\delta)+2\log k\}.
\end{aligned}
\end{equation}
Here $h\le k\le p$, $G\le p$, $\sum_{g\in\mathcal H}(k_g+1)=k+h\le2k$, and
\begin{equation}
\log(1/\eta)=\log\{128(1+\sigma^{2\star})\}+2\log(1/\delta)+2\log k+\log p
\le C\log(ep/\delta).
\end{equation}
Using these inequalities, every term is bounded by a constant times $k\log(ep/\delta)$, which proves Equation~\eqref{eq:prediction_prior_mass}. We have used independence of the coefficients conditional on the latent variables; the bound allows their dependence after the scales and allocations are integrated out. \qed

\paragraph{Proof of Theorem~\ref{thm:consistency_prediction}}
We are interested in the posterior probability that the prediction distance exceeds $L\epsilon_n$. As in the preceding consistency proof, we first bound the integrated likelihood and then construct a test for coefficients outside this neighbourhood. For the integrated likelihood, we use a smaller neighbourhood with prediction distance at most $\epsilon_n^2$.

\paragraph{Prior probability near the true mean.}
Recall that $q_n=k_n\log(e p_n n)$ and $\epsilon_n=(q_n/n)^{1/4}$. Set $K_0=\max(1,K)$ and define
\begin{equation}
\begin{aligned}
\delta_n &=\epsilon_n^2/K_0,\\
H_n&=\{b:\|b-b_n^\star\|_1\le\delta_n\}.
\end{aligned}
\end{equation}
Since $q_n=o(n)$, we have $\delta_n\le1$ eventually. For every $b\in H_n$, the column-norm condition and triangle inequality imply
\begin{equation}
\frac{\|X_n(b-b_n^\star)\|_2}{\sqrt n}
\le\sum_j|b_j-b_{n,j}^\star|\frac{\|X_{n,j}\|_2}{\sqrt n}
\le K\delta_n\le\epsilon_n^2.
\end{equation}
Also $\log(1/\delta_n)=\log K_0+\tfrac12\log(n/q_n)\le C+\tfrac12\log n$, since $q_n\ge1$. Equation~\eqref{eq:prediction_prior_mass} therefore gives
\begin{equation}
\Pi_n(H_n)\ge\exp\{-Ck_n\log(ep_n/\delta_n)\}
\ge e^{-C_0q_n}
\end{equation}
with $C_0$ independent of the group partition. On $H_n$, we also have $\|X_n(b-b_n^\star)\|_2^2\leq n\epsilon_n^4=q_n$. These two bounds allow us to control the prior probability and the likelihood on the same neighbourhood.

\paragraph{Integrated likelihood.}
Set $h_b=X_n(b-b_n^\star)$. Since the residual variance is fixed at $\sigma^{2\star}$, the likelihood ratio in Equation~\eqref{eq:consistency_likelihood_notation} becomes
\begin{equation}
L_n(b)=\exp\left\{\frac{\varepsilon_n'h_b}{\sigma^{2\star}}
-\frac{\|h_b\|_2^2}{2\sigma^{2\star}}\right\}.
\end{equation}
To bound its integral over $H_n$, we first average the log likelihood with respect to the prior restricted to this neighbourhood. This avoids using the bound $|\varepsilon_n'h_b|\le\|\varepsilon_n\|_2\|h_b\|_2$, whose order can be too large when $q_n\ll n$. Define
\begin{equation}
\begin{aligned}
Q_H(B)&=\frac{\Pi_n(B\cap H_n)}{\Pi_n(H_n)},\\
E_Hg(b)&=\int g(b)\,dQ_H(b),\\
\bar h&=E_Hh_b.
\end{aligned}
\end{equation}
Here $E_H$ denotes expectation under the prior conditioned on $H_n$. Since $H_n$ depends on the fixed truth and prior, but not on the observations, $\bar h$ is deterministic under the true model. For every fixed observation vector, $\|h_b\|_2^2\le q_n$ on $H_n$ implies
\begin{equation}
|\log L_n(b)|
\leq\frac{\|\varepsilon_n\|_2\sqrt{q_n}}{\sigma^{2\star}}
+\frac{q_n}{2\sigma^{2\star}}<\infty.
\end{equation}
Hence the log likelihood ratio is integrable under $Q_H$. Jensen's inequality applied to the convex exponential function yields
\begin{equation}
\begin{aligned}
D_n:=\int L_n\,d\Pi_n
&\ge\Pi_n(H_n)E_HL_n\\
&\ge\Pi_n(H_n)\exp\left\{
\frac{\varepsilon_n'\bar h}{\sigma^{2\star}}
-\frac{E_H\|h_b\|_2^2}{2\sigma^{2\star}}\right\}.
\end{aligned}
\end{equation}
The deterministic vector $\bar h$ satisfies $\|\bar h\|_2^2\le E_H\|h_b\|_2^2\le q_n$. Under the true experiment, the remaining random term $Z_n=\varepsilon_n'\bar h/\sigma^{2\star}$ is centered Gaussian with variance at most $q_n/\sigma^{2\star}$. The Gaussian tail inequality gives
\begin{equation}
\Pr_{b^\star}\{Z_n<-(\sqrt{8}/\sigma^\star)\,q_n\}
\le\exp\left\{-\frac{(8/\sigma^{2\star})q_n^2}{2q_n/\sigma^{2\star}}\right\}
=e^{-4q_n}.
\end{equation}
If $\bar h=0$, the left probability is zero, so this bound still holds. Combining it with the prior probability and quadratic-term bounds proves
\begin{equation}
\Pr_{b^\star}\{D_n<e^{-C_Dq_n}\}\le e^{-4q_n},
\label{eq:prediction_denominator}
\end{equation}
where $C_D=C_0+\frac1{2\sigma^{2\star}}+(\sqrt{8}/\sigma^\star).$
\paragraph{Testing prediction error.}
We use the squared norm of the residuals relative to the true mean. Under a coefficient vector $b$, its expectation is
\begin{equation}
E_b\!\left\{\frac{\|y_n-X_nb_n^\star\|_2^2}{n}\right\}
=\sigma^{2\star}+\frac{\|h_b\|_2^2}{n}
=\sigma^{2\star}+d_n(b,b_n^\star)^2.
\end{equation}
Thus a prediction distance of at least $r$ increases this expectation by at least $r^2$. We use a threshold halfway between the expectation under the truth and the smallest expectation under these alternatives. For $0<r\le\sigma^\star$, define
\begin{equation}
\psi_{n,r}=\mathbf1\left\{\frac{\|y_n-X_nb_n^\star\|_2^2}{n}
>\sigma^{2\star}+\frac{r^2}{2}\right\}.
\end{equation}
We will show that both testing errors are bounded by $e^{-c_0nr^4}$. Write $u=r^2/\sigma^{2\star}\in(0,1]$ and $V=\|y_n-X_nb_n^\star\|_2^2/\sigma^{2\star}$. 

Under the null, $V\sim\chi_n^2$. Equation~\eqref{eq:proof_chisq_bounds}, with $a=1+u/2$, gives
\begin{equation}
E_{\star}\psi_{n,r}\le
\exp\left[-\frac n2\{u/2-\log(1+u/2)\}\right]
\le e^{-nu^2/32}.
\end{equation}
For the last inequality, note that for $0\le x\le1$,
\begin{equation}
x-\log(1+x)=\int_0^x\frac{t}{1+t}\,dt
\ge\int_0^x\frac t2\,dt=\frac{x^2}{4},
\end{equation}
and take $x=u/2$.

Under any $b$ with $\|h_b\|_2/\sqrt n\ge r$, $V$ has a noncentral chi-squared law with noncentrality $\lambda=\|h_b\|_2^2/\sigma^{2\star}\ge nu$. To bound its lower tail, write $V=\sum_{i=1}^nZ_i^2$, where the $Z_i$ are independent $\normal(\mu_i,1)$ variables with $\mu_i=h_{b,i}/\sigma^\star$ and $\sum_i\mu_i^2=\lambda$. For $t>0$, completing the square gives the one-coordinate integral
\begin{equation}
\begin{aligned}
E_b(e^{-tZ_i^2})
&=\frac1{\sqrt{2\pi}}\int_{\mathbb R}
 e^{-tz^2-(z-\mu_i)^2/2}\,dz\\
&=\frac{e^{-t\mu_i^2/(1+2t)}}{\sqrt{2\pi}}
\int_{\mathbb R}\exp\!\left\{-\frac{1+2t}{2}
\left(z-\frac{\mu_i}{1+2t}\right)^2\right\}\,dz\\
&=(1+2t)^{-1/2}e^{-t\mu_i^2/(1+2t)}.
\end{aligned}
\end{equation}
Multiplying over the independent coordinates therefore yields
\begin{equation}
E_b(e^{-tV})=(1+2t)^{-n/2}\exp\left\{-\frac{\lambda t}{1+2t}\right\}.
\end{equation}
Applying Markov's inequality to $e^{-tV}$ and using $\lambda\ge nu$, we obtain
\begin{equation}
\begin{aligned}
E_b(1-\psi_{n,r})
&=\Pr_b\{V\le n(1+u/2)\}\\
&\le\exp\left(n\left[t(1+u/2)-\tfrac12\log(1+2t)
-\frac{tu}{1+2t}\right]\right).
\end{aligned}
\end{equation}
The elementary inequalities $\log(1+2t)\ge2t-2t^2$ and $(1+2t)^{-1}\ge1-2t$ hold for $t\ge0$. They bound the bracket by
\begin{equation}
t^2(1+2u)-\frac{tu}{2}.
\end{equation}
Choose $t=u/\{4(1+2u)\}$. The bound becomes $-u^2/\{16(1+2u)\}\le-u^2/48$, since $u\le1$. Together with the null bound this proves the uniform testing inequalities
\begin{equation}
E_{\star}\psi_{n,r}\le e^{-c_0nr^4},\qquad
\sup_{\|h_b\|_2/\sqrt n\ge r}E_b(1-\psi_{n,r})\le e^{-c_0nr^4},
\label{eq:prediction_tests}
\end{equation}
where $ c_0=\frac{1}{48}\left[\frac{1}{\sigma^{2\star}}\right]^2$.
Note that these bounds depend on $b$ only through $\|h_b\|_2$, so they hold uniformly over all coefficients at prediction distance at least $r$, even if $X_n$ is rank deficient. Likewise, the assumption of known $\sigma^{2\star}$ plays a role here, since the threshold uses it value.

\paragraph{Posterior concentration.}
We now combine Equation~\eqref{eq:prediction_denominator} with the testing bounds in Equation~\eqref{eq:prediction_tests}. Set $r_n=L\epsilon_n=L(q_n/n)^{1/4}$, so that $nr_n^4=L^4q_n$, and choose a fixed $L$ large enough that $c_0L^4\ge C_D+4$. Since $\epsilon_n\to0$, the condition $r_n\le\sigma^\star$ required for the test holds for all sufficiently large $n$. Define
\begin{equation}
A_n=\{b:\|X_n(b-b_n^\star)\|_2/\sqrt n>r_n\},\qquad
N_n=\int_{A_n}L_n(b)\,d\Pi_n(b).
\end{equation}
As in the preceding consistency proof, nonnegativity permits Tonelli's theorem, and the likelihood ratio converts expectation under the truth into expectation under $b$:
\begin{equation}
\begin{aligned}
E_{\star}\{(1-\psi_{n,r_n})N_n\}
&=\int_{A_n}E_{\star}\{(1-\psi_{n,r_n})L_n(b)\}\,d\Pi_n(b)\\
&=\int_{A_n}E_b(1-\psi_{n,r_n})\,d\Pi_n(b)
\le e^{-c_0L^4q_n}.
\end{aligned}
\end{equation}
Here $nr_n^4=L^4q_n$. Markov's inequality consequently gives
\begin{equation}
\Pr_{\star}\{(1-\psi_{n,r_n})N_n>e^{-(C_D+1)q_n}\}
\le e^{-(c_0L^4-C_D-1)q_n}\le e^{-3q_n}.
\end{equation}
The other two probabilities are $\Pr_{\star}(\psi_{n,r_n}=1)\le e^{-c_0L^4q_n}\le e^{-4q_n}$ and the denominator bound in Equation~\eqref{eq:prediction_denominator}. All three are summable, because
\begin{equation}
q_n=k_n\log(e p_n n)\ge\log(en),\qquad
\sum_{n=1}^\infty e^{-3q_n}\le e^{-3}\sum_{n=1}^\infty n^{-3}<\infty.
\end{equation}
Borel--Cantelli implies that, under the true data-generating sequence, only finitely many of these failures occur. Eventually almost surely, $\psi_{n,r_n}=0$, $D_n\geq e^{-C_Dq_n}$, and $(1-\psi_{n,r_n})N_n\leq e^{-(C_D+1)q_n}$. 

Substituting these three bounds into Equation~\eqref{eq:consistency_posterior_decomposition} gives
\begin{equation}
\Pi_n(A_n\mid y_n)=\frac{N_n}{D_n}
\le\psi_{n,r_n}+\frac{(1-\psi_{n,r_n})N_n}{D_n}
\le e^{-q_n}\longrightarrow0.
\end{equation}
This proves the result. The choice $\epsilon_n=(q_n/n)^{1/4}$ follows from comparing the testing exponent $nr^4$ with the exponents of order $q_n$ in the prior and integrated-likelihood bounds. \qed

\section{Additional results}

In the following we present additional results for the simulations. Section~\ref{subsec:meff_additional} shows the prior of $m_{\text{eff},g}$ for a smaller value of $a_2$, and further results of the comparison between grouped and nongrouped priors. Section~\ref{subsec:external_comparators} compares the Group-R2 prior with five further grouped priors. Section~\ref{subsec:gigg_calibration} shows the prior on $R^2$ that the GIGG prior implies, and describes how we calibrate the GIGG to the Group-R2 prior. Section~\ref{subsec:gigg_diagnostics} reports the sampling diagnostics of the comparison with the GIGG.

\subsection{Effective number of nonzero coefficients $\mathrm{m}_{\mathrm{eff},g}$ additional simulations}
\label{subsec:meff_additional}

\begin{figure}[t!]
    \centering
    \includegraphics[width=\linewidth]{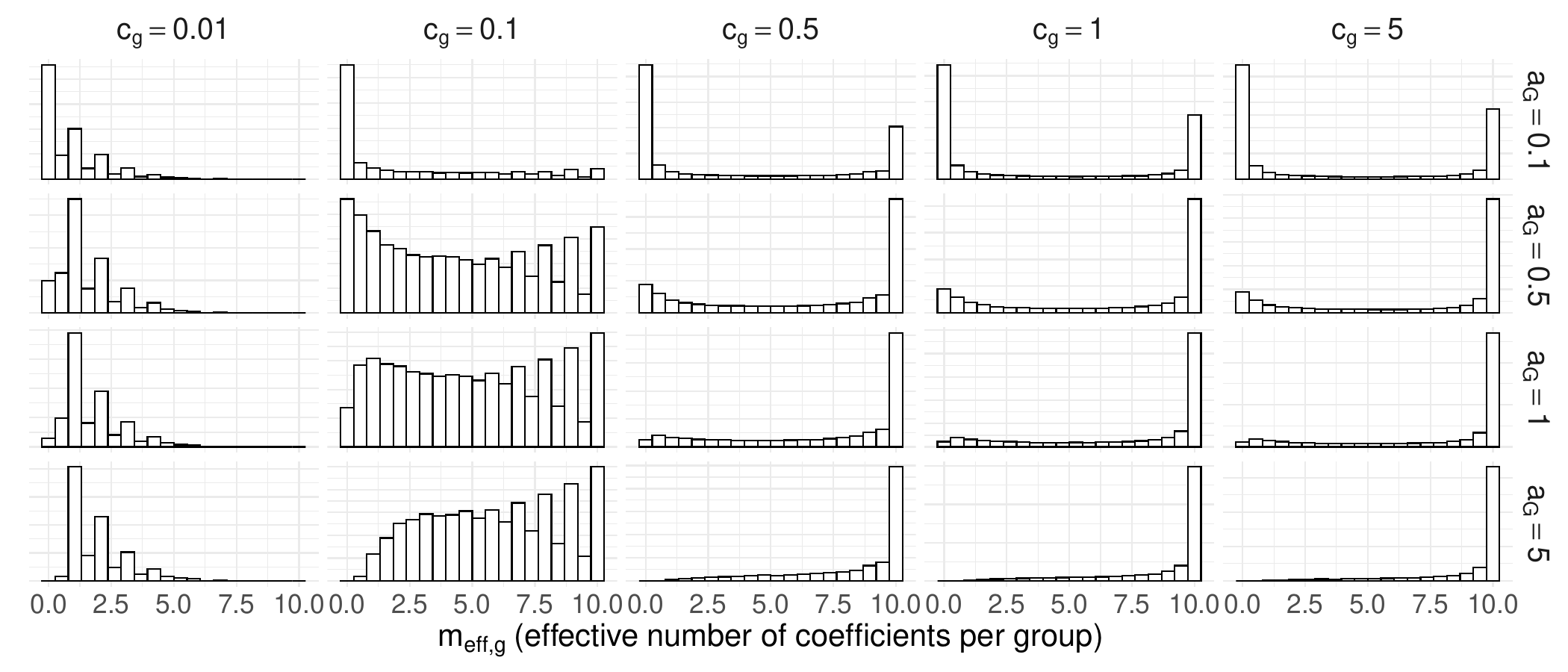}
    \caption{Prior predictive distributions of the group-wise effective number of nonzero coefficients ($\text{m}_{\text{eff},g}$) under different combinations of $a_G, c_g$ for groups of size 10. Rows vary $a_G$, the concentration parameter of the group-level Dirichlet prior, and columns vary $c_g$, the within-group concentration. We set $a_2=0.1$, $G=10$, $p_g=20, \forall g$.
 }
 \label{fig:meff_aditional}
\end{figure}

We present in Figure~\ref{fig:meff_aditional} the prior predictive distributions for $\text{m}_{\text{eff},g}$ under the alternative that $a_2=0.1$, This configuration encourages fatter tails of the marginal prior, leading to a stronger tendency towards variable selection type behavior, where most panels show relatively more mass at the edges of the shrinkage coefficient space. 

\subsubsection{Simulations: Lower-dimensional results}

\begin{figure}[t!]%
\includegraphics[width=\linewidth]{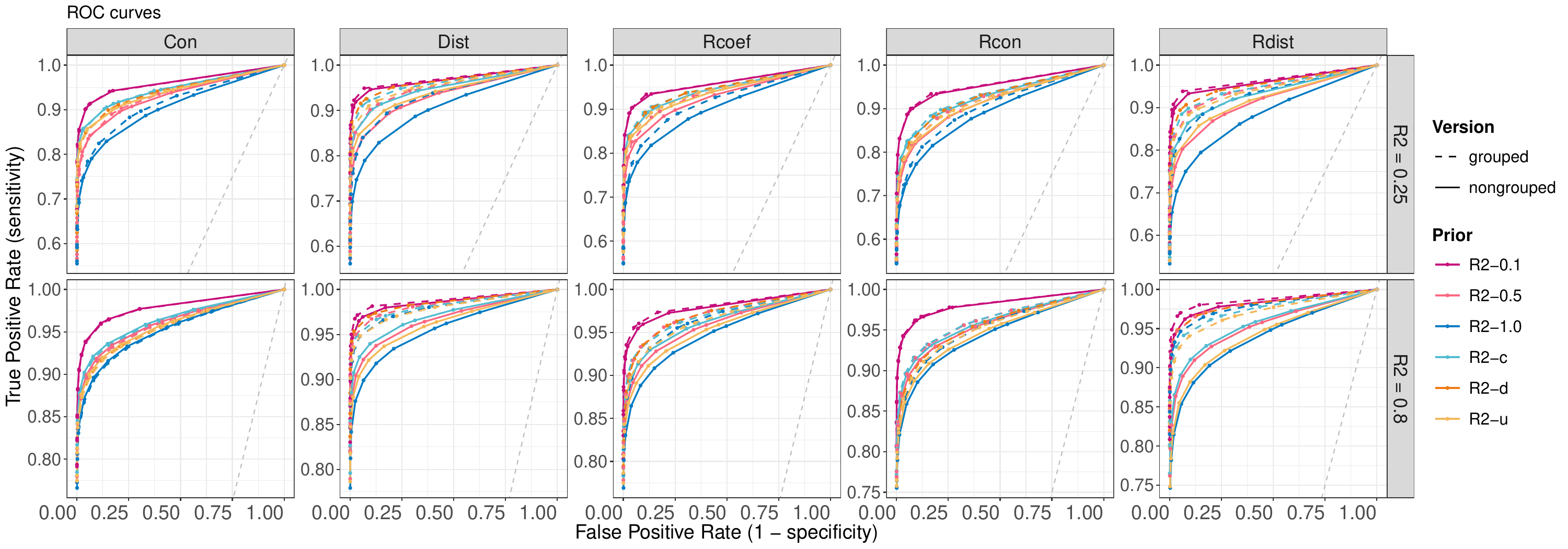}
\caption{ \textbf{ROC curves for the lower-dimensional scenario ($p = 100, n= 200$).} Grouped and non-grouped $R^2$ priors are compared. The grey line shows the diagonal of the unit square. ``Con” denotes the concentrated signal, ``Dist” the distributed signal, ``Rcon” and ``Rdist” their random counterparts, and ``Rcoef” the random coefficients setting.}
\label{fig:roc_low_dim}
\end{figure}%

Figure \ref{fig:roc_low_dim} displays ROC curves for the various priors across the considered low-dimensional scenarios. Overall, Group-R2 priors tend to outperform their non-grouped counterparts, particularly in settings where the signal is concentrated within specific groups. For example, this advantage is clearly visible for the \texttt{R2-}$1.0$ prior, which shows consistently better performance across all data-generating processes. However, as the concentration parameter $a_G$ decreases the performance gap narrows. For instance, under the \texttt{R2-}$0.1$ prior, the benefit of grouping becomes less pronounced.


\subsubsection{Simulations: High-dimensional results}

\begin{figure}[t]%
\includegraphics[width=\linewidth]{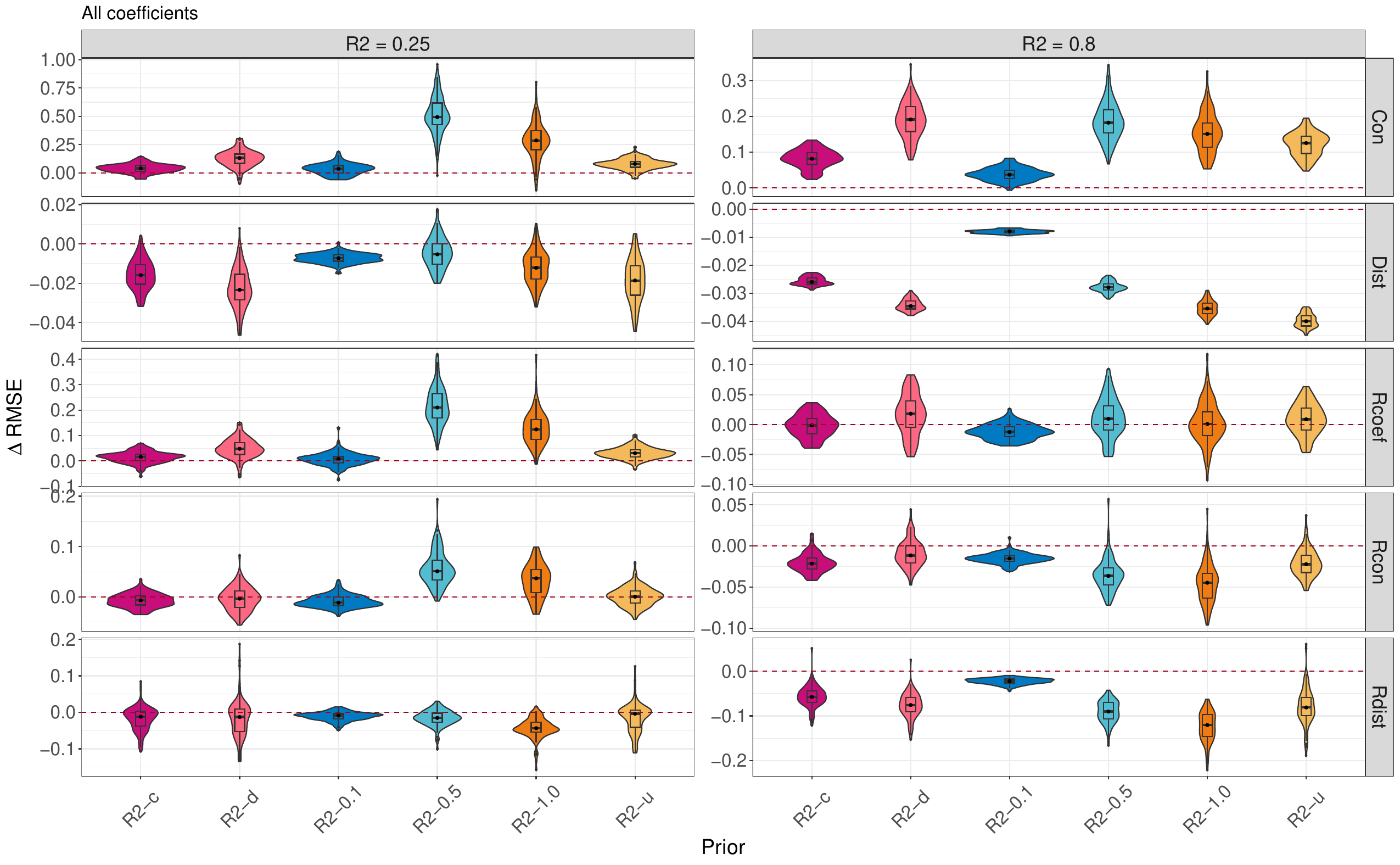}
  \caption{ \textbf{$\Delta \rmse$  for the high-dimensional scenario ($p = 500$, $n = 200$).} Values below the line indicate improvement in parameter recovery when considering groups. ``Con” denotes the concentrated signal, ``Dist” the distributed signal, ``Rcon” and ``Rdist” their random counterparts, and ``Rcoef” the random coefficients setting. Group-R2 priors improve parameter recovery across distributed and random signal settings, especially as $a_G$ increases. In the concentrated case, no substantial improvement or deterioration is observed. 
 }
  \label{fig:rmse_all_high_dimension}
\end{figure}

We show $\Delta\rmse$ in Figure~\ref{fig:rmse_all_high_dimension}. The results show that the Group-R2 priors consistently improve $\rmse$ across distributed signal settings in the high-dimensional case. In contrast, no improvement is achieved under the Random Coefficients case. However performance does improve in this scenario as $R^2$ increases. Even though improvement is not strong in this case, including grouping structures in the prior is not detrimental either, as the distributions of $\Delta \elpd$ are centered around zero. Concentrated signals are again challenging, however the loss involved by including groups is small.

\begin{figure}[!ht]%
\includegraphics[width=\linewidth]{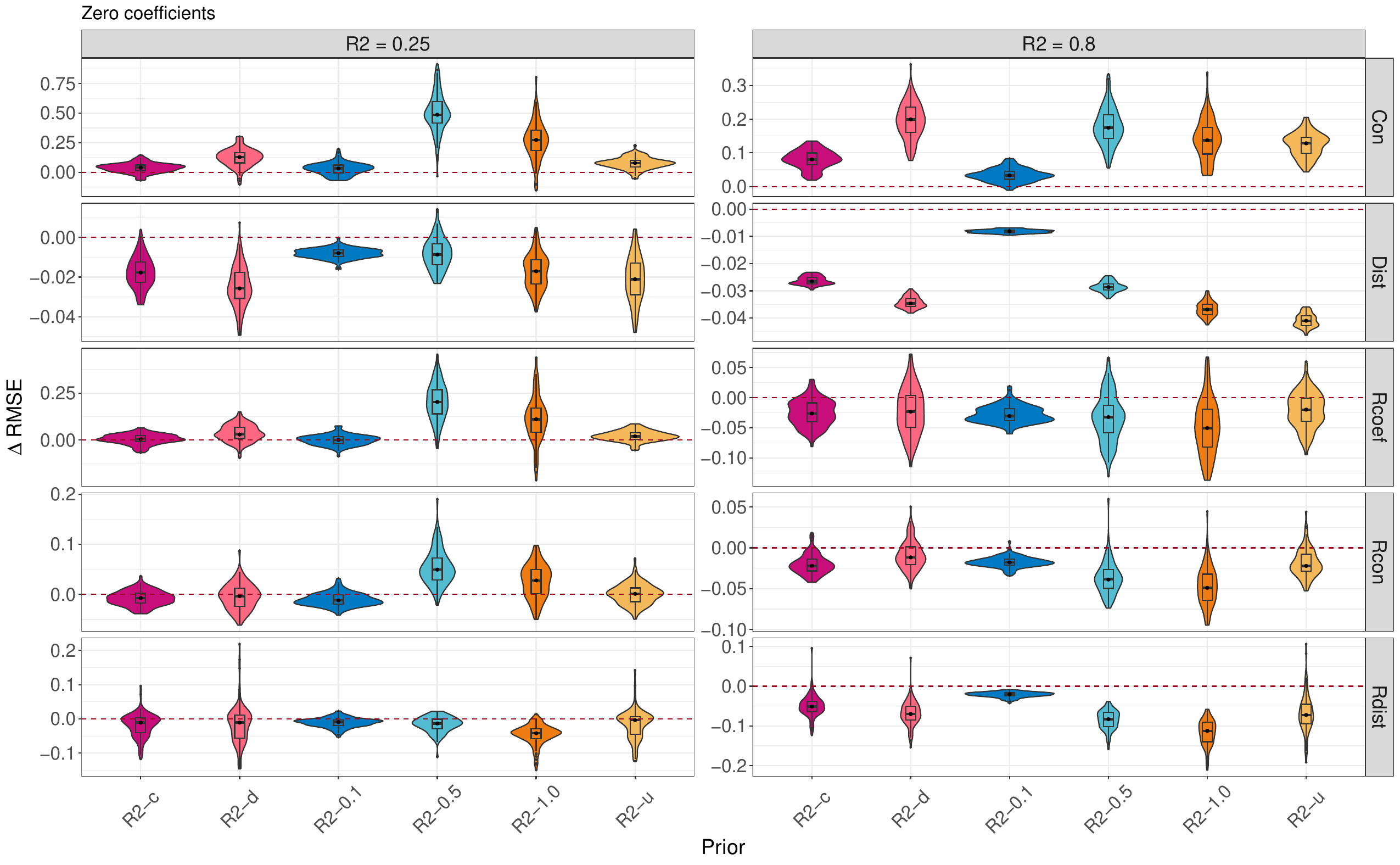}
  \caption{ \textbf{$\Delta \rmse$ for truly zero coefficients for the high-dimensional scenario ($p = 500$, $n = 200$).} Values below the line indicate improvement in parameter recovery when considering groups. ``Con” denotes the concentrated signal, ``Dist” the distributed signal, ``Rcon” and ``Rdist” their random counterparts, and ``Rcoef” the random coefficients setting. 
 }
\label{fig:rmse_zero_high_dimension}
\end{figure}

\begin{figure}[!ht]%
\includegraphics[width=\linewidth]{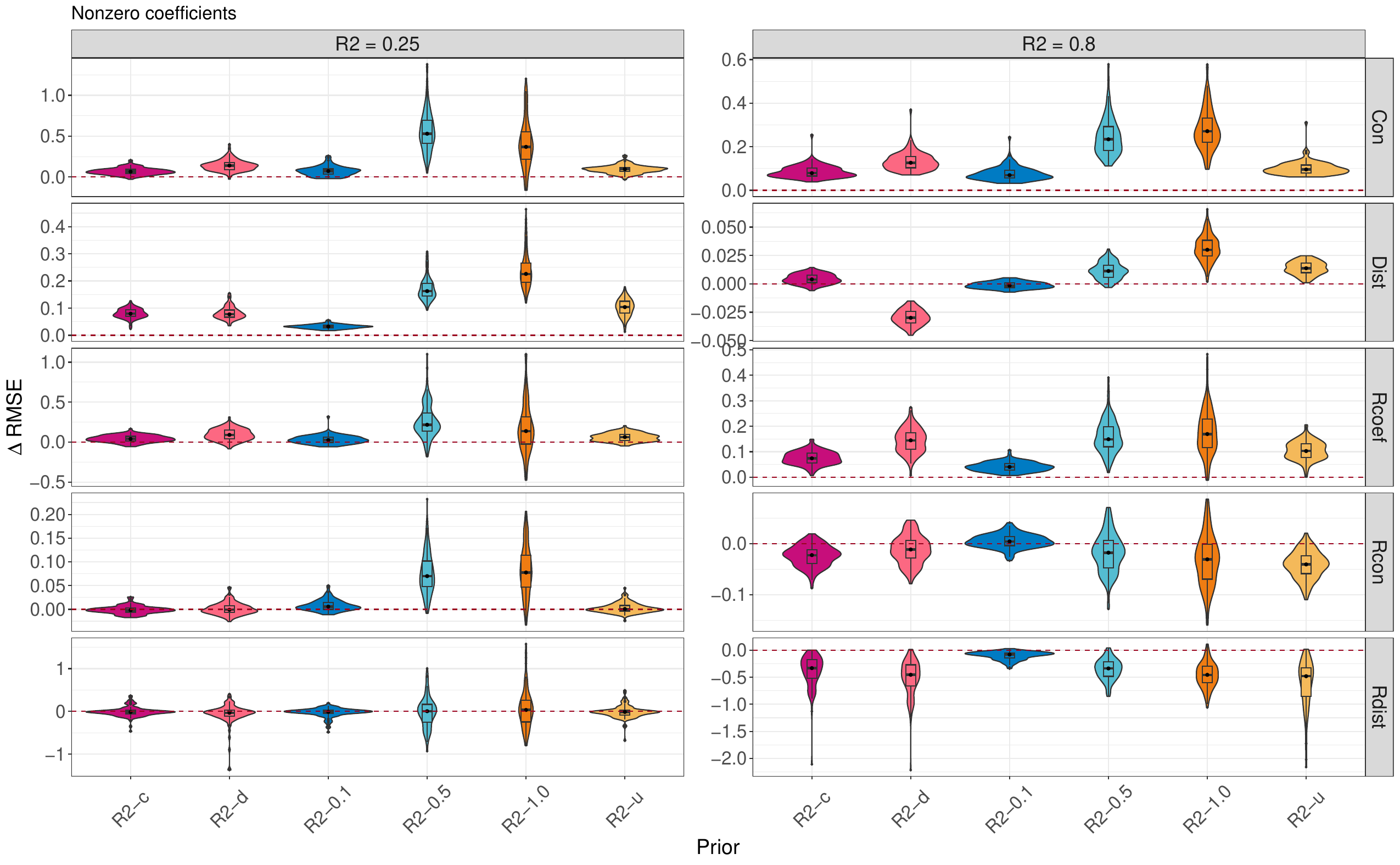}
  \caption{ \textbf{$\Delta \rmse$ for truly nonzero coefficients for the high-dimensional scenario ($p = 500$, $n = 200$).} Values below the line indicate improvement in parameter recovery when considering groups. ``Con” denotes the concentrated signal, ``Dist” the distributed signal, ``Rcon” and ``Rdist” their random counterparts, and ``Rcoef” the random coefficients setting. 
 } 
\label{fig:rmse_signal_high_dimension}
\end{figure}

We present the partitioned $\Delta\rmse$ results in Figures \ref{fig:rmse_zero_high_dimension} and \ref{fig:rmse_signal_high_dimension}, corresponding to truly zero and truly nonzero coefficients, respectively. The overall trends mirror those observed in the low-dimensional setting. Grouped versions of the R2 prior consistently outperform their non-grouped counterparts in detecting noise, with the exception of the concentrated signal scenario. In terms of signal recovery, certain priors lead to improved performance. As discussed earlier, the uniform R2 prior appears to offer a robust default across scenarios. However, we believe that incorporating well-informed prior knowledge about the grouping structure or sparsity pattern could lead to further gains.

\begin{figure}[h!]%
\includegraphics[width=\linewidth]{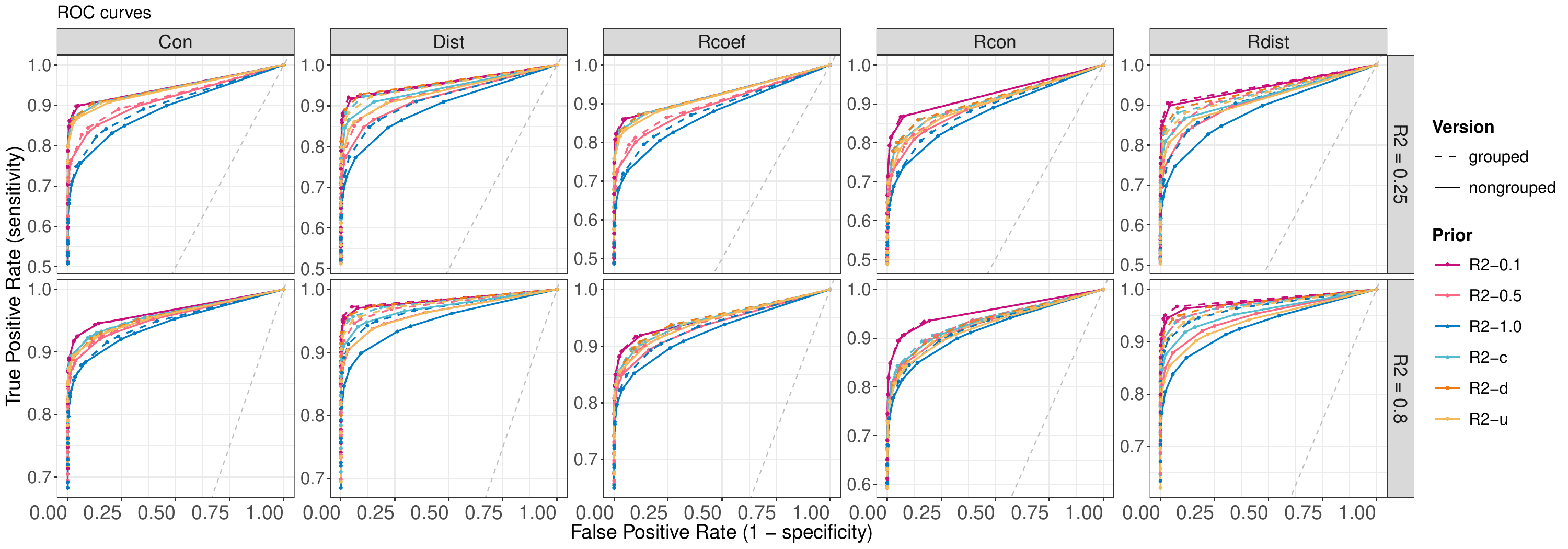}
 \caption{ \textbf{ROC curves for the high-dimensional scenario ($p = 500, n= 200$).} Grouped and non-grouped $R^2$ priors are compared. The grey line shows the diagonal of the unit square. ``Con” denotes the concentrated signal, ``Dist” the distributed signal, ``Rcon” and ``Rdist” their random counterparts, and ``Rcoef” the random coefficients setting.}
\label{fig:roc_high_dim}
\end{figure}

Figure~\ref{fig:roc_high_dim} shows the ROC curves for the high-dimensional case. The results again parallel the low-dimensional setting: under high noise and sparse signals, grouping structures help distinguish signal from noise more accurately on average, improving overall classification performance.

\subsection{Comparison with further grouped shrinkage priors}
\label{subsec:external_comparators}

The experiments of Section~\ref{sec:experiments} isolate the effect of grouping by contrasting each Group-$R^2$ prior with its own nongrouped counterpart. This appendix instead compares the Group-$R^2$ prior against five grouped shrinkage priors proposed elsewhere in the literature. 
These are the Bayesian group lasso with spike and slab (BGL-SS) and the Bayesian sparse group selection with spike and slab (BSGS-SS) of \citet{xu_bayesian_2015}, the group horseshoe (GHS) of \citet{xu_bayesian_2016}, the grouped regression with adaptive shrinkage priors (GRASP) of \citet{graspgroupedregressionadaptive2025}, and the group-informed $g$-prior of \citet{agarwal_bayesian_2025}, which we fit with the authors' \texttt{GiVSA} sampler and refer to as the $g$-prior throughout.

Writing $\mathcal{Q}$ for a quantity of interest, we report $\Delta \mathcal{Q} = \mathcal{Q}(\text{Group-}R^2) - \mathcal{Q}(M_{\text{ext}})$, where $M_{\text{ext}}$ is one of the competing group-priors. Because the two Group-$R^2$ configurations of Section~\ref{sec:experiments} encode opposite assumptions about how signal is spread within a group, we let the Group-$R^2$ side follow the data generating mechanism: \texttt{R2-c} ($a_G = 1$, $c_g = 0.5$) is used on the concentrated scenarios (``Con”, ``Rcon”, ``Rcoef”) and \texttt{R2-d}
($a_G = 0.5$, $c_g = 1$) on the distributed ones (``Dist”, ``Rdist”). This gives each competing model the more favorable of the two matched alternatives in every panel.
As before, $\Delta \elpd > 0$ and $\Delta \rmse < 0$ favor the Group-$R^2$ prior.

\subsubsection{Competing models}
\label{subsubsec:external_models}

Tables~\ref{tab:external_comparators} and \ref{tab:external_implementations} collect the five priors in the notation of Section~\ref{subsec:group_r2_intro}. All of them can be written as a scale mixture $b_{gl} \given V_{gl} \sim \normal(0, V_{gl})$, and they differ in how the variance $V_{gl}$ is built from a global, a group-level and a coefficient-level component, and in whether that construction admits an exact zero.

\begin{table}[t]
\footnotesize
\centering
\setlength{\tabcolsep}{4pt}
\begin{threeparttable}
\caption{Prior specification of the competing group-priors, in the notation of
Section~\ref{subsec:group_r2_intro}. Every prior is a scale mixture
$b_{gl} \given V_{gl} \sim \normal(0, V_{gl})$; $\delta_{0}$ denotes a point mass at
zero and $\mathrm{C}^{+}$ a half-Cauchy distribution. Implementation details are in
Table~\ref{tab:external_implementations}.}
\label{tab:external_comparators}
\begin{tabular}{>{\raggedright\arraybackslash}p{1.4cm}
                >{\raggedright\arraybackslash}p{5.0cm}
                >{\raggedright\arraybackslash}p{4.2cm}}
\toprule
Prior & Coefficient variance $V_{gl}$ & Hyperparameters \\
\midrule
BGL-SS &
A whole group is zero or not: $b_g \sim (1-\pi_0)\normal_{p_g}(0, \sigma^2\tau_g^2 I) + \pi_0 \delta_{0}$ &
$\tau_g^2 \sim \gammadist\!\left(\tfrac{p_g+1}{2}, \tfrac{\eta^2}{2}\right)$;
$\eta$ estimated by Monte Carlo EM \\
\addlinespace
BSGS-SS &
As BGL-SS, with a second spike and slab on each $b_{gl}$ inside an active group &
As BGL-SS \\
\addlinespace
GHS &
$\sigma^2 \tau^2 \delta_g^2 \lambda_{gl}^2$ &
$\delta_g \sim \mathrm{C}^{+}(0,1)$, $\lambda_{gl} \sim \mathrm{C}^{+}(0,1)$ \\
\addlinespace
GRASP &
$\sigma^2 \tau^2 \delta_g^2 \lambda_{gl}^2$ &
$\dfrac{\delta_g^2}{1+\delta_g^2} \sim \betadist(\alpha_D, \beta_D)$,
$\dfrac{\lambda_{gl}^2}{1+\lambda_{gl}^2} \sim \betadist(\alpha_G, \beta_G)$;
all four shapes learned \\
\addlinespace
$g$-prior (GiVSA) &
$g\,\sigma^2 G_{\mathcal{M}}^{-1}$ for a selected model $\mathcal{M}$ and $0$
otherwise, with $G_{\mathcal{M}} = \diag(X_g'X_g)_{g \in \mathcal{M}}$ &
$g = c_0\, p^{2(1+\varsigma)}$ with $(c, c_0) = (0.01, 0.0025)$ and
$\varsigma = 10^{-2}$ \\
\bottomrule
\end{tabular}
\end{threeparttable}
\end{table}

\begin{table}[t]
\footnotesize
\centering
\setlength{\tabcolsep}{3pt}
\begin{threeparttable}
\caption{Implementations used for the competing group-priors. Each prior is fitted
with its authors' own code rather than a reimplementation. The last column records
the step we add where the released code does not return posterior draws of all the
quantities the evaluation of Section~\ref{sec:eval_metrics} requires.}
\label{tab:external_implementations}
\begin{tabular}{>{\raggedright\arraybackslash}p{1.15cm}
                >{\raggedright\arraybackslash}p{1.35cm}
                >{\raggedright\arraybackslash}p{2.45cm}
                >{\raggedright\arraybackslash}p{2.95cm}
                l}
\toprule
Prior & Package & Sampler & Completion & Reference \\
\midrule
BGL-SS\tnote{c}  & \texttt{MBSGS}   & Gibbs & $\sigma^2$ retained from the sweep in which $b$ is stored, and $b_0$ drawn from its exact flat-prior conditional & \citet{xu_bayesian_2015} \\
\addlinespace
BSGS-SS & \texttt{MBSGS}   & Gibbs & As BGL-SS & \citet{xu_bayesian_2015} \\
\addlinespace
GHS     & \texttt{bayesreg}\tnote{a} & Gibbs & $b_0$ drawn from its exact flat-prior conditional & \citet{xu_bayesian_2016} \\
\addlinespace
GRASP\tnote{c}   & \texttt{bayesreg}\tnote{a} & Gibbs with Metropolis steps for the shape parameters & As GHS & \citet{graspgroupedregressionadaptive2025} \\
\addlinespace
$g$-prior (GiVSA) & \texttt{GiVSA}\tnote{b} & Population-based multiple-try Metropolis over model space & $(b, \sigma^2)$ drawn from the closed-form Normal--Inverse-Gamma conditional of the $g$-prior given each visited model, then $b_0$ as above & \citet{agarwal_bayesian_2025} \\
\bottomrule
\end{tabular}
\begin{tablenotes}\footnotesize
\item[a] The grouped fork of \texttt{bayesreg} released with
\citet{graspgroupedregressionadaptive2025}. The CRAN version has no group priors.
\item[b] The authors' own code. It samples over model space only and returns no
draws of $b$ or $\sigma^2$.
\item[c] GRASP is reported for $p = 100$ only, and BGL-SS is omitted from the
$p = 500$ figures, as described in Section~\ref{subsubsec:external_elpd}.
\end{tablenotes}
\end{threeparttable}
\end{table}

All five priors are fitted to a standardized design and a centered response, and are run as a single chain of $2000$ draws after $2000$ burn-in iterations, which matches the $4 \times 1000$ post-warmup draws used for the Group-$R^2$ models. Neither BGL-SS, BSGS-SS nor the $g$-prior included an intercept in their package implementations, so for these $b_0$ is drawn from its exact conditional $b_0 \given b, \sigma^2 \sim \normal(\bar{y} - \bar{x}'b, \sigma^2/n)$ under a flat prior. For the $g$-prior, which samples over model space alone, $b$ and $\sigma^2$ are drawn from the Normal--Inverse-Gamma conditional implied by the $g$-prior at each visited model, which is available in closed form. All coefficients are returned to the original scale of $X$ before any comparison is made.

\subsubsection{Functional differences}
\label{subsubsec:external_discussion}

Three differences are relevant for what follows.

First, and most importantly, the competing models divide into \emph{selection} priors and \emph{continuous shrinkage} priors. BGL-SS, BSGS-SS and the $g$-prior place positive probability on $b_{gl} = 0$ exactly, either for a whole group or for individual coefficients, whereas GHS, GRASP and the Group-$R^2$ prior shrink continuously and set nothing exactly to zero. Since the data generating processes of Section~\ref{sec:experiments} are genuinely sparse, we expect the selection priors to be more accurate on the truly zero coefficients, and the continuous priors to pay less of a shrinkage penalty on the truly nonzero ones. Aggregate measures of parameter recovery are dominated by the zero coefficients and should therefore favor the selection priors.

Second, the priors differ in \emph{where} the variance budget is controlled. The Group-$R^2$ prior is the only one that places an interpretable prior directly on $R^2$ and then partitions it, first across groups through $\phi \sim \dirichlet s(a_G)$ and then within groups through $\varphi_g \sim \dirichlet s(c_g)$. The simplex constraint means the groups compete for a fixed budget $\tau^2$, so evidence for one group is evidence against the others. GHS and GRASP have a group scale $\delta_g^2$ but no such constraint: their group scales are independent, and the total explained variance is whatever the product of the scales implies. 

To anticipate the results, note that a symmetric $\dirichlet s(a_G)$ has mean $1/G$ in every coordinate, so an allocation in which every group carries a comparable share of the variance is close to the mean of the distribution, whereas an allocation in which a single group carries nearly all of it sits far in the tail, and increasingly so as $G$ grows. We therefore expect the Group-$R^2$ prior to do comparatively well on the concentrated scenarios, where each of the $G$ groups contains one signal, and comparatively less well on the distributed ones, where a single active group must absorb the entire variance budget.

Third, the $g$-prior of \citet{agarwal_bayesian_2025} is the only competing model whose informativeness grows with the design: $g = c_0 p^{2(1+\varsigma)}$ increases faster than $p$, so the prior is substantially more concentrated at $p = 500$ than at $p = 100$ and visits smaller models. Its two blocks should therefore not be read as one behavior observed twice.

\subsubsection{Predictive performance}
\label{subsubsec:external_elpd}

\begin{figure}[t]%
\includegraphics[width=\linewidth]{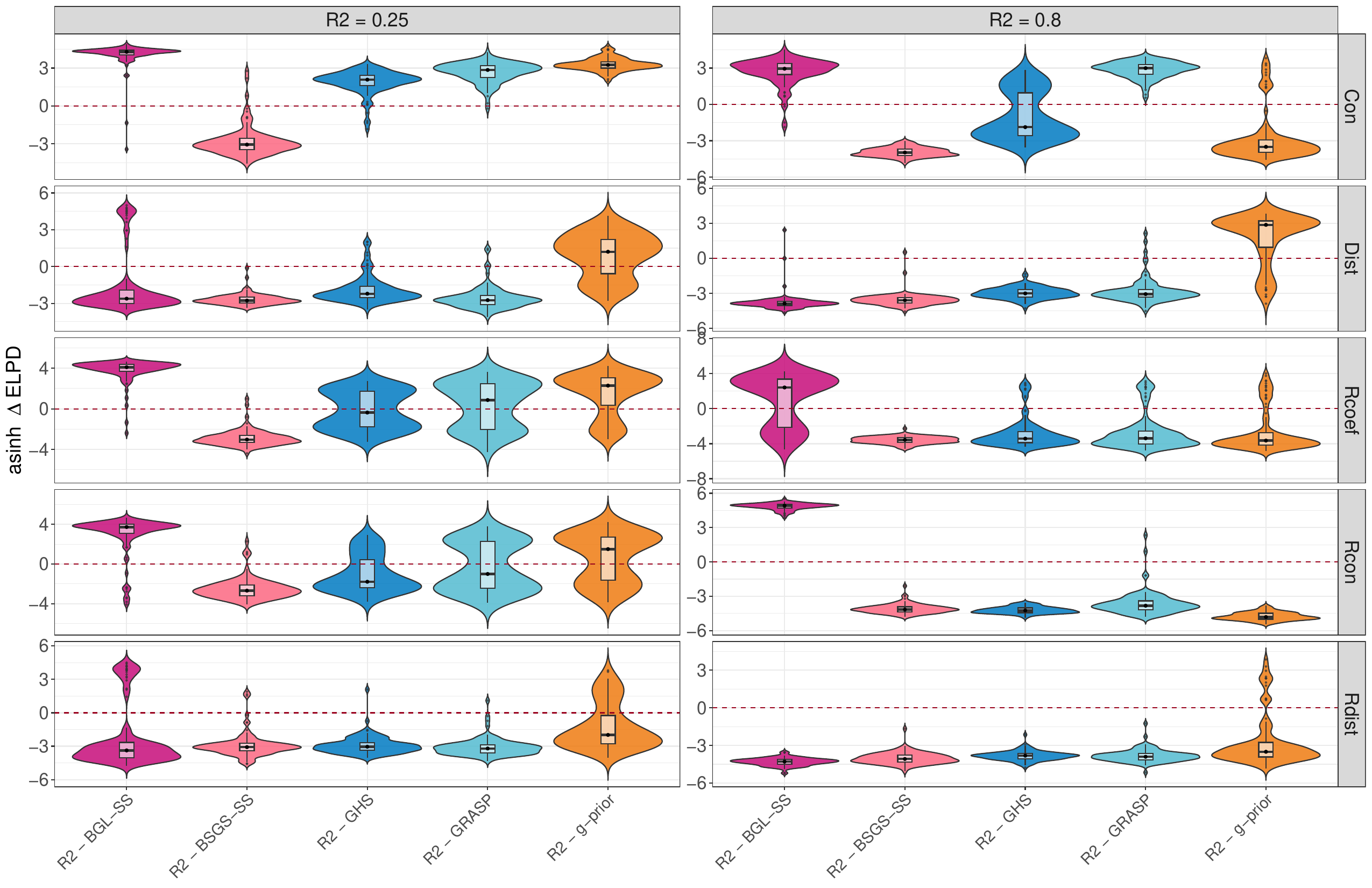}
  \caption{ \textbf{$\Delta \elpd$ (asinh-transformed) against the competing group-priors for the lower-dimensional scenario ($p = 100$, $n = 200$)}. Values above the line indicate better predictive performance of the Group-$R^2$ prior. The Group-$R^2$ side is \texttt{R2-c} on the concentrated scenarios and \texttt{R2-d} on the distributed ones. ``Con” denotes the concentrated signal, ``Dist” the distributed signal, ``Rcon” and ``Rdist” their random counterparts, and ``Rcoef” the random coefficients setting. The Group-$R^2$ prior is competitive with, but does not dominate, the competing group-priors.}
  \label{fig:external_elpd_low_dimension}
\end{figure}

\begin{figure}[t]%
\includegraphics[width=\linewidth]{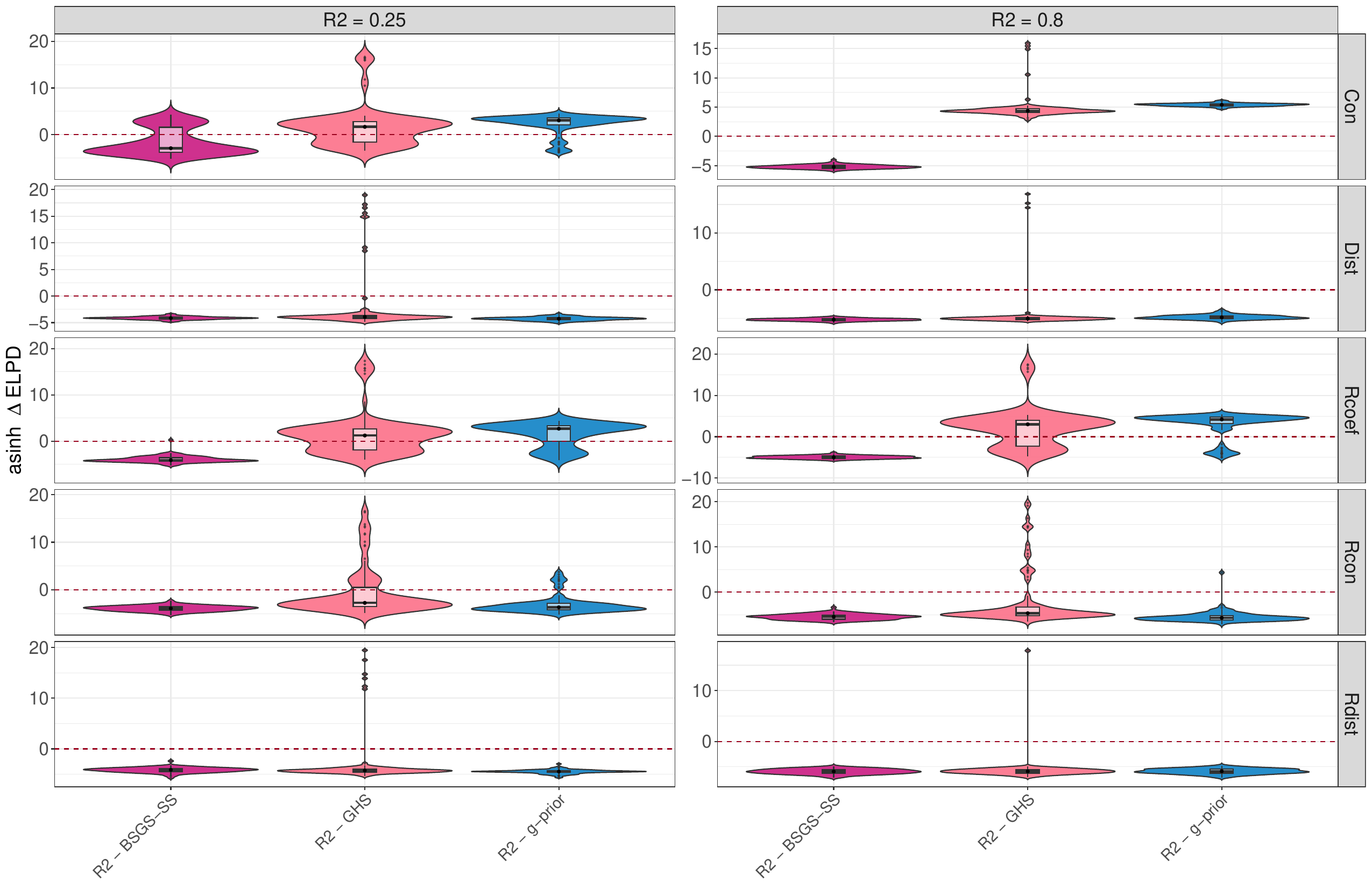}
  \caption{ \textbf{$\Delta \elpd$ (asinh-transformed) against the competing group-priors for the high-dimensional scenario ($p = 500$, $n = 200$)}. Values above the line indicate better predictive performance of the Group-$R^2$ prior. GRASP and BGL-SS are omitted from this block because their samplers break down when $p > n$, as described in Section~\ref{subsubsec:external_elpd}. ``Con” denotes the concentrated signal, ``Dist” the distributed signal, ``Rcon” and ``Rdist” their random counterparts, and ``Rcoef” the random coefficients setting. The upper outliers in the GHS panels are replications in which its sampler becomes unstable.}
  \label{fig:external_elpd_high_dimension}
\end{figure}

Figures~\ref{fig:external_elpd_low_dimension} and \ref{fig:external_elpd_high_dimension} report $\Delta \elpd$ for the two lower and higher-dimensional DGPs respectively. In the lower-dimensional setting the Group-$R^2$ prior is competitive but not dominant: it improves on BGL-SS (median $\Delta \elpd = 8.2$, favorable in $58\%$ of replications), is indistinguishable from the $g$-prior ($-1.6$, $45\%$), and is outperformed by GHS ($-6.8$, $22\%$), GRASP ($-7.3$, $32\%$) and most clearly by BSGS-SS ($-15.4$, $1\%$). In the high-dimensional setting the ordering is preserved and the gaps widen: GHS ($-14.4$, $36\%$), the $g$-prior ($-28.5$, $36\%$) and BSGS-SS ($-53.5$, $3\%$) all predict better than the Group-$R^2$ prior.

Consistent with the budget argument above, the Group-$R^2$ prior is at its strongest on the concentrated scenarios, where it improves on every competing model except BSGS-SS at $p = 100$, and at its weakest on the distributed scenarios, where it is behind almost uniformly. Pooling over the competing models, the median $\Delta \elpd$ is $-1.1$ on the concentrated scenarios against $-11.4$ on the distributed ones at $p = 100$, and $-13.9$ against $-58.4$ at $p = 500$. The widening of the gap with $G$ is what the simplex argument predicts, since the distributed design requires one of $G$ groups to hold essentially the whole budget and that configuration moves further into the tail of $\dirichlet s(a_G)$ as $G$ grows. The advantage of BSGS-SS is likewise concentrated where its extra within-group spike and slab is most useful.

The comparison also deteriorates as the signal strengthens. Moving from $R^2 = 0.25$ to $R^2 = 0.80$, the median $\Delta \elpd$ falls against every competing model, from $18.6$ to $2.7$ (BGL-SS), $3.5$ to $-17.5$ ($g$-prior), $-2.5$ to $-15.8$ (GHS), $-3.2$ to $-13.2$ (GRASP) and $-9.2$ to $-23.8$ (BSGS-SS) at $p = 100$, with the same pattern at $p = 500$. This is a property of the hyperparameters used here rather than of the prior class. Both Group-$R^2$ configurations in this comparison set $R^2 \sim \betadist(1, 2)$, with prior mean $1/3$ and $\Pr(R^2 > 0.8) = 0.04$, so the prior is close to the truth when $R^2 = 0.25$ and places almost no mass near it when $R^2 = 0.80$, and $\tau^2$ is shrunk towards too small a total variance. None of the competing priors constrains the total explained variance in this way. The scales of GHS and GRASP are independent and heavy-tailed, which allows better insulation in our simulations against high true signal strength. The same mechanism appears with the opposite sign in Section~\ref{subsec:gigg_comparison}, where the rule $a_1 = G a_G$ places the prior mean of $R^2$ well above the truth and the Group-$R^2$ prior is then the over-optimistic one. Predictive performance relative to these models therefore tracks how well the prior on $R^2$ is centered, which is the quantity the prior makes available for the user to set.

Two competing models fail numerically at $p = 500$ and are excluded from that block rather than reported. GRASP aborts $746$ of $1000$ fits: its adaptive update of the beta shape parameters fits a gamma proposal whose scale can become non-positive, after which the Metropolis acceptance ratio is not defined. BGL-SS completes every fit but returns a degenerate posterior in $86\%$ of replications, with the residual scale overestimated by more than an order of magnitude. The resulting $\Delta \elpd$ would be of order $10^{5}$ and would reflect the breakdown of the sampler rather than the predictive performance of the Group-$R^2$ prior. On the other hand, both behave well at $p = 100$. GHS shows a milder version of the same instability, with $5\%$ of its $p = 500$ fits producing extreme values. These are retained and are visible as the upper outliers in Figure~\ref{fig:external_elpd_high_dimension}.

\subsubsection{Parameter recovery}
\label{subsubsec:external_rmse}

\begin{figure}[!ht]%
\includegraphics[width=\linewidth]{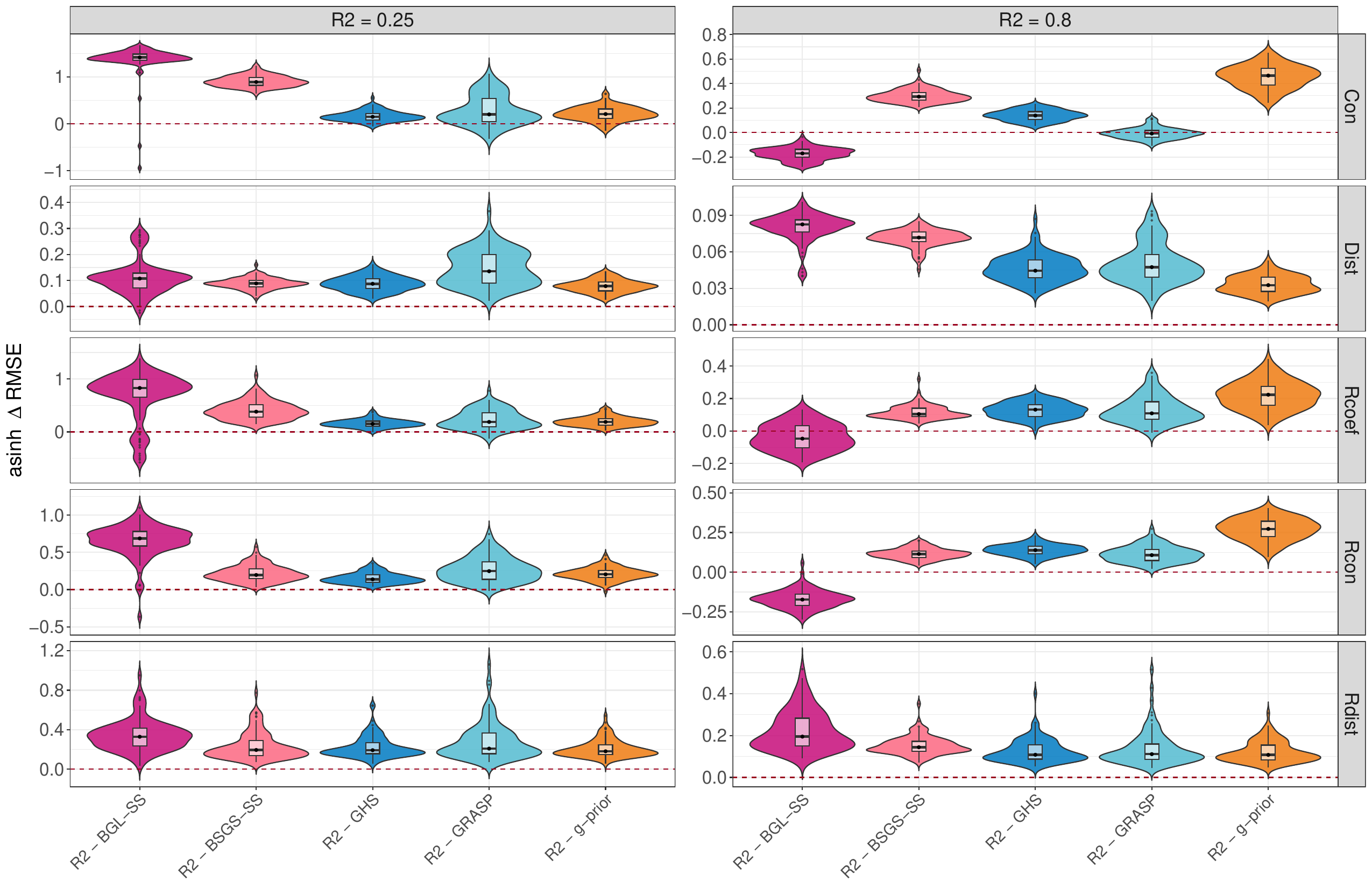}
  \caption{ \textbf{$\Delta \rmse$ against the competing group-priors for the lower-dimensional scenario ($p = 100$, $n = 200$).} Values below the line indicate better parameter recovery of the Group-$R^2$ prior. ``Con” denotes the concentrated signal, ``Dist” the distributed signal, ``Rcon” and ``Rdist” their random counterparts, and ``Rcoef” the random coefficients setting. Aggregate recovery favors the competing models, and the decomposition in Figures~\ref{fig:external_rmse_zero_low_dimension} and \ref{fig:external_rmse_signal_low_dimension} shows why.}
  \label{fig:external_rmse_all_low_dimension}
\end{figure}

\begin{figure}[!ht]%
\includegraphics[width=\linewidth]{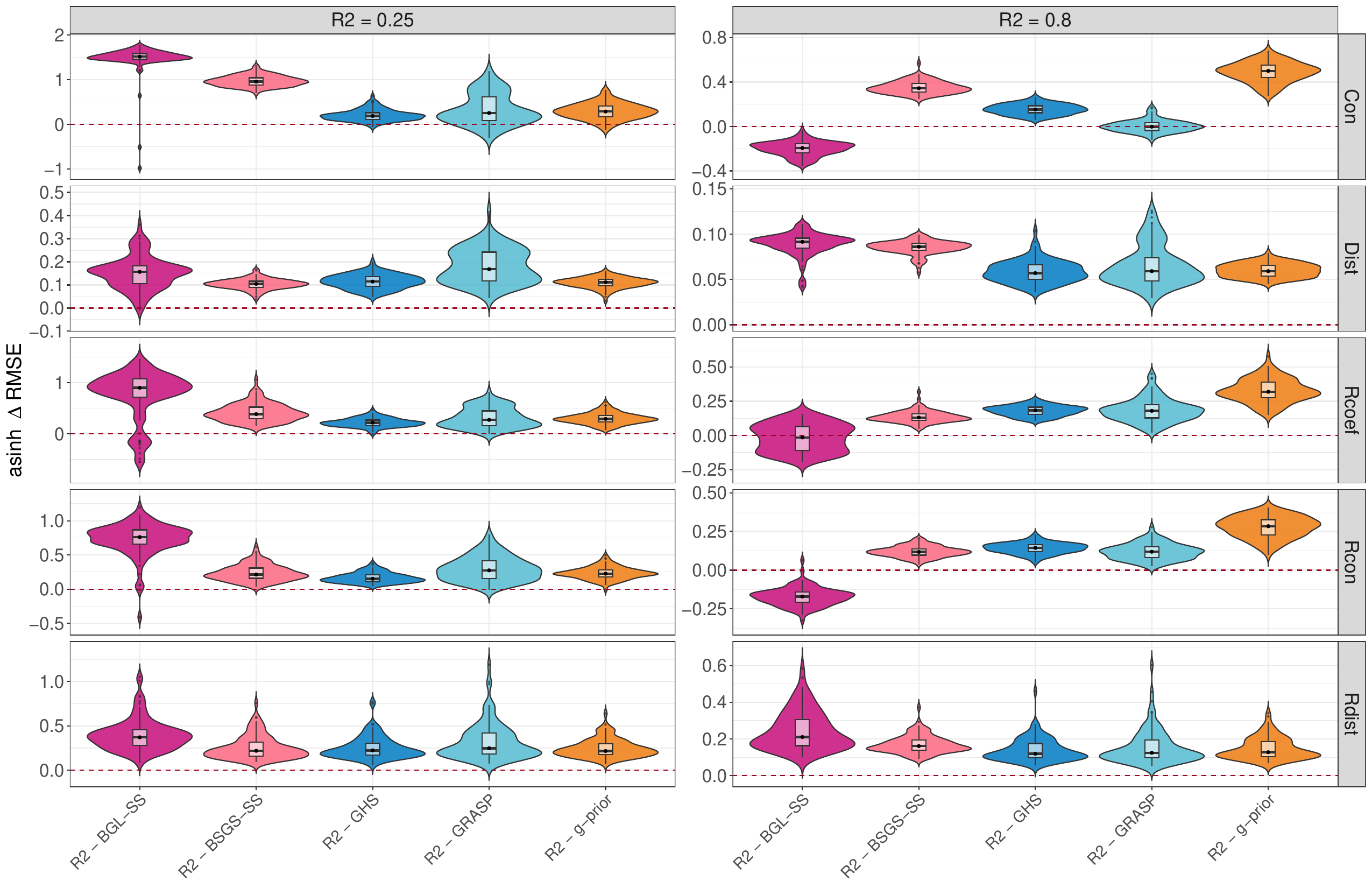}
  \caption{ \textbf{$\Delta \rmse$ for truly zero coefficients against the competing group-priors ($p = 100$, $n = 200$).} Values below the line indicate better parameter recovery of the Group-$R^2$ prior. ``Con” denotes the concentrated signal, ``Dist” the distributed signal, ``Rcon” and ``Rdist” their random counterparts, and ``Rcoef” the random coefficients setting. Priors that place mass at exactly zero recover the zero coefficients better.}
  \label{fig:external_rmse_zero_low_dimension}
\end{figure}

\begin{figure}[!ht]%
\includegraphics[width=\linewidth]{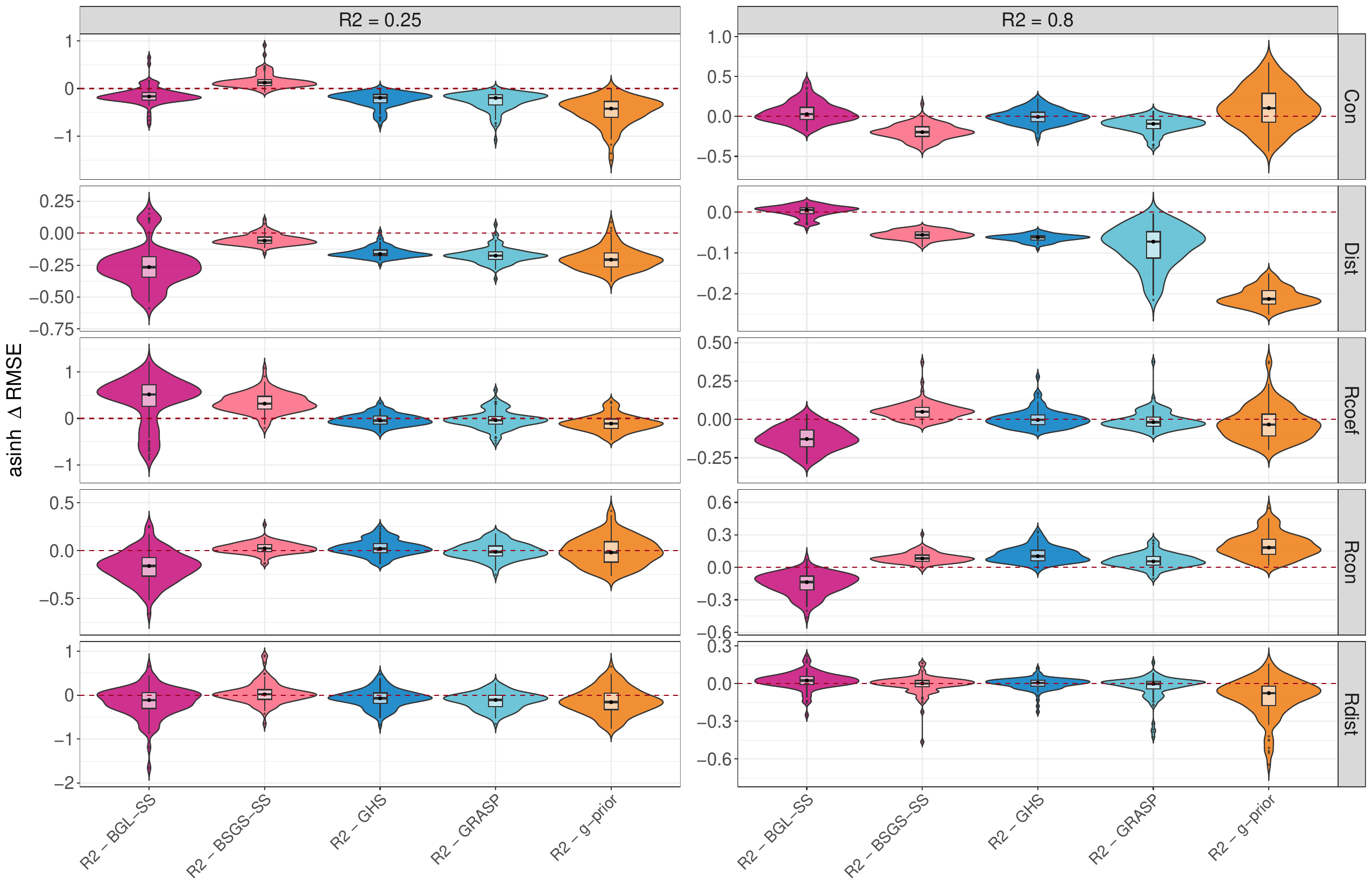}
  \caption{ \textbf{$\Delta \rmse$ for truly nonzero coefficients against the competing group-priors ($p = 100$, $n = 200$).} Values below the line indicate better parameter recovery of the Group-$R^2$ prior. ``Con” denotes the concentrated signal, ``Dist” the distributed signal, ``Rcon” and ``Rdist” their random counterparts, and ``Rcoef” the random coefficients setting. The ordering of Figure~\ref{fig:external_rmse_zero_low_dimension} reverses against all competing models except BSGS-SS.}
  \label{fig:external_rmse_signal_low_dimension}
\end{figure}

\begin{figure}[!ht]%
\includegraphics[width=\linewidth]{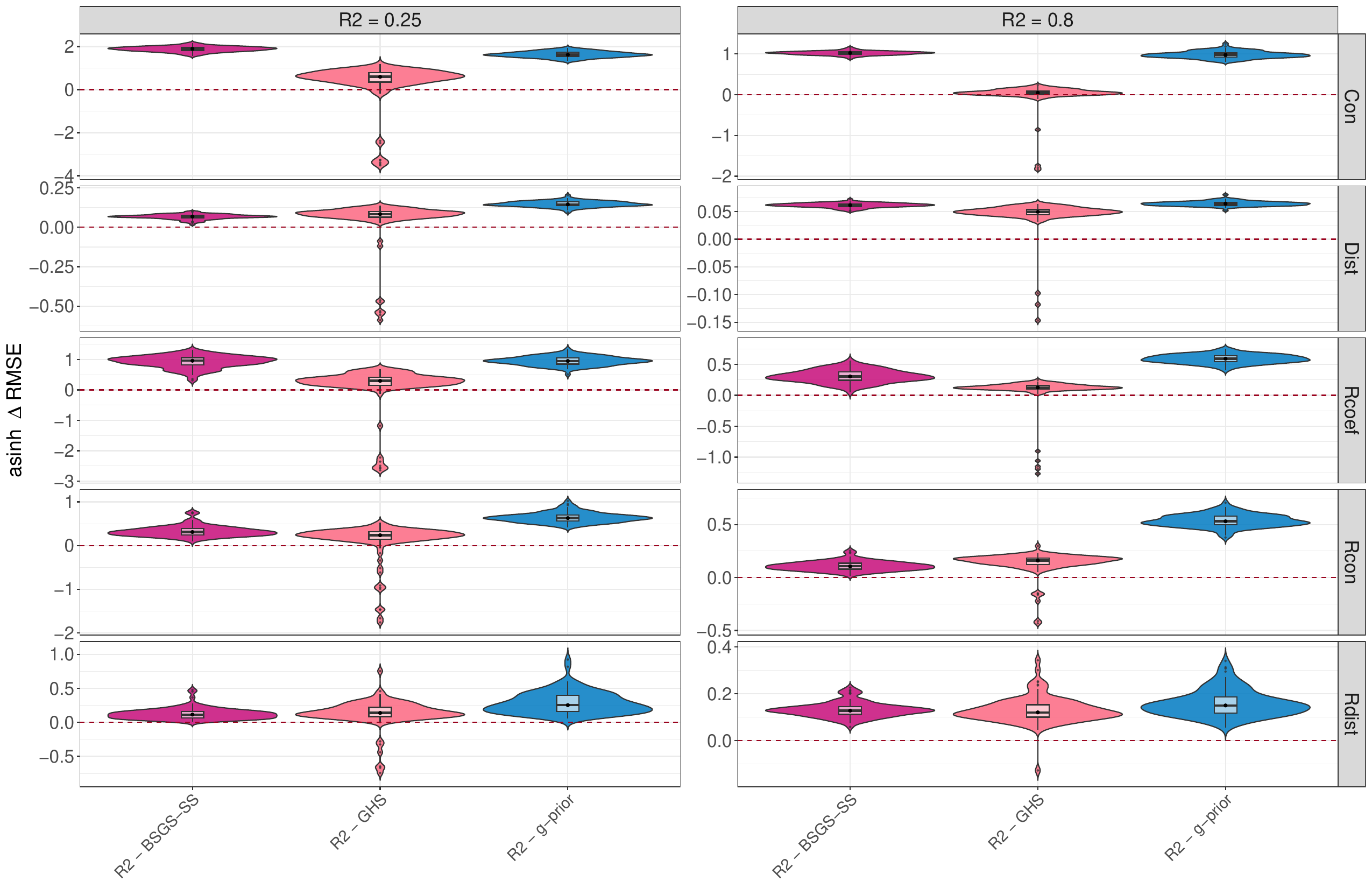}
  \caption{ \textbf{$\Delta \rmse$ against the competing group-priors for the high-dimensional scenario ($p = 500$, $n = 200$).} Values below the line indicate better parameter recovery of the Group-$R^2$ prior. GRASP and BGL-SS are omitted from this block, as described in Section~\ref{subsubsec:external_elpd}. ``Con” denotes the concentrated signal, ``Dist” the distributed signal, ``Rcon” and ``Rdist” their random counterparts, and ``Rcoef” the random coefficients setting.}
  \label{fig:external_rmse_all_high_dimension}
\end{figure}

\begin{figure}[!ht]%
\includegraphics[width=\linewidth]{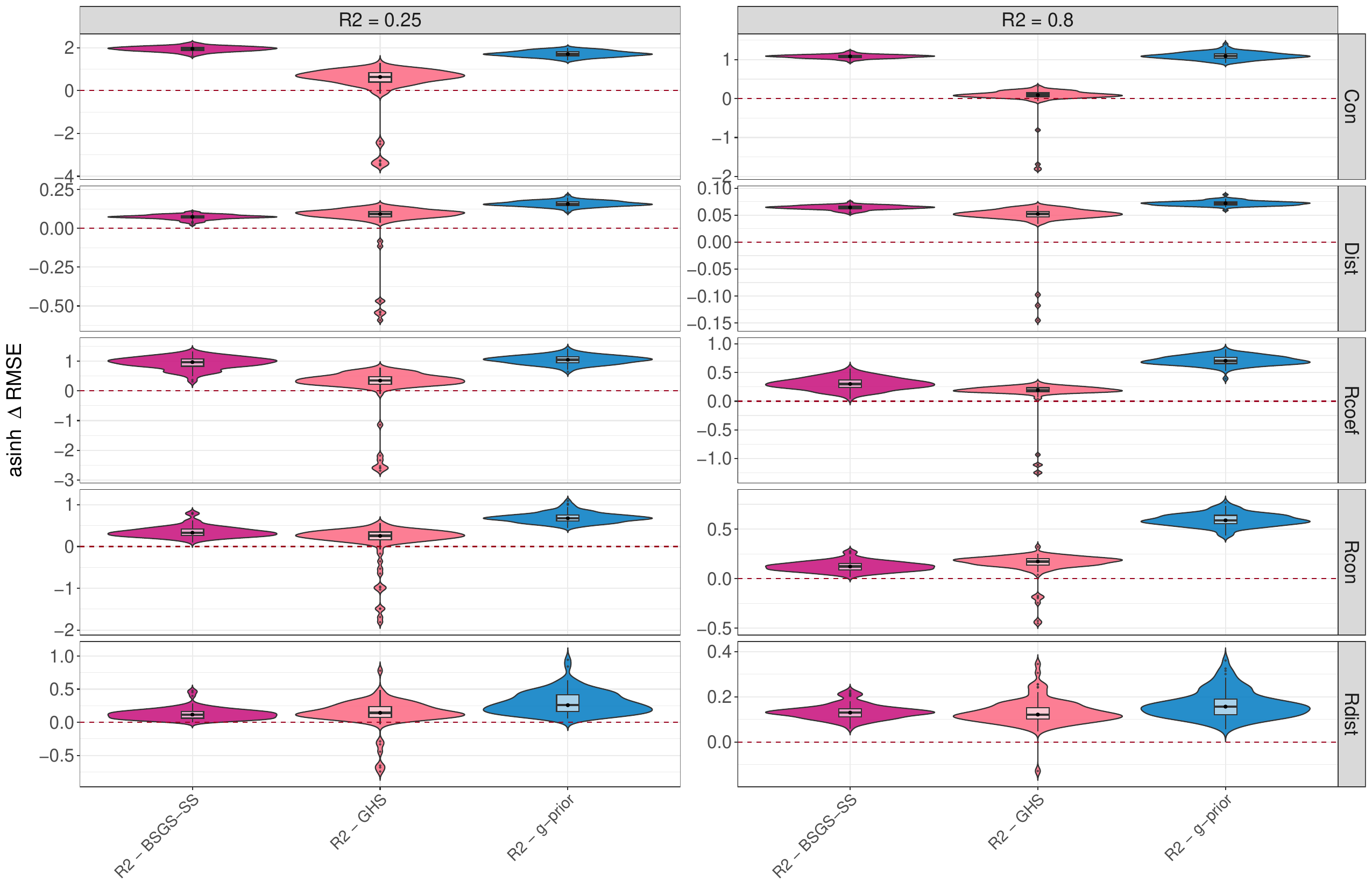}
  \caption{ \textbf{$\Delta \rmse$ for truly zero coefficients against the competing group-priors ($p = 500$, $n = 200$).} Values below the line indicate better parameter recovery of the Group-$R^2$ prior. ``Con” denotes the concentrated signal, ``Dist” the distributed signal, ``Rcon” and ``Rdist” their random counterparts, and ``Rcoef” the random coefficients setting.}
  \label{fig:external_rmse_zero_high_dimension}
\end{figure}

\begin{figure}[!ht]%
\includegraphics[width=\linewidth]{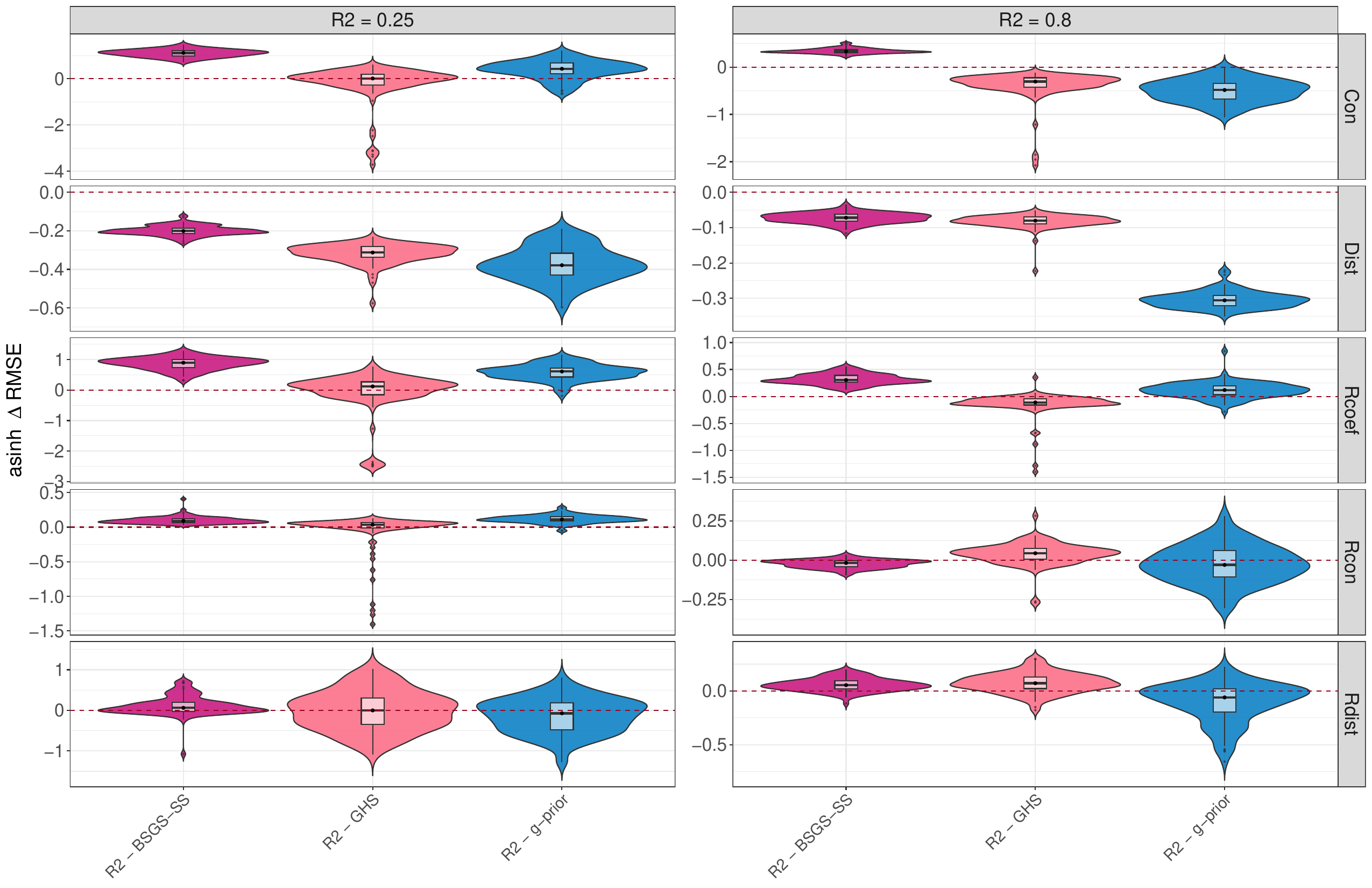}
  \caption{ \textbf{$\Delta \rmse$ for truly nonzero coefficients against the competing group-priors ($p = 500$, $n = 200$).} Values below the line indicate better parameter recovery of the Group-$R^2$ prior. ``Con” denotes the concentrated signal, ``Dist” the distributed signal, ``Rcon” and ``Rdist” their random counterparts, and ``Rcoef” the random coefficients setting.}
  \label{fig:external_rmse_signal_high_dimension}
\end{figure}

Figures~\ref{fig:external_rmse_all_low_dimension} and \ref{fig:external_rmse_all_high_dimension} report $\Delta \rmse$ over all coefficients, and Figures~\ref{fig:external_rmse_zero_low_dimension}--\ref{fig:external_rmse_signal_high_dimension} split it into the truly zero and the truly nonzero coefficients.

On aggregate the Group-$R^2$ prior recovers coefficients less accurately than every competing model, in both blocks, with median $\Delta \rmse$ between $0.12$ and $0.57$. The decomposition shows that this is driven entirely by the zero coefficients, where the median difference is positive throughout ($0.15$ to $0.66$): the selection priors place mass at zero and the continuous Group-$R^2$ prior does not, so a coefficient that is truly zero is estimated exactly by the former and only shrunk by the latter. On the truly nonzero coefficients the ordering reverses against four of the five competing models, with median differences of $-0.11$ ($g$-prior), $-0.07$ (BGL-SS), $-0.06$ (GRASP) and $-0.04$ (GHS) at $p = 100$: the Group-$R^2$ prior over-shrinks genuine signal less than they do. BSGS-SS is the exception and recovers the nonzero coefficients better as well ($+0.02$ at $p = 100$, $+0.11$ at $p = 500$), which is consistent with its within-group spike and slab being the only competing prior that can keep a group active while zeroing individual coefficients inside it.

This is the familiar trade-off between selection and continuous shrinkage rather than a deficiency specific to the Group-$R^2$ prior, and it has a direct practical implication: if the goal is to recover a sparse coefficient vector, a prior with an exact zero is preferable, and if the goal is to estimate the magnitude of the coefficients that are present, continuous shrinkage of the Group-$R^2$ form is competitive. Since the aggregate is dominated by the zeros in these designs, the aggregate alone gives a misleading picture of either.

\subsubsection{Sampling diagnostics}
\label{subsubsec:external_diagnostics}

Tables~\ref{tab:external_diag_low} and \ref{tab:external_diag_high} report convergence diagnostics for the competing group-priors and for the Group-$R^2$ side of each panel, constructed as in Section~\ref{subsec:gigg_diagnostics}. For every fit we take the largest $\widehat{R}$ and the smallest bulk effective sample size over the coefficients of the relevant type, and report the median of those quantities over the $100$ replications. Two differences from Section~\ref{subsec:gigg_diagnostics} should be kept in mind. The five competing models are Gibbs samplers run as a single chain, so $\widehat{R}$ is computed by splitting that chain rather than across independent chains, and none of them produces the divergence diagnostics available for the Hamiltonian sampler used for the Group-$R^2$ models.

The Group-$R^2$ posterior is sampled more reliably than any of the competing models in every scenario and both dimensions, with $\widehat{R}$ at or below $1.004$ and median effective sample sizes above $2000$ and no divergent transitions. Among the competing models the two spike and slab priors mix well, with $\widehat{R}$ below $1.011$ throughout, while GHS, GRASP and the $g$-prior do not. At $p = 100$ their median $\widehat{R}$ reaches $1.40$ and their median effective sample sizes fall to between $7$ and $180$ out of $2000$ draws, and at $p = 500$ the $g$-prior returns effective sample sizes in the single digits. For the $g$-prior this is expected rather than anomalous, since it samples over model space and its marginal for a coefficient is a mixture of a point mass and a continuous component, which is strongly autocorrelated by construction. For GHS and GRASP it reflects the difficulty of their posterior geometry at these dimensions. The consequence for the comparison is that the deficits reported against GHS, GRASP and the $g$-prior in Section~\ref{subsubsec:external_elpd} rest on far fewer effective draws than the Group-$R^2$ side, and should be read as indicative rather than precise. The comparison against BGL-SS and BSGS-SS does not carry this caveat.

A further point concerns BGL-SS. In $27\%$ of its fits at $p = 100$ and $6\%$ at $p = 500$ at least one coefficient is exactly constant over the whole chain, so $\widehat{R}$ and the effective sample size are not defined for it. This is a property of a group spike and slab rather than a failure, since it means the group in question was never selected, and those coefficients are excluded from the medians reported below.

\begin{table}[!ht]
\footnotesize
\centering
\setlength{\tabcolsep}{4pt}
\caption{Sampling diagnostics for the lower-dimensional scenario ($p = 100$, $n = 200$).
Median over replications of the per-fit maximum $\widehat{R}$ and minimum bulk effective
sample size, split by whether the coefficient is truly zero or truly nonzero, together
with the median number of divergent transitions per fit. The Group-$R^2$ row is
\texttt{R2-c} on the concentrated scenarios and \texttt{R2-d} on the distributed ones.
Divergences are reported for the Group-$R^2$ models only, since the competing models are
Gibbs samplers.}
\label{tab:external_diag_low}
\begin{tabular}{llcccccc}
\toprule
 & & \multicolumn{2}{c}{$\widehat{R}$} & \multicolumn{2}{c}{ESS (bulk)} & \\
\cmidrule(lr){3-4}\cmidrule(lr){5-6}
Scenario & Model & zero & nonzero & zero & nonzero & Divergences \\
\midrule
Con & Group-$R^2$ & 1.003 & 1.002 & 2,184 & 2,083 & 0 \\
 & BGL-SS & 1.004 & 1.002 & 1,081 & 999 & -- \\
 & BSGS-SS & 1.006 & 1.003 & 775 & 638 & -- \\
 & GHS & 1.051 & 1.028 & 107 & 123 & -- \\
 & GRASP & 1.179 & 1.090 & 40 & 69 & -- \\
 & $g$-prior & 1.155 & 1.106 & 7 & 12 & -- \\
\addlinespace
Dist & Group-$R^2$ & 1.003 & 1.002 & 2,375 & 3,435 & 0 \\
 & BGL-SS & 1.003 & 1.002 & 1,170 & 1,625 & -- \\
 & BSGS-SS & 1.004 & 1.006 & 922 & 442 & -- \\
 & GHS & 1.101 & 1.012 & 176 & 187 & -- \\
 & GRASP & 1.397 & 1.068 & 117 & 37 & -- \\
 & $g$-prior & 1.066 & 1.051 & 24 & 25 & -- \\
\addlinespace
Rcoef & Group-$R^2$ & 1.003 & 1.002 & 2,246 & 2,177 & 0 \\
 & BGL-SS & 1.004 & 1.003 & 1,064 & 1,033 & -- \\
 & BSGS-SS & 1.006 & 1.005 & 786 & 543 & -- \\
 & GHS & 1.071 & 1.026 & 133 & 114 & -- \\
 & GRASP & 1.280 & 1.092 & 53 & 41 & -- \\
 & $g$-prior & 1.086 & 1.109 & 16 & 11 & -- \\
\addlinespace
Rcon & Group-$R^2$ & 1.003 & 1.002 & 2,095 & 2,064 & 0 \\
 & BGL-SS & 1.005 & 1.003 & 834 & 810 & -- \\
 & BSGS-SS & 1.009 & 1.005 & 487 & 460 & -- \\
 & GHS & 1.061 & 1.033 & 105 & 120 & -- \\
 & GRASP & 1.292 & 1.178 & 30 & 43 & -- \\
 & $g$-prior & 1.121 & 1.074 & 10 & 17 & -- \\
\addlinespace
Rdist & Group-$R^2$ & 1.003 & 1.002 & 2,114 & 3,339 & 0 \\
 & BGL-SS & 1.002 & 1.002 & 1,147 & 1,649 & -- \\
 & BSGS-SS & 1.004 & 1.006 & 956 & 457 & -- \\
 & GHS & 1.107 & 1.008 & 156 & 292 & -- \\
 & GRASP & 1.353 & 1.036 & 93 & 53 & -- \\
 & $g$-prior & 1.072 & 1.041 & 22 & 33 & -- \\
\bottomrule
\end{tabular}
\end{table}

\begin{table}[!ht]
\footnotesize
\centering
\setlength{\tabcolsep}{4pt}
\caption{Sampling diagnostics for the high-dimensional scenario ($p = 500$, $n = 200$).
Entries are defined as in Table~\ref{tab:external_diag_low}. BGL-SS and GRASP are omitted
from this block, as described in Section~\ref{subsubsec:external_elpd}.}
\label{tab:external_diag_high}
\begin{tabular}{llcccccc}
\toprule
 & & \multicolumn{2}{c}{$\widehat{R}$} & \multicolumn{2}{c}{ESS (bulk)} & \\
\cmidrule(lr){3-4}\cmidrule(lr){5-6}
Scenario & Model & zero & nonzero & zero & nonzero & Divergences \\
\midrule
Con & Group-$R^2$ & 1.004 & 1.003 & 2,655 & 2,965 & 0 \\
 & BSGS-SS & 1.007 & 1.005 & 1,098 & 1,260 & -- \\
 & GHS & 1.162 & 1.115 & 38 & 79 & -- \\
 & $g$-prior & 1.349 & 1.132 & 3 & 7 & -- \\
\addlinespace
Dist & Group-$R^2$ & 1.004 & 1.002 & 2,857 & 3,090 & 0 \\
 & BSGS-SS & 1.006 & 1.009 & 568 & 416 & -- \\
 & GHS & 1.232 & 1.016 & 88 & 107 & -- \\
 & $g$-prior & 1.031 & 1.058 & 31 & 19 & -- \\
\addlinespace
Rcoef & Group-$R^2$ & 1.004 & 1.003 & 2,616 & 2,628 & 0 \\
 & BSGS-SS & 1.007 & 1.006 & 1,098 & 1,074 & -- \\
 & GHS & 1.160 & 1.120 & 45 & 48 & -- \\
 & $g$-prior & 1.263 & 1.211 & 3 & 4 & -- \\
\addlinespace
Rcon & Group-$R^2$ & 1.004 & 1.003 & 2,180 & 2,008 & 0 \\
 & BSGS-SS & 1.011 & 1.008 & 659 & 567 & -- \\
 & GHS & 1.167 & 1.124 & 29 & 37 & -- \\
 & $g$-prior & 1.231 & 1.117 & 4 & 7 & -- \\
\addlinespace
Rdist & Group-$R^2$ & 1.004 & 1.002 & 2,400 & 2,893 & 0 \\
 & BSGS-SS & 1.005 & 1.007 & 651 & 448 & -- \\
 & GHS & 1.217 & 1.009 & 92 & 248 & -- \\
 & $g$-prior & 1.027 & 1.024 & 41 & 39 & -- \\
\bottomrule
\end{tabular}
\end{table}

\clearpage
\subsection{Calibration of the GIGG hyperparameters}
\label{subsec:gigg_calibration}

This section first shows the prior on $R^2$ that the GIGG implies before calibration. It then gives the computational detail behind the calibration defined in
\eqref{eq:gigg_calibration} and reports the results relegated from
Section~\ref{subsec:gigg_comparison}.

\begin{figure}[t!]
    \centering
    \includegraphics[width=\linewidth]{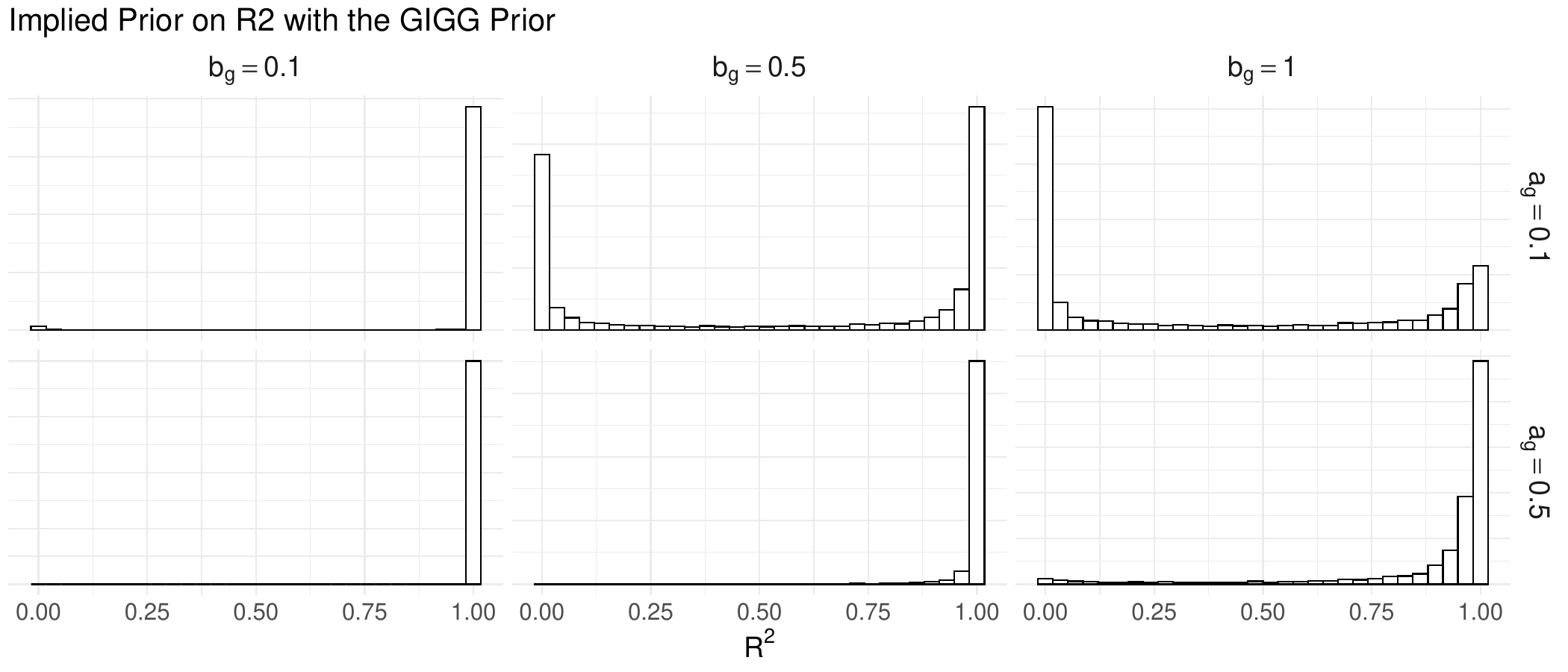}
    \caption{Implied prior distributions on $R^2$ under the GIGG prior for various combinations of $a_g, b_g$. We obtain the distributions via Monte Carlo and consider $G= 10$ groups, $p = 50$ covariates and we have fixed $\sigma = 1$. The hyperparameter $a_g$ governs the strength of thresholding, since it controls the spike at zero in the marginal distribution of $b_{gl}$. In contrast, $b_g$ regulates the degree of shrinkage dependence within each group, influencing how strongly individual coefficients borrow strength from one another. The induced $R^2$ distribution varies substantially with these parameters.}
    \label{fig:gigg_implied_r2}
\end{figure}

\paragraph{Implied prior on $R^2$.}
Any prior on the coefficients implies a prior on $R^2$. Under the GIGG prior, this prior follows from $(a_g, b_g)$ through \eqref{eq:gigg_implied_r2} and has no closed form. Figure~\ref{fig:gigg_implied_r2} shows it for several values of $(a_g, b_g)$. For most values, almost all of its mass lies near one. For $a_g = 0.1$ and $b_g \geq 0.5$, it has modes at both zero and one. Moderate values of $R^2$ receive little mass in all cases. A target prior on $R^2$, such as the prior of a Group-R2 model, can therefore be matched only numerically, which is the purpose of the calibration below. Under the Group-R2 prior, the prior on $R^2$ is set directly.

\paragraph{Objective.}
The distribution of $R^2_{\text{GIGG}}$ in \eqref{eq:gigg_implied_r2} is available only by simulation. We draw $\tau^2$, $\gamma_g^2$ and $\lambda_{gl}^2$ from the GIGG hierarchy and form $V$ directly, leaving $b$ marginalized rather than drawing it and using $\sum_{gl} b_{gl}^2$. While the  two agree in distribution, the former removes a layer of Monte Carlo noise.  In one dimension the optimal transport plan is the monotone rearrangement, so for two samples of equal size the objective in \eqref{eq:gigg_calibration} reduces to a mean absolute difference of order statistics,
\begin{equation}
  W_1 = \frac{1}{S}\sum_{s=1}^{S}
    \left| R^2_{(s)} - \tilde{R}^2_{(s)} \right|,
\end{equation}
with $R^2_{(s)}$ the sorted draws from Equation~\eqref{eq:gigg_implied_r2} and $\tilde{R}^2_{(s)}$ the sorted draws from the target $\betadist(G a_G, a_2)$.

\paragraph{Optimization.}
Every candidate $(a_g, b_g)$ is evaluated on the same fixed matrix of uniform random numbers, pushed through the relevant quantile functions. Using common random numbers in this way makes the objective a deterministic and smooth function of $(a_g, b_g)$ rather than a noisy one, which is what allows a derivative free optimizer to converge. We minimize over the box $a_g \in [10^{-3}, 0.5]$ and $b_g \in [10^{-3}, 10]$. The upper bound on $a_g$ keeps the pole of the marginal prior at the origin, which is the property that makes the GIGG a shrinkage prior, and the bound on $b_g$ is a numerical guard. The search proceeds in two stages, a coarse grid over the box followed by a Nelder-Mead refinement from the best grid point, carried out on a logit reparameterization so that the constraints are respected without a penalty term. The grid stage is retained because the objective is flat over large parts of the box, since once $b_g$ is small enough the implied $R^2$ is pinned near one whatever $a_g$ does.

\paragraph{Reported fit.}
The search uses $S = 4000$ draws. The distance reported in
Table~\ref{tab:gigg_calibration} is recomputed at the returned optimum on
$S = 20{,}000$ independently seeded draws, so it is an out of sample measure of fit
rather than the minimized value.

\begin{figure}[!ht]%
\includegraphics[width=\linewidth]{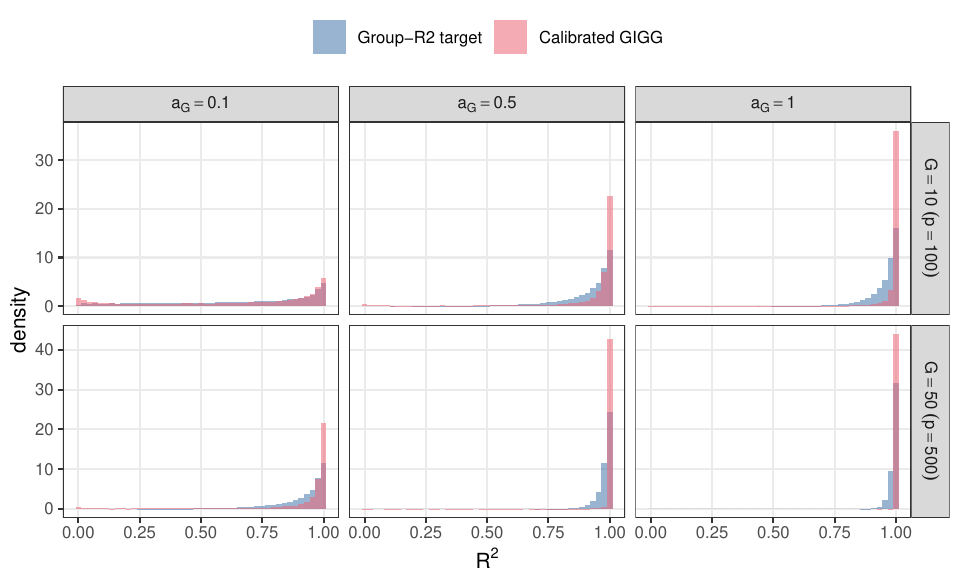}
  \caption{ \textbf{Implied prior on $R^2$ under the calibrated GIGG against the Group-R2 target.} The target is the $\betadist(G a_G, a_2)$ prior that the Group-R2 model places on $R^2$ directly. The calibrated GIGG uses the shapes of Table~\ref{tab:gigg_calibration}, which are the values fitted in the study. Both are drawn from $20{,}000$ prior samples.}
  \label{fig:gigg_implied_r2_calibration}
\end{figure}

Figure~\ref{fig:gigg_implied_r2_calibration} shows the resulting distributions. At
$a_G = 0.1$ and $G = 10$ the target retains mass across the interior of the unit
interval and the calibrated GIGG follows it reasonably closely. In every other panel both distributions concentrate near one, and the calibrated GIGG is the more
concentrated of the two. This is the sense in which the small $W_1$ values of
Table~\ref{tab:gigg_calibration} record two distributions meeting at the boundary
rather than agreeing in shape.

\begin{figure}[!ht]%
\includegraphics[width=\linewidth]{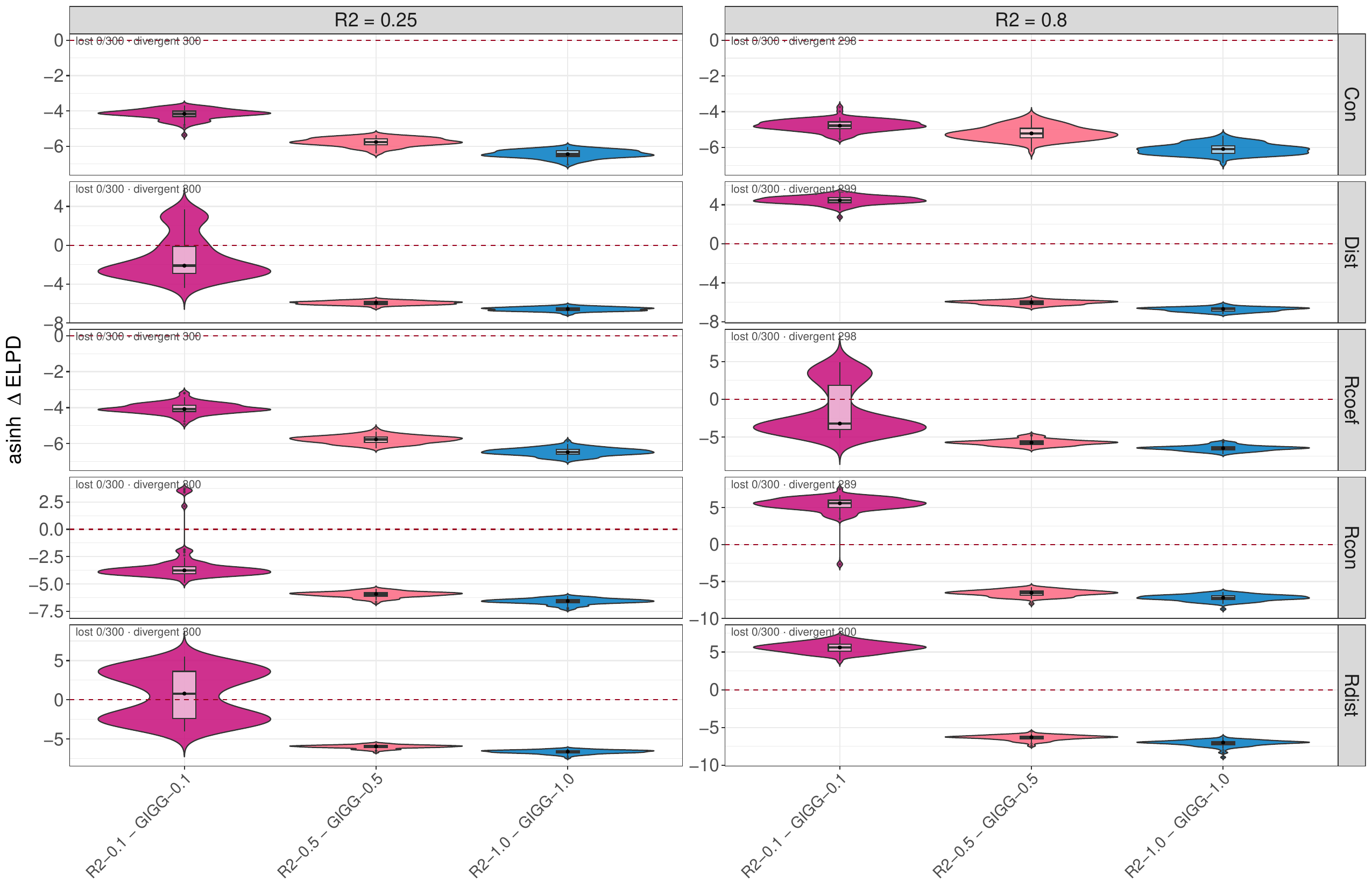}
  \caption{ \textbf{$\Delta \elpd$ (asinh-transformed) against the calibrated GIGG for the high-dimensional scenario ($p = 500$, $n = 200$)}. Values above the line indicate better predictive performance of the Group-R2 prior. ``Con” denotes the concentrated signal, ``Dist” the distributed signal, ``Rcon” and ``Rdist” their random counterparts, and ``Rcoef” the random coefficients setting.}
  \label{fig:gigg_elpd_high_dimension}
\end{figure}

\begin{figure}[!ht]%
\includegraphics[width=\linewidth]{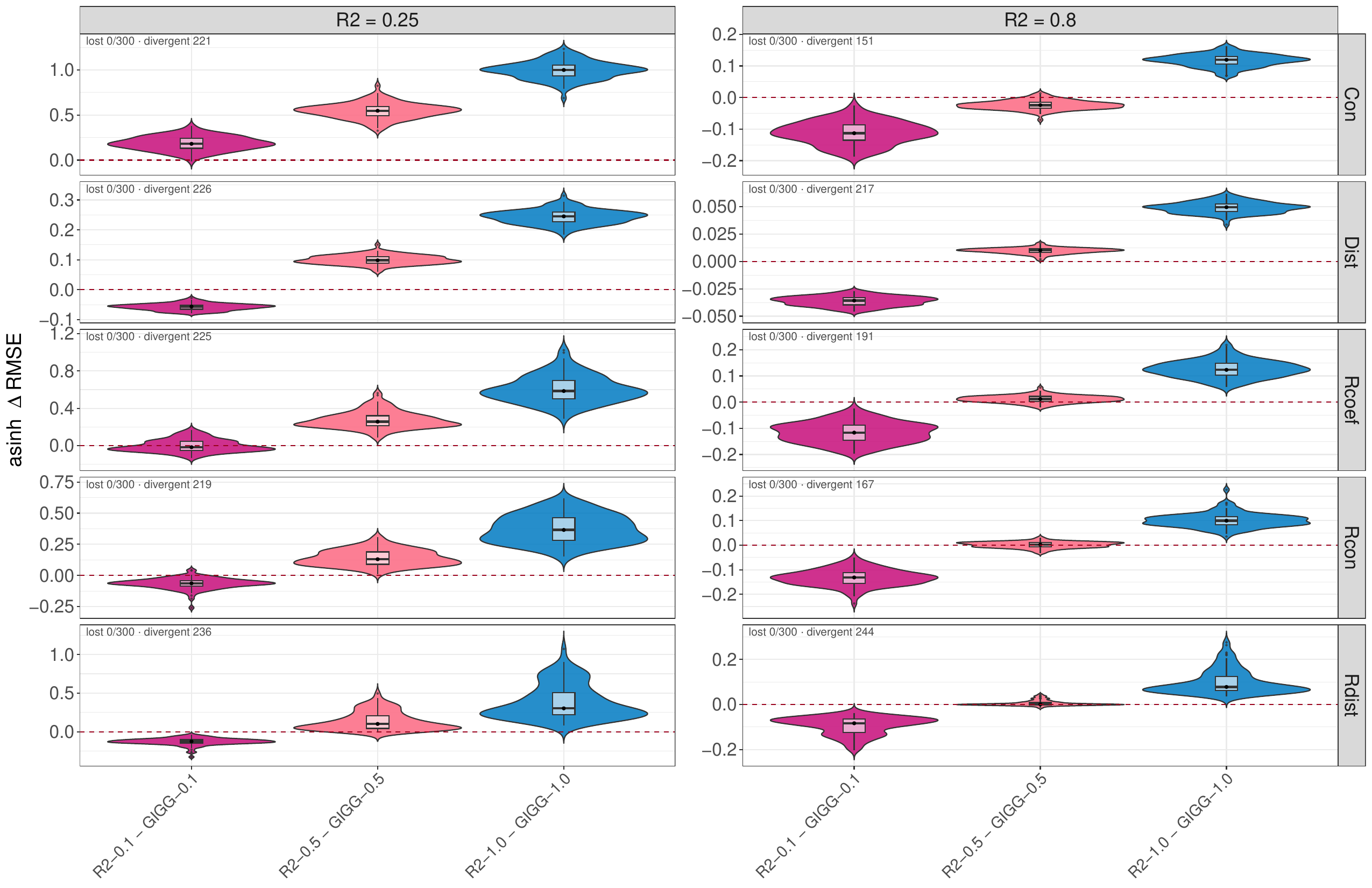}
  \caption{ \textbf{$\Delta \rmse$ against the calibrated GIGG for the lower-dimensional scenario ($p = 100$, $n = 200$).} Values below the line indicate better parameter recovery of the Group-R2 prior. ``Con” denotes the concentrated signal, ``Dist” the distributed signal, ``Rcon” and ``Rdist” their random counterparts, and ``Rcoef” the random coefficients setting.}
  \label{fig:gigg_rmse_low_dimension}
\end{figure}

\begin{figure}[!ht]%
\includegraphics[width=\linewidth]{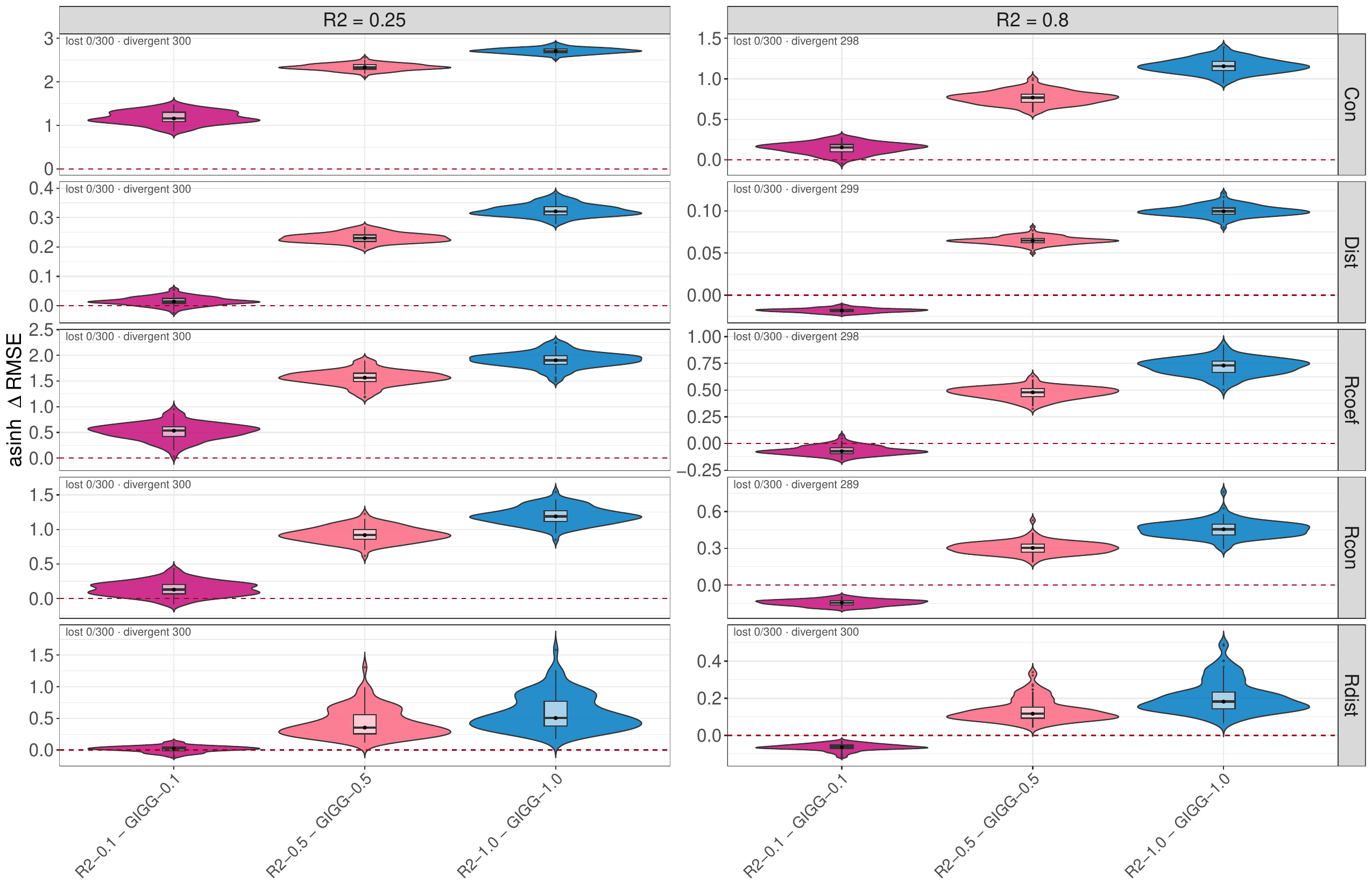}
  \caption{ \textbf{$\Delta \rmse$ against the calibrated GIGG for the high-dimensional scenario ($p = 500$, $n = 200$).} Values below the line indicate better parameter recovery of the Group-R2 prior. ``Con” denotes the concentrated signal, ``Dist” the distributed signal, ``Rcon” and ``Rdist” their random counterparts, and ``Rcoef” the random coefficients setting.}
  \label{fig:gigg_rmse_high_dimension}
\end{figure}

\paragraph{Relegated results.}
Figure~\ref{fig:gigg_elpd_high_dimension} gives the predictive comparison for the
high-dimensional setting, and Figures~\ref{fig:gigg_rmse_low_dimension} and
\ref{fig:gigg_rmse_high_dimension} give parameter recovery over all coefficients for both settings. The ordering across $a_G$ is the same as in Figure~\ref{fig:gigg_elpd_low_dimension} in every case.

\begin{figure}[!ht]%
\includegraphics[width=\linewidth]{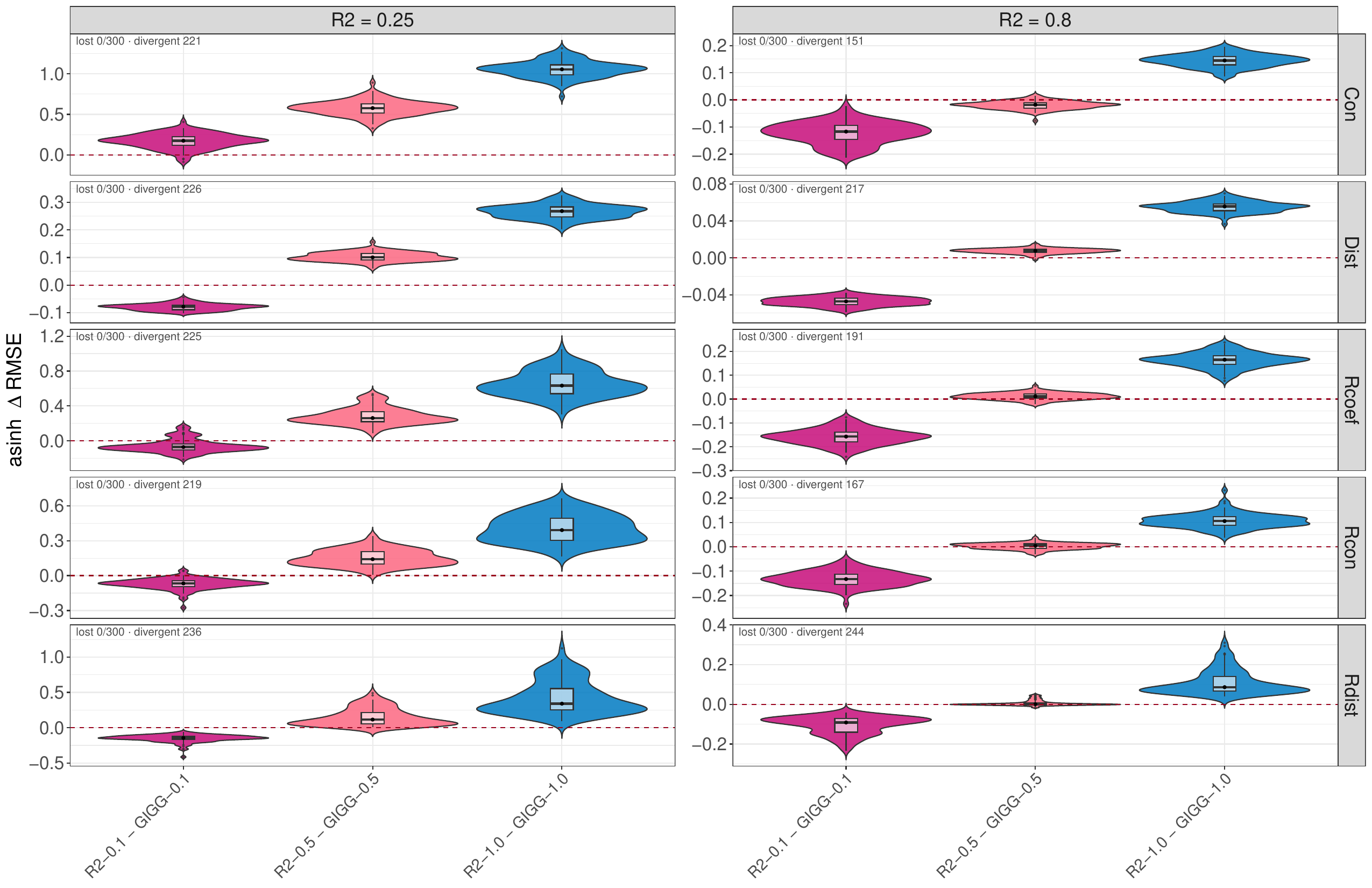}
  \caption{ \textbf{$\Delta \rmse$ on the truly zero coefficients against the calibrated GIGG for the lower-dimensional scenario ($p = 100$, $n = 200$).} Values below the line indicate better parameter recovery of the Group-R2 prior. ``Con” denotes the concentrated signal, ``Dist” the distributed signal, ``Rcon” and ``Rdist” their random counterparts, and ``Rcoef” the random coefficients setting.}
  \label{fig:gigg_rmse_zero_low}
\end{figure}

\begin{figure}[!ht]%
\includegraphics[width=\linewidth]{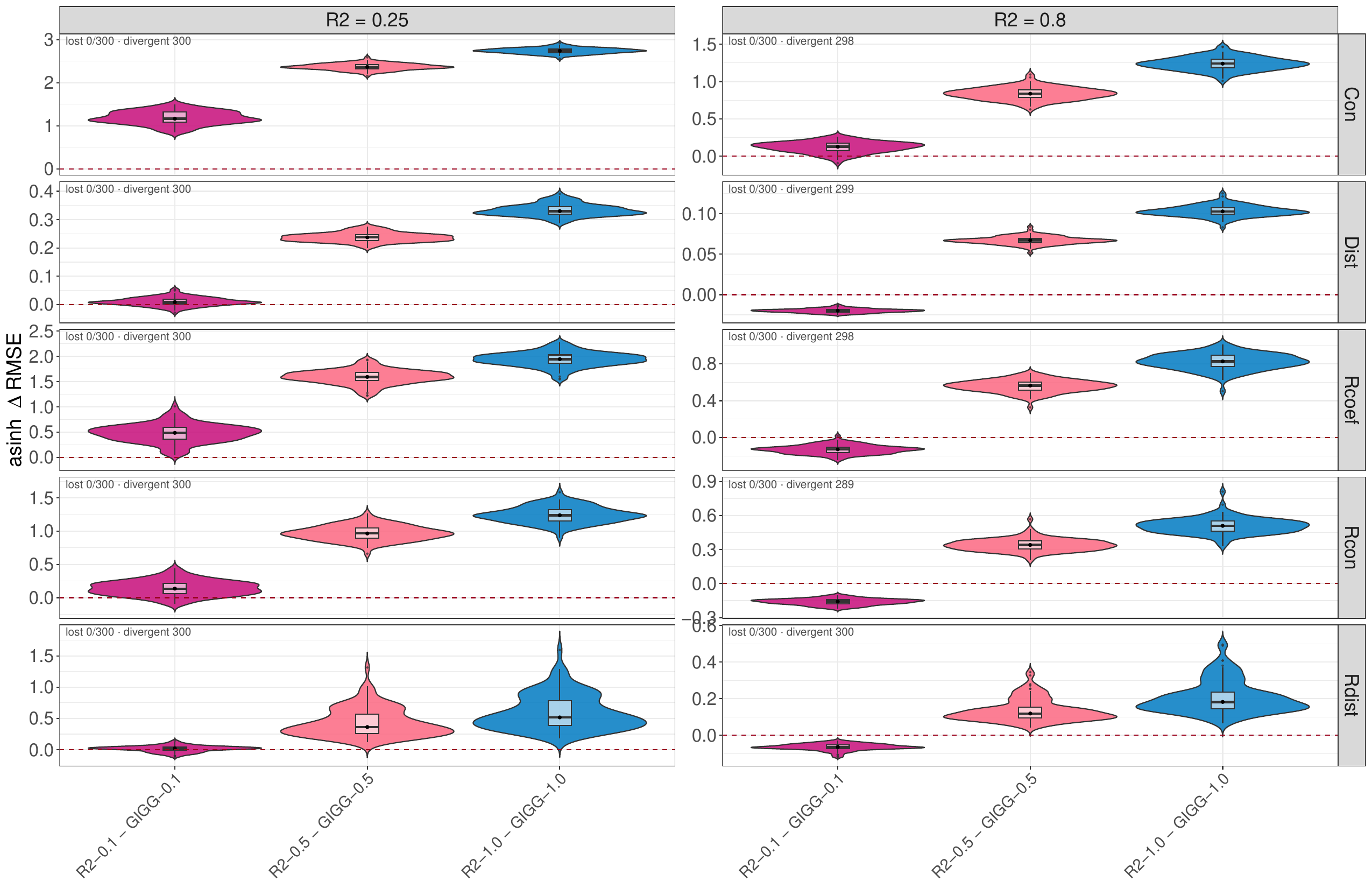}
  \caption{ \textbf{$\Delta \rmse$ on the truly zero coefficients against the calibrated GIGG for the high-dimensional scenario ($p = 500$, $n = 200$).} Entries are defined as in Figure~\ref{fig:gigg_rmse_zero_low}.}
  \label{fig:gigg_rmse_zero_high}
\end{figure}

\begin{figure}[!ht]%
\includegraphics[width=\linewidth]{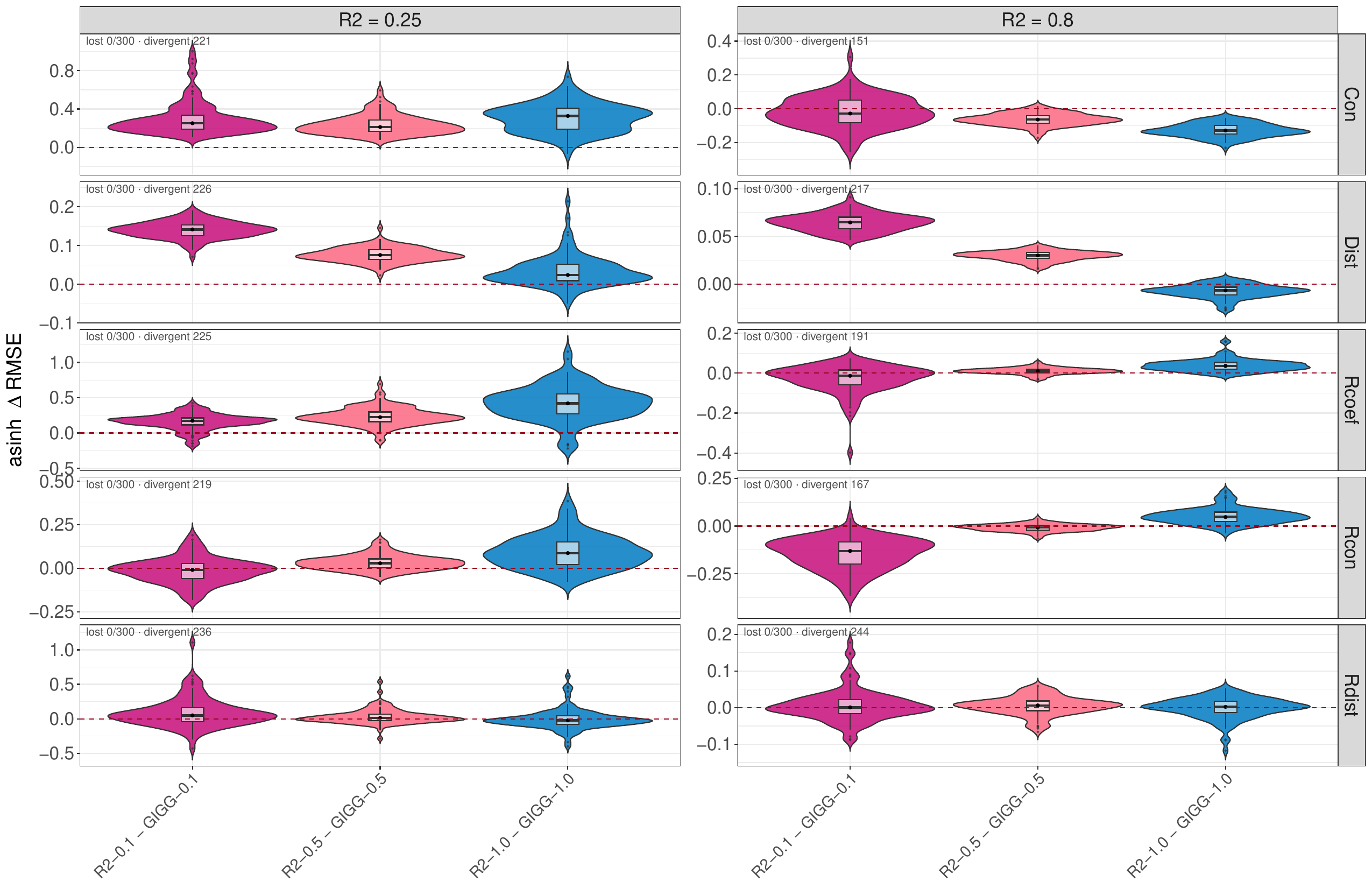}
  \caption{ \textbf{$\Delta \rmse$ on the truly nonzero coefficients against the calibrated GIGG for the lower-dimensional scenario ($p = 100$, $n = 200$).} Entries are defined as in Figure~\ref{fig:gigg_rmse_zero_low}.}
  \label{fig:gigg_rmse_nonzero_low}
\end{figure}

\begin{figure}[!ht]%
\includegraphics[width=\linewidth]{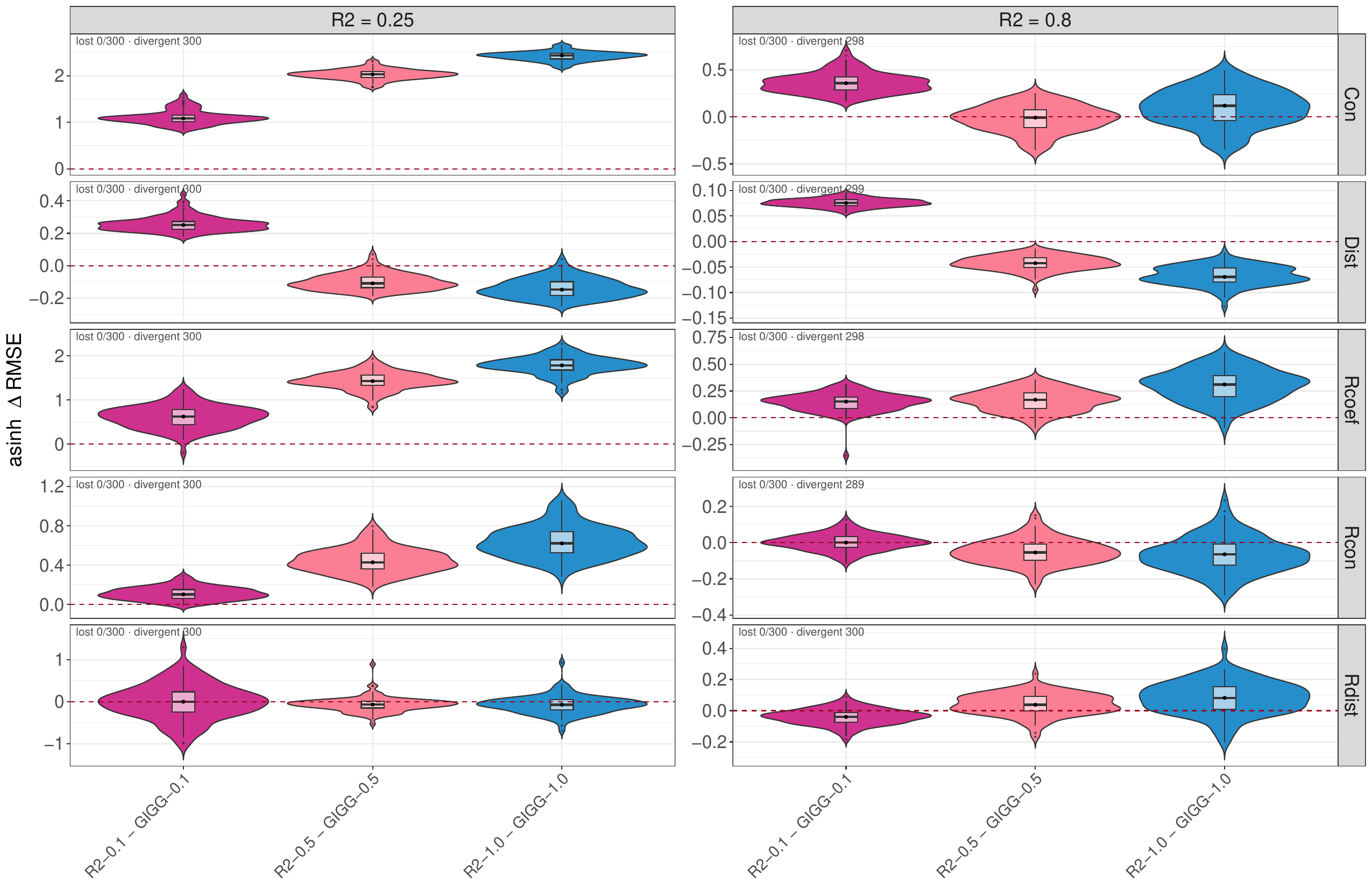}
  \caption{ \textbf{$\Delta \rmse$ on the truly nonzero coefficients against the calibrated GIGG for the high-dimensional scenario ($p = 500$, $n = 200$).} Entries are defined as in Figure~\ref{fig:gigg_rmse_zero_low}.}
  \label{fig:gigg_rmse_nonzero_high}
\end{figure}

Figures~\ref{fig:gigg_rmse_zero_low} to~\ref{fig:gigg_rmse_nonzero_high} split the same comparison by coefficient type, separating the coefficients that are truly zero from those that are truly nonzero. The two types behave differently. On the truly zero coefficients the aggregate ordering is reproduced and the differences grow with both $a_G$ and the dimension, with median $\Delta \rmse$ of $0.030$ and $0.210$ for $a_G = 0.5$ and $a_G = 1$ at $p = 100$ and $0.555$ and $0.856$ at $p = 500$. On the truly nonzero coefficients the corresponding medians are $0.026$ and $0.025$ at $p = 100$ and $0.039$ and $0.104$ at $p = 500$, so the difference stays small irrespective of $a_G$ and of the dimension. Most of the aggregate difference in recovery is therefore attributable to the coefficients that are truly zero, which is consistent with the account of Section~\ref{subsec:gigg_comparison}, since the allocation of the variance across groups determines how strongly the null coefficients in inactive groups are shrunk. The exception is $a_G = 0.1$ at $p = 100$, where the Group-R2 prior recovers the zero coefficients better in $88\%$ of replications, with a median $\Delta \rmse$ of $-0.090$, but the nonzero coefficients slightly worse, with a median of $0.046$.

\subsection{Sampling diagnostics for the GIGG comparison}
\label{subsec:gigg_diagnostics}

Tables~\ref{tab:gigg_diag_low} and~\ref{tab:gigg_diag_high} report convergence
diagnostics for the six models entering the comparison of
Section~\ref{subsec:gigg_comparison}, separately for the truly zero and the truly
nonzero coefficients and for each data generating scenario. For every fit we take the largest $\widehat{R}$ and the smallest bulk effective sample size over the coefficients of the relevant type, and report the median of those quantities over the $100$ replications. The last column gives the median number of divergent transitions per fit, out of the $4{,}000$ post-warmup draws, which is a property of the fit rather than of a coefficient.

Three points stand out. First, the dependence on $a_G$ runs in opposite directions for the two priors. Mixing in the GIGG deteriorates as $a_G$ increases, whereas in the Group-R2 prior it is $a_G = 0.1$ that is the difficult case and the larger values that sample cleanly. Second, the deterioration in the GIGG is confined to the high-dimensional setting. At $p = 100$ all six models attain $\widehat{R}$ below $1.01$
and effective sample sizes above $1{,}400$, and the GIGG in fact mixes somewhat better
than its Group-R2 counterpart. At $p = 500$ the picture changes, and
\texttt{GIGG-}$1.0$ reaches a median $\widehat{R}$ of up to $1.064$ with a median
effective sample size near $120$ on the zero coefficients, values at which the posterior
cannot be regarded as adequately explored. Third, divergent transitions are not aligned
with the $\widehat{R}$ and effective sample size ordering. \texttt{R2-}$0.1$ records the
largest counts anywhere in the table, up to a median of $130$ per fit, while its
$\widehat{R}$ and effective sample sizes remain acceptable, and \texttt{GIGG-}$1.0$ at
$p = 100$ records fewer than $10$ per fit. The models with the weakest diagnostics are
therefore \texttt{GIGG-}$1.0$ and \texttt{GIGG-}$0.5$ at $p = 500$, which are also the
two configurations with the largest apparent predictive advantage over the Group-R2
prior in Section~\ref{subsec:gigg_comparison}, and the differences reported there should
be read with that in mind.

\begin{table}[!ht]
\footnotesize
\centering
\setlength{\tabcolsep}{4pt}
\caption{Sampling diagnostics for the lower-dimensional scenario ($p = 100$, $n = 200$).
Median over replications of the per-fit maximum $\widehat{R}$ and minimum bulk effective
sample size, split by whether the coefficient is truly zero or truly nonzero, together
with the median number of divergent transitions per fit out of $4{,}000$ post-warmup draws.}
\label{tab:gigg_diag_low}
\begin{tabular}{llcccccc}
\toprule
 & & \multicolumn{2}{c}{$\widehat{R}$} & \multicolumn{2}{c}{ESS (bulk)} & \\
\cmidrule(lr){3-4}\cmidrule(lr){5-6}
Scenario & Model & zero & nonzero & zero & nonzero & Divergences \\
\midrule
Con & \texttt{R2-0.1} & 1.007 & 1.004 & 1,492 & 1,406 & 14.5 \\
 & \texttt{GIGG-0.1} & 1.004 & 1.002 & 2,824 & 2,659 & 0 \\
 & \texttt{R2-0.5} & 1.003 & 1.002 & 1,913 & 1,856 & 0 \\
 & \texttt{GIGG-0.5} & 1.003 & 1.002 & 2,439 & 2,429 & 0 \\
 & \texttt{R2-1.0} & 1.003 & 1.002 & 1,968 & 2,031 & 0 \\
 & \texttt{GIGG-1.0} & 1.003 & 1.002 & 1,909 & 1,959 & 9 \\
\addlinespace
Dist & \texttt{R2-0.1} & 1.005 & 1.002 & 1,853 & 2,357 & 127 \\
 & \texttt{GIGG-0.1} & 1.004 & 1.003 & 2,349 & 5,188 & 0 \\
 & \texttt{R2-0.5} & 1.003 & 1.002 & 1,740 & 2,406 & 0 \\
 & \texttt{GIGG-0.5} & 1.003 & 1.002 & 2,079 & 3,836 & 0 \\
 & \texttt{R2-1.0} & 1.003 & 1.002 & 1,815 & 2,329 & 0 \\
 & \texttt{GIGG-1.0} & 1.004 & 1.002 & 1,893 & 2,740 & 7 \\
\addlinespace
Rcoef & \texttt{R2-0.1} & 1.005 & 1.003 & 1,760 & 1,919 & 113.5 \\
 & \texttt{GIGG-0.1} & 1.004 & 1.003 & 2,626 & 3,446 & 0 \\
 & \texttt{R2-0.5} & 1.003 & 1.002 & 1,830 & 2,109 & 0 \\
 & \texttt{GIGG-0.5} & 1.003 & 1.002 & 2,252 & 2,868 & 0 \\
 & \texttt{R2-1.0} & 1.003 & 1.002 & 1,921 & 2,153 & 0 \\
 & \texttt{GIGG-1.0} & 1.003 & 1.002 & 1,807 & 2,055 & 8 \\
\addlinespace
Rcon & \texttt{R2-0.1} & 1.006 & 1.004 & 1,530 & 1,553 & 44 \\
 & \texttt{GIGG-0.1} & 1.004 & 1.002 & 2,310 & 2,471 & 0 \\
 & \texttt{R2-0.5} & 1.003 & 1.002 & 1,800 & 1,891 & 0 \\
 & \texttt{GIGG-0.5} & 1.003 & 1.002 & 2,204 & 2,486 & 0 \\
 & \texttt{R2-1.0} & 1.003 & 1.002 & 1,907 & 2,009 & 0 \\
 & \texttt{GIGG-1.0} & 1.003 & 1.002 & 1,747 & 1,909 & 10 \\
\addlinespace
Rdist & \texttt{R2-0.1} & 1.005 & 1.002 & 1,707 & 2,301 & 129 \\
 & \texttt{GIGG-0.1} & 1.004 & 1.002 & 2,106 & 4,008 & 0 \\
 & \texttt{R2-0.5} & 1.004 & 1.002 & 1,544 & 2,415 & 0 \\
 & \texttt{GIGG-0.5} & 1.004 & 1.002 & 1,833 & 3,437 & 0 \\
 & \texttt{R2-1.0} & 1.003 & 1.002 & 1,800 & 2,344 & 0 \\
 & \texttt{GIGG-1.0} & 1.004 & 1.002 & 1,700 & 2,532 & 9 \\
\bottomrule
\end{tabular}
\end{table}

\begin{table}[!ht]
\footnotesize
\centering
\setlength{\tabcolsep}{4pt}
\caption{Sampling diagnostics for the high-dimensional scenario ($p = 500$, $n = 200$).
Entries are defined as in Table~\ref{tab:gigg_diag_low}.}
\label{tab:gigg_diag_high}
\begin{tabular}{llcccccc}
\toprule
 & & \multicolumn{2}{c}{$\widehat{R}$} & \multicolumn{2}{c}{ESS (bulk)} & \\
\cmidrule(lr){3-4}\cmidrule(lr){5-6}
Scenario & Model & zero & nonzero & zero & nonzero & Divergences \\
\midrule
Con & \texttt{R2-0.1} & 1.009 & 1.006 & 875 & 1,333 & 117 \\
 & \texttt{GIGG-0.1} & 1.005 & 1.003 & 2,903 & 3,322 & 0 \\
 & \texttt{R2-0.5} & 1.005 & 1.003 & 1,439 & 1,863 & 0 \\
 & \texttt{GIGG-0.5} & 1.016 & 1.012 & 536 & 814 & 33 \\
 & \texttt{R2-1.0} & 1.004 & 1.003 & 1,556 & 2,001 & 0 \\
 & \texttt{GIGG-1.0} & 1.064 & 1.045 & 117 & 221 & 73 \\
\addlinespace
Dist & \texttt{R2-0.1} & 1.007 & 1.002 & 1,167 & 2,101 & 129.5 \\
 & \texttt{GIGG-0.1} & 1.005 & 1.002 & 2,762 & 4,189 & 0 \\
 & \texttt{R2-0.5} & 1.005 & 1.002 & 1,344 & 2,046 & 0 \\
 & \texttt{GIGG-0.5} & 1.009 & 1.003 & 1,084 & 1,460 & 29 \\
 & \texttt{R2-1.0} & 1.004 & 1.002 & 1,502 & 2,071 & 0 \\
 & \texttt{GIGG-1.0} & 1.013 & 1.006 & 637 & 710 & 87 \\
\addlinespace
Rcoef & \texttt{R2-0.1} & 1.008 & 1.008 & 1,002 & 1,136 & 119 \\
 & \texttt{GIGG-0.1} & 1.004 & 1.004 & 2,666 & 2,971 & 0 \\
 & \texttt{R2-0.5} & 1.005 & 1.004 & 1,428 & 1,675 & 0 \\
 & \texttt{GIGG-0.5} & 1.011 & 1.011 & 679 & 742 & 32 \\
 & \texttt{R2-1.0} & 1.004 & 1.003 & 1,558 & 1,766 & 0 \\
 & \texttt{GIGG-1.0} & 1.049 & 1.049 & 149 & 188 & 70.5 \\
\addlinespace
Rcon & \texttt{R2-0.1} & 1.010 & 1.007 & 822 & 860 & 122 \\
 & \texttt{GIGG-0.1} & 1.005 & 1.003 & 2,171 & 2,263 & 0 \\
 & \texttt{R2-0.5} & 1.005 & 1.004 & 1,152 & 1,293 & 0 \\
 & \texttt{GIGG-0.5} & 1.016 & 1.013 & 516 & 486 & 33 \\
 & \texttt{R2-1.0} & 1.004 & 1.003 & 1,326 & 1,471 & 0 \\
 & \texttt{GIGG-1.0} & 1.053 & 1.044 & 171 & 149 & 71 \\
\addlinespace
Rdist & \texttt{R2-0.1} & 1.008 & 1.002 & 1,038 & 2,128 & 129.5 \\
 & \texttt{GIGG-0.1} & 1.005 & 1.002 & 2,226 & 3,273 & 0 \\
 & \texttt{R2-0.5} & 1.005 & 1.002 & 1,263 & 2,083 & 0 \\
 & \texttt{GIGG-0.5} & 1.009 & 1.003 & 957 & 1,487 & 33 \\
 & \texttt{R2-1.0} & 1.004 & 1.002 & 1,365 & 1,971 & 0 \\
 & \texttt{GIGG-1.0} & 1.018 & 1.007 & 522 & 672 & 81.5 \\
\bottomrule
\end{tabular}
\end{table}

\end{appendix}

\begin{acks}[Acknowledgments]

Funded by Deutsche Forschungsgemeinschaft (DFG, German Research Foundation) Project 500663361. Javier Enrique Aguilar and Paul-Christian Bürkner acknowledge the computing time provided on the Linux HPC cluster at Technical University Dortmund (LiDO3), partially funded in the course of the Large-Scale Equipment Initiative by DFG Project 271512359. Paul-Christian Bürkner further acknowledges support of the Deutsche Forschungsgemeinschaft (DFG, German Research Foundation) via the Collaborative Research Center 391 (Spatio-Temporal Statistics for the Transition of Energy and Transport) – 520388526. Javier Enrique Aguilar further acknowledges travel support from the European Union’s Horizon 2020 research and innovation programme under grant agreement No 951847. David Kohns acknowledges the computational resources provided by the Aalto Science-IT project, and the support of the Research Council of Finland Flagship programme: Finnish Center for Artificial Intelligence, Research Council of Finland project (340721), and the Finnish Foundation for Technology Promotion.

\end{acks}



\section*{Statement on AI use}
We used OpenAI's Codex for proofreading, language editing, and improving mathematical exposition. This assistance included improving the clarity, organization, and readability of the mathematical arguments. The authors directed the research and retain full responsibility for the method, mathematical results, proofs, references, and final manuscript.

\bibliographystyle{imsart-nameyear}
\bibliography{bib/refs}

\end{document}